\documentclass[oneside,a4paper,11pt]{article}

\usepackage[margin=1in]{geometry}  
\usepackage{setspace}              

\usepackage{amsmath}
\usepackage{amssymb}
\usepackage{comment}
\usepackage{caption}
\usepackage{graphicx}
\usepackage{natbib}
\usepackage{subcaption}
\usepackage{color}
\usepackage{float}
\usepackage{hyperref}
\usepackage{booktabs}
\usepackage{ctable}
\usepackage{colortbl}
\usepackage{xcolor}

\newtheorem{Definition}{\bf \large Definition}[section]
\newtheorem{Theorem}{\bf \large Theorem}[section]
\newtheorem{Lemma}[Theorem]{\bf \large Lemma}
\newtheorem{Corollary}[Theorem]{\bf \large Corollary}
\newtheorem{Remark}{\bf \large Remark}[section]

\newcommand{\proof}{\noindent \textbf{Proof.} \hspace {0.1cm}}
\renewcommand{\arraystretch}{1.2}
\newcommand{\pr}{{\mathbb P}}
\newcommand{\E}{{\mathbb E}}
\newcommand{\dif}{\,\mathrm{d}}

\DeclareMathOperator*{\argmax}{argmax}
\newcommand{\keywords}[1]{\vspace{1em}\noindent\textbf{Keywords:} #1}

\usepackage{fancyhdr}      
\begin{document}
\title{On effects of present-bias on carbon emission patterns\\ towards a net zero target}

\numberwithin{equation}{section}
\author{
  Hansj\"org Albrecher\footnote{Department of Actuarial Science, Faculty of Business and Economics, and Expertise Center for Climate Extremes, University of Lausanne, and Swiss Finance Institute, Switzerland. Email: \texttt{hansjoerg.albrecher@unil.ch}} 
  \and
  Jinxia Zhu\footnote{School of Risk and Actuarial Studies, University of New South Wales, Sydney, Australia. 
  Email: \texttt{jinxia.zhu@unsw.edu.au}. (Corresponding author).}
}

\date{}
\maketitle
\begin{abstract}
This paper explores the optimal policy for using an allocated carbon emission budget over time with the objective to maximize profit, by explicitly taking into account present-biased 
preferences of decision-makers, accounting for time-inconsistent preferences. The setup can be adapted to be applicable for either a (present-biased) individual or also for a company which seeks a balance between production and emission schedules. In particular, we use and extend stochastic control techniques developed for optimal dividend strategies in insurance risk theory for the present purpose. The approach enables a quantitative analysis to assess the effects of present-bias, of sustainability awareness, and the efficiency of a potential carbon tax in a simplified model. In some numerical implementations, we illustrate in what way a higher degree of present-bias leads to excess emission patterns, while placing greater emphasis on sustainability reduces carbon emissions. Furthermore, we show that for low levels of carbon tax, its increase has a positive effect on curbing emissions, while beyond a certain threshold that marginal impact gets considerably weaker.
\end{abstract}


\section{Introduction}
Following the Paris agreement within the United Nations Framework Convention on Climate Change (UNFCCC) aiming to reduce global greenhouse gas emissions (see e.g.\ \cite{popovski2018implementation}), many countries have recently committed to set a time horizon until which a net zero target in terms of carbon emissions should be achieved. In order to reach such a target in a realistic way, companies and individuals will have to (and in some cases already do) receive carbon emission budgets over a given time horizon, and it is an interesting question how such budgets can and will influence the behaviour of individuals and companies which are a priori profit-maximizing entities. Concretely, how will a firm determine its production and -- correspondingly -- carbon emission schedule, if it is given an aggregate emission budget constraint? How will a (rational) individual decide on its carbon-intensive consumption pattern, if it is given an aggregate emission budget constraint? And how does a certain level of carbon tax enforced by a government steer or incentivize this process towards a net-zero target? There are many directions from which answers to such questions may be sought, and there is a strong need to increase the understanding of underlying mechanisms and incentives, see e.g.\ \cite{saleh2025estimating,chekriy2025probabilistic}. In this paper we would like to contribute to this topic by relating it to optimal dividend/consumption problems in insurance and corporate finance, and adopt and extend techniques developed there for the present purpose. In particular, we want to address the above questions in a relatively simple model to better understand the effects of present-biasedness in this context. We provide a framework that allows to accommodate both the situation of a company (where monetary profitability is linked to the amount of carbon-intensive production) and the situation of an individual (where profitability is rather measured in terms of carbon-intensive consumption). In either case the decision-maker  is given a finite 'budget' of aggregate available carbon emissions and is subject to a certain degree of present-biasedness. By establishing and exploiting a link to stochastic control techniques for optimal dividend problems in insurance risk theory, we develop a framework that leads to quantitative results in terms of optimal behavior given the objective function and (carbon budget) constraint. Within the chosen simple model assumptions, this also allows to study the effect of the governmental measure of a carbon tax towards an aggregate carbon emission target. In particular, we intend to quantify the sensitivity of the results with respect to the degree of present-bias of the decision-maker, and compare it to the values in the absence of present-bias.  
In terms of the maximization criterion, we consider the maximization of profit (or consumption benefit in the case of an individual), but also allow for a term in the objective function that reflects a certain degree of social responsibility and sustainability awareness (see e.g.\ \cite{korn2025framework,korn20252} for other ways to incorporate sustainability considerations in profitability criteria). 

In any inter-temporal choice problem, discounting is one of the key factors that influence the optimal strategy. Traditionally, exponential discounting is used, where the time preference for a payment (or consumption token) occurring at time $ t $ can be fully captured by a single discount rate at that time. In such a case, the optimization problem is time-consistent, and the optimal decision regarding actions at time $ t $ will only depend on available resources at that moment, regardless of the time $s<t$ at which this decision was taken. That is, an 'optimal' decision made at time $ s_1 $ for an action at a future time $ t >s_1$ will be preferred by the decision maker at any later time $ s_2>s_1 $ as well.
Most of the literature in optimal control relies on exponential discounting to calculate the present value of future payments (or monetary translations of consumption opportunities). This is usually done by assuming time-consistent preferences and employing a constant discount rate (see e.g.\ \cite{schmidli2007stochastic,albrecher2009optimality,azcue2014stochastic} for surveys in the field of dividend strategies). In rare cases, a stochastic discount rate (\cite{eisenberg2015optimal, reppen2020optimal}) is employed. 
 However, empirical studies often observe patterns of preference reversals. \cite{Laibson1998} describes ``a conflict between today's preferences and the preferences which will be held in the future",  and the exponential discount functions cannot capture such a tendency. In many practical situations, decision-makers are present-biased, preferring smaller but earlier rewards to larger but later ones, in particular when such earlier rewards are near. Only when the time until such rewards is far distant, such preferences may be flipped,  cf.\ for instance (\cite{Palacios-HuertaPerez-Kakabadse2011}). \cite{Laibson1998} noted already that  hyperbolic discounting  may be used to model  such preferences, and it has been shown to outperform exponential discounting in certain empirical studies. 
 \cite{Laibson1998} used a quasi-hyperbolic discount function for discrete-time models, where an additional constant discount factor is introduced in the utility of all cash-flows in future periods regardless of timing, see also \cite{PhelpsPollak1968}.
 Such a quasi-hyperbolic discount function  mimics the quantitative properties of a hyperbolic discount function while maintaining analytical tractability. \cite{HarrisLaibson2013} then proposed a stochastic quasi-hyperbolic discount function as a continuous-time model of non-exponential discounting, where an additional discount factor is added for future periods. It is this latter model that we will adopt for our present study.

In this paper, we propose a simple continuous-time dynamic model framework based on a general linear diffusion to investigate the optimal production and carbon emission strategy for a firm with an allocated emission budget, taking present-bias into account. As mentioned above, the setup can also be interpreted for the situation of an individual taking decisions on carbon-intensive consumption, but we will formulate the paper for the context of a firm, and only add some interpretations for the situation of an individual in the concluding section.

Concretely, the remaining emission budget will be modeled as a stochastic process over time with a given initial allocation, a drift term representing a trend (e.g., projected unavoidable minimal emissions in the production process),
and a volatility term reflecting fluctuations  (which may e.g.\ stem from remaining uncertainties in establishing the present carbon balance). Effective carbon emissions act as a deduction term from this process in the dynamics, reducing the available remaining budget as production continues. In the paper we consider a general form for the drift and volatility terms, allowing for various scenarios. The company develops a production/emission schedule aimed at maximizing the present value of total expected future profit up to a fixed terminal time, plus a term that rewards for not yet having used up the budget at any time until depletion, which can be interpreted as a contribution to sustainability considerations of 
society. In order to account for present-biased preferences, we use stochastic quasi-hyperbolic discounting rather than standard exponential discounting for determining the present value of future profits and the reward term. The weighting of the reward term in the objective function then formalizes the balancing between profit maximization and sustainability considerations.
 
Under time-inconsistent preferences (such as the  stochastic quasi-hyperbolic discounting considered in this paper), there are typically two alternative assumptions about decision-makers: naive agents and sophisticated agents (\cite{GrenadierWang2007}). The naive agent assumes that future selves act according to the preferences of the current self, which is possible if there is a commitment mechanism that ensures that future selves commit to the strategy chosen by the current self. The sophisticated agent, in contrast, ``correctly foresees that their future selves act according to their own preferences” (\cite{GrenadierWang2007}), see also {\cite{FrederickLoewensteinODonoghue2002}. In the sophisticated agent case, there is no optimal solution, as a solution that is optimal in the eyes of the decision-maker at time $t$
 will not remain optimal later.
Conflicts arise unless there are pre-commitment mechanisms ensuring that an optimal decision made earlier will be upheld by future decision-makers, even if it is no longer optimal for them to do so (\cite{Strotz1956}). In this paper, we consider sophisticated decision makers and assume that there are no commitment devices. In other words, we consider scenarios where the early selves do not have a technology to commit the actions of later selves. This is a common and realistic scenario, see e.g.\  \cite{IversonKarp2021}. 
Correspondingly, we formulate the control problem as an intra-personal subgame and seek equilibrium solutions (\cite{HarrisLaibson2013} and \cite{MaskinTirole2001}). 
 We establish the existence of equilibrium strategies and equilibrium solutions theoretically and provide a procedure for determining an equilibrium solution and the associated equilibrium strategy. We further investigate the impact of the degree of present-bias on the agent's behavior and the respective carbon emission consequences. We find the intuitive result that a higher degree of present-bias leads to higher emissions, and earlier depletion of the carbon allowance. Additionally, we analyze the role of the sustainability term in shaping the company's carbon emission behavior and explore how a carbon tax might affect the company's emission decisions.
Naturally, higher sustainability awareness curbs production and reduces carbon emissions, just as imposing a carbon tax does. The results of this paper contribute to understand how the two effects are related, and which level of carbon tax replaces which level of sustainability awareness to lead to the same result. In addition, it will turn out that once the carbon tax reaches a certain threshold, its effectiveness begins to decline. 

We consider the carbon tax as being determined exogeneously by policymakers (social planners), and then the individual firms' behavior is studied in response to that. The carbon tax can then be interpreted as the social cost of carbon. Various suggestions exist in the literature for determining the appropriate or optimal social cost of carbon, with a common approach based on well-established Integrated Assessment Models (IAMs), which integrate climate and economic systems. For example, one of the earliest and most frequently used IAMs for climate change is the DICE/RICE family of models (see \cite{Nordhaus2018} for details on its development). Most of these IAM studies assume exponential discounting, typically at a constant discount rate. \cite{FriesQuante2024} extended the DICE model to incorporate a stochastic discount rate, which technically remains a form of exponential discounting. See also \cite{colaneri2024random} for another type of stochastic control problem where a company decides on investments in carbon abatement technologies in view of carbon tax costs, and \cite{bourgey2024} for optimizing the emission level alongside constraints on emission mitigation scenarios, additionally taking into account credit risk. Our approach based on stochastic quasi-hyperbolic discounting can therefore also be interpreted as an extension of certain aspects of that literature to the explicit consideration of present-biasedness. Since stochastic quasi-hyperbolic discounting approximates hyperbolic discounting, and the latter is often empirically found to better represent individuals' true time preferences, cf.\ \cite{FrederickLoewensteinODonoghue2002}, a contribution of this paper is also to offer a quantitative approach to systematically understand the effects of that deviation from exponential discounting for questions of that part of social planning.
 Note that the choice of appropriate discount rates related to climate policy is a subject of on-going political debate, all the way since the Stern Review (\cite{stern2006review}). We will not delve into respective discussions here, and the discount rate values applied in the numerical section are only for illustrative purposes. Our aim is to contribute -- in a simplified model with explicit formulas for the optimal strategies -- to the understanding of how present-biasedness affects the decision-making of profit-maximizing rational agents (who are not social planners), which  may also provide insights for social planners to develop effective policies towards specific targets. 

On the technical side, we deal with a regular control equilibrum problem with stochastic quasi-hyperbolic discounting under a general linear diffusion framework. For a similar problem with capital injections, but without a sustainability component and for a different type of discounting (pseudo-exponential discounting) as well as constant drift and volatility coefficients only,  see \cite{hu2025equilibrium}. For singular control problems under other non-exponential discounting functions, see also \cite{ZhaoWeiWang2014} and \cite{LiLiZeng2015}, and for another way to formalize time-inconsistency see \cite{strini2023time}. Finally, equilibrium strategies for singular (rather than regular as in this paper) control under stochastic quasi-hyperbolic discounting have been identified in \cite{ChenLiZeng2014} and \cite{ChenWangDengZeng2016} for a compound Poisson model with negative jumps of exponential type, in \cite{LiChenZeng2015} for a Brownian risk model, and in \cite{ZhuSiuYang2020} for a linear growth restricted  diffusion process.
 
The remainder of the paper is organized as follows. Section \ref{Formulation} defines the model setup, introduces exponential and quasi-hyperbolic discounting and defines the type of optimal strategies we are investigating. Section \ref{exponential} derives these optimal strategies for exponential discounting, and spells out the explicit formulas for the Brownian model in more detail. Section \ref{sol-hyber-sec} then establishes the equlibrium solution for the case of stochastic quasi-hyperbolic discounting, which is the core for the study of the present-bias effects considered in this paper. Section \ref{expectation} deals with the determination of the probability of early depletion of the carbon budget when following the optimal strategy. Section \ref{brownian} is then dedicated to numerical illustrations of the impact of present-bias, the level of social responsability and the amount of carbon tax on the emission schedule for a Brownian motion model with constant diffusion coefficients. Detailed interpretations of the interplay of various factors are given. In Section \ref{surp}, it is then shown that the analysis can also be extended to more general diffusion models, including an Ornstein-Uhlenbeck type process for the time-development of the available carbon budget. Finally, Section \ref{conclu} concludes. 
All mathematical derivations and proofs are moved to the Appendix.

\section{Problem Formulation }\label{Formulation}
Let $(\Omega,\mathcal{F},\{\mathcal{F}_t;t\ge 0\},\pr)$ be a filtered complete probability space with a right-continuous filtration $\{\mathcal{F}_t;t\ge 0\}$. Consider a firm whose production depends on its energy consumption, and let $P_t$ represent the (monetary) instantaneous production profit at time $t$. We assume that $P_t=\gamma (l_t+\underline{l})$ for some $\gamma>0$, where $ l_t + \underline{l} $ is the instantaneous emission rate at time $ t $, with $ \underline{l} $ representing the baseline emission rate required to maintain minimal production activity, and $ l_t $ representing the excess emission rate for additional production above that minimum level. Since the baseline emission rate $ \underline{l} $ can not be avoided in any case, the control to be considered in this paper is the excess carbon emission $L=\{L_t=\int_0^t  l_s\,\dif s;t\ge 0\}$ up to time $t$. 
  Let  $c_{ind}\, P_t$ represent the physical cost for producing $P_t$ and $c_{tax}(l_t+\underline{l})$ the carbon tax paid for the resulting emission. Then the total cost, which equals the sum of production cost and carbon tax, is $(c_{ind}\,\gamma+c_{tax})(l_t+\underline{l})$. Assume now that the firm is allocated with a total (CO$_2$) emission allowance (endowment)\footnote{The total emission allowance $x_0$ could for instance be the total emission budget allocated to the company according to a net-zero target around 2050 set by the  Intergovernmental Panel on Climate Change.} $x_0$ and let $X_t^L$ represent the remaining emission allowance at time $t$ according to the general diffusion dynamics
 \begin{align}
   X_t^L&:=x_0 + \int_0^t\overline{\mu} (X_s) \dif s+ \int_0^t
  \sigma(X_s) \dif W_s- \int_0^t  (\underline{l}+l_s)\dif s, \ \ t\ge 0.\label{dynamics-000}
 \end{align} 
  The drift term $ \overline{\mu}(\cdot) $ could be zero, or negative in a deteriorating situation, or also positive, e.g.\ due to technological advances and increasing carbon capture capabilities over time. Here, $W_t$ is a standard Brownian motion. Let $\mathcal{F}^W$ represent the filtration generated by $\{W_t;t\ge 0\}$.
 
The functions $ \overline{\mu}(\cdot) $ and the volatility $ \sigma(\cdot) $ are assumed to be Lipschitz continuous, satisfying a linear  condition, that is, there exists a constant $C>0$ such that $\overline{\mu}^2(x)+\sigma^2(x)\le C(1+x^2)$ for all $x$. As proven in \cite{GikhmanSkorokhod1972}, these conditions guarantee the existence and uniqueness of a strong solution to \eqref{dynamics-000} for each $x_0$ and each $\mathcal{F}^W$-adapted, nondecreasing, left-continuous process $L$,  see also \cite{ShreveLehoczkyGaver1984}.
We further assume  that $ \sigma(\cdot) $ is non-vanishing  and $0\le l_s\le \bar{l}$, where $\bar{l}$ is a positive constant. Additionally, we impose the restriction $\overline{\mu}^\prime(x)\le \delta$ for $x\ge 0$, where $\delta>0$ is the exponential discount rate discussed further below. This latter assumption will serve mathematical tractability,  but it also has practical relevance, as one would not expect the growth rate of available carbon allowance to increase significantly over time. 

With 
 $\mu(x)=\overline{\mu}(x)-\underline{l}$
     we can simplify \eqref{dynamics-000} to 
\begin{align}
   X_t^L&:=x_0 + \int_0^t{\mu} (X_s) \dif s+ \int_0^t
  \sigma(X_s) \dif W_s- L_t, \ \ t\ge 0.\label{dynamics-00s}
 \end{align} 
\noindent Let
\begin{align}
\tau^{L}=    \inf\{t\ge 0: X_t^L\le 0\}\label{ruin-time}
\end{align}
denote the emission allowance depletion time when applying emission schedule $L$.
In this paper we are interested in the optimal emission schedule for the company that maximizes the expected present value of profit. At time $t$, the objective function to be maximized therefore is
\begin{align}
\mathcal{P}^E(x,t;L)=&\,\E\left[\left.
\int_t^{\tau^{L}} D(t,s) \big[(\gamma-\beta) ( l_s+\underline{l}) + \overline{\Lambda} \big]  \dif s \right|X_t=x\right],
\label{edynamic}
\end{align}
where $D(t,s)$ is the discounting function for calculating the present value at time $t$ of cashflows at future times $s\ge t$ and $\beta=c_{ind}\,\gamma+c_{tax}$. 
While the focus is on maximizing profit, we also introduce a constant rate $\overline{\Lambda}>0$ that rewards for the carbon allowance to not be depleted early (i.e., having the depletion time $\tau^{L}$ being larger). It can be interpreted as an intangible utility term (e.g., sustainability value). This will allow to consider the tradeoff between profitability, costs and `social responsibility' represented through $\overline{\Lambda}$ (cf.\ \cite{ThAl07} for the introduction of such a term for dividend problems). With $\Lambda:=\overline{\Lambda}+(\gamma-\beta)\underline{l}$, 
we can simplify \eqref{edynamic} to 
\begin{align}
\mathcal{P}^E(x,t;L)=&\,\E\left[\left.
\int_t^{\tau^{L}} D(t,s) \big[(\gamma-\beta)  l_s + \Lambda \big]  \dif s \right|X_t=x\right].
\label{edynamic2}
\end{align}

\subsection{Exponential discounting} As a benchmark model, and also as an intermediate result needed in the derivations, we will first consider exponential discounting, that is $D(t,s)=e^{-\delta (s-t)}$ for some constant rate $\delta>0$.  In this case the objective function in \eqref{edynamic2} at time 0 reads as 
\begin{align}
\mathcal{P}^E(x;L)
=&\E\bigg[\left.\int_0^{\tau^{L}} e^{-\delta s}((\gamma-\beta) l_s+\Lambda)\dif s\right|X_0=x\bigg].\label{exponential-objective}
\end{align}
Correspondingly, the optimization goal is to look for a schedule of excess emissions $L$ that maximizes $\mathcal{P}^E(x;L)$, leading to the value function \begin{align}
V^E(x)=\sup_{L\in\Pi}\mathcal{P}^E(x;L)
=&\sup_{L\in\Pi}\E\bigg[\left.\int_0^{\tau^{L}} e^{-\delta s}((\gamma-\beta) l_s+\Lambda)\dif s\right|X_0=x\bigg],\label{Value-Exponential}
\end{align}
where  $\Pi$  denotes the set of \textit{admissible strategies}, which will be specified later.

\subsection{Stochastic quasi-hyperbolic discounting}
If the decision maker is present-biased, we use the following stochastic quasi-hyperbolic discount function  introduced in  \cite{HarrisLaibson2013}: 
\begin{align}\label{stochdisc}
    D(t,s)=
    \begin{cases}
    e^{-\delta (s-t)},& t<s< t+ \eta,\\
     \alpha\, e^{-\delta (s-t)},&s\ge t+\eta,
     \end{cases}     \end{align}
     where $\eta$ represents the (random) duration of the current regime and $0\le \alpha\le 1$ is a constant. That is, cashflows during the present period are valued using exponential discounting at force $\delta$, while cashflows emerging in the future period are discounted by a smaller value. 

One can interpret this stochastic discounting framework in the way that there is a sequence of decision makers, to each of whom time is divided into two intervals, the present and the future, and each decision maker is present-biased. The present will last for a random length of time which we model as an exponential random variable with parameter $\lambda>0$, independent of the current carbon allowance. All cash-flows in the present period are discounted exponentially with force $\delta$ and the cash-flows in the future period are then discounted more strongly with additional factor $\alpha$.  Assume that the decision maker at time $0$ is called ``self $0$". The present period for ``self $0$" starts at time $0$ and ends at time $\eta_0$. ``Self $0$"  exercises control for her present period and is present-biased. At the end of ``self $0$"'s present period, a new self, ``self $1$", starts to take over decision making. ``Self $1$" is also present-biased and she can only exercise control during her own present period, which lasts from time $s_1:=\eta_0$ to $s_1+\eta_1$. Acoordingly, the present period of ``self $n$" ($n=1,2,3,\cdots$), also present-biased, is from time $s_{n}$ to $s_{n}+\eta_n$. Each self takes decisions according to $D(s,t)$ given in \eqref{stochdisc}. 
More specifically, if we use  $D_n(t)$ to represent the present value  at time $s_n$ of one dollar payable at time $t$  from ``self $n$"'s perspective, then $D_n(t)=e^{-\delta (t-s_n)}$ for $s_n\le t<s_n+\eta_n$ and $D_n(t)=\alpha e^{-\delta (t-s_n)}$ for $t\ge s_n+\eta_n$.  

Although each self controls the emission schedule only during her present period, she does so keeping in mind the total production profit, i.e.\ the profit of the present period as well as the one in future periods.  Different selves have conflicting preferences as they value the production profit and survival utility during a particular period differently.
We assume that there are no commitment mechanisms (in the sense that later selves are not committed to what earlier selves considered optimal), 
and that the decision-maker is sophisticated and rational. In addition, she can correctly foresee her future actions. For this intra-personal game we will consider Markov policies only, and seek a Markov-perfect equilibrium (MPE).
That is, we restrict the admissible strategies to stationary  Markov-perfect equilibrium (MPE) policies. An emission schedule $L=\{l_t;t\ge 0\}$ is said to be \textit{admissible} if it is a Markov policy with $0\le l_s \le \bar{l}$.   We use $\Pi$ to denote the set of \textit{admissible strategies}.

Let $\pi^{(n,\rightarrow)(L,\tilde{L})}$ represent the strategy where ``self $n$" adopts $L$ and the future selves adopt $\tilde{L}$. Let $\pi^{(n,\rightarrow)(L,\tilde{L})}_t$ represent the cumulative amount of emissions from time $s_n$ to  $t$ under $\pi^{(n,\rightarrow)(L,\tilde{L})}$. Then,  $\pi^{(n,\rightarrow)(L,\tilde{L})}_{s_n-}=0$, $\dif \pi^{(n,\rightarrow)(L,\tilde{L})}_{t}=l_t\dif t$ for $t\in[s_n,s_{n+1})$ and  $\dif \pi^{(n,\rightarrow)(L,\tilde{L})}_{t}=\tilde{l}_t \dif t $ for $t\ge s_{n+1}$.
The reward to ``self $n$" is  the  expected present value at time $s_n$ of the entire future net production profit up to the time of depletion plus the reward from the $\Lambda$-term. Given $X_{s_n-}^L=x$, for any $x\ge 0$, the objective function for ``self $n$" with pair $(L,\tilde{L})$ is 
\begin{align}
&\mathcal{P}_n(x;L,\tilde{L})\nonumber\\
=\;&\E_{s_n,x}\bigg[\int^{\tau^{\pi^{(n,\rightarrow)(L,\tilde{L})}}_{s_n}\wedge {(s_n+\eta_n)}}_{s_n} e^{-\delta (t-s_n)}(\gamma-\beta) l_t\dif
t\nonumber\\&
+I\{s_n+\eta_n\le \tau^{\pi^{(n,\rightarrow)(L,\tilde{L})}}_{s_n}\}
\int_{s_n+\eta_n}^{\tau^{\pi^{(n,\rightarrow)(L,\tilde{L})}}_{s_n}}\alpha e^{-\delta (t-s_n)}(\gamma-\beta)\tilde{l}_t\dif
t\nonumber\\
&+\int^{\tau^{\pi^{(n,\rightarrow)(L,\tilde{L})}}_{s_n}\wedge {(s_n+\eta_n)}}_{s_n} e^{-\delta (t-s_n)}\Lambda\dif
t
+I\{s_n+\eta_n\le \tau^{\pi^{(n,\rightarrow)(L,\tilde{L})}}_{s_n}\}
\int_{s_n+\eta_n}^{\tau^{\pi^{(n,\rightarrow)(L,\tilde{L})}}_{s_n}}\alpha e^{-\delta (t-s_n)}\Lambda\dif
t\bigg],
\end{align}
where $I\{\cdot\}$ is the indicator function and $\tau^{\pi^{(n,\rightarrow)(L,\tilde{L})}}_{s_n}$ is the (potential) depletion time during the active period of ``self $n$" when following the strategy $(L,\tilde{L})$.
The first term inside the expectation above represents the discounted net amount of production profit in the present period, where all the cashflows are discounted with force $\delta$,  and the second term is the total discounted  net amount of production profits in all  the future periods up to the time of depletion, where all the cashflows are discounted by the force $\delta$ and then further discounted by the factor $\alpha$. The last two terms represent the benefit of surviving up to depletion time represented through the reward rate $\Lambda$.

Let $\mathcal{P}(x;L,\tilde{L}):=\mathcal{P}_0(x;L,\tilde{L})$ and $\pi^{L,\tilde{L}}:=\pi^{(0,\rightarrow)(L,\tilde{L})}$. The objective of ``self $n$" is to find a Markov strategy $L^*$ (a \textit{MPE policy} that maximizes the above expected reward  with respect to $L$ in the sense that}
\begin{align}
\mathcal{P}_n(x;L^*,L^*)=\sup_{L\in\Pi}\mathcal{P}_n(x;L,L^*)\label{eq: MPE}
\end{align}
subject to the production constraint). Note the problem is stationary, although the preferences of the decision makers are time-inconsistent. Then we only need to solve the game problem based on the reward $\mathcal{P}=\mathcal{P}_0$. That is, we are seeking an admissible strategy $L^*$ such that 
\begin{align}\label{31525-1}
    \mathcal{P}(x;L^*,L^*)=\sup_{L\in\Pi} \mathcal{P}(x;L,L^*).
\end{align} 

Indeed, if a MPE strategy exists that satisfies \eqref{eq: MPE} for all $n$, then no self has an incentive to deviate from it, given that all future selves adopt it as well. 

\section{Optimal Solutions under Exponential Discounting}\label{exponential}

Under exponential discounting, the mathematical formulation of the optimization problem is similar to (and a slight extension of) the problem in \cite{Zhu2015a}, where there was no $\Lambda$ term. In the following we accordingly adapt the technique developed in that reference. One easily derives the Hamilton-Jacobi-Bellman (HJB) equation
\begin{align}
&\sup_{l\in[0,\bar{l}]}\left(\frac{\sigma^2(x)}{2}g^{\prime\prime}(x) +
\mu(x)g^\prime(x)-\delta g(x)+{l}(\gamma-\beta-g^\prime(x))+\Lambda\right)=0, \label{hjb0}
\quad  g(0)=0. 
 \end{align}
If the value function $V^E$ is sufficiently smooth, then a standard verification theorem shows that $V^E$ is a classical solution to the HJB equation \eqref{hjb0}. 
We establish the existence of a classical solution by constructing one explicitly, using a class of auxiliary functions defined below. To that end, let us consider a threshold strategy for any given threshold $b\ge 0$ as
\begin{align}
    L^{b}:=\{\bar{l}\cdot I\{X_t\ge b\}; t\ge 0\}\label{Lb-25}
\end{align} and denote its corresponding value function under exponential discounting as
\begin{align}
    V_b^E(x):=\mathcal{P}^E(x;L^b).\label{VbE-25}
\end{align} 

The function $V_b^E(x)$ will be instrumental in searching for a solution, both under exponential and stochastic quasi-hyperbolic discounting considered later. 

\begin{Lemma}\label{28525-1}
The function $V_b^E(x)$ solves the initial value problem
\begin{align}
&\frac{\sigma^2(x)}{2}g^{\prime\prime}(x) +
\mu(x)g^\prime(x)-\delta g(x)+\Lambda=0 \mbox{ for $0< x< b$},\label{1819-1}\\
&\frac{\sigma^2(x)}{2}g^{\prime\prime}(x) +
(\mu(x)-\bar{l})g^\prime(x)-\delta g(x)+\bar{l}(\gamma-\beta)+\Lambda=0 \mbox{ for $ x\ge b$}, \label{1819-2}\\
& g(0)=0,\label{251025-1}
 \end{align}
 and has the form
  \begin{eqnarray}
V_b^E(x)=\left\{\begin{array}{ll}
C_{1}(b)(v_1(x)-v_2(x))+B_1(x),&0\le x< b,\\
C_3(b) v_3(x)+u(x),&x\ge b,
\end{array}\right.
\label{14225-5}
\end{eqnarray}
where
\begin{align}
&C_1\,(b)=\frac{(B_1(b)-u(b))v_3^\prime(b)-(B_1^\prime(b)-u^\prime(b))v_3(b)}{(v_1^\prime(b)-v_2^\prime(b))v_3(b)-(v_1(b)-v_2(b)) v_3^\prime(b)},\\
&C_3(b)=\frac{(u^\prime(b)-B_1^\prime(b))(v_1(b)-v_2(b))-(u(b)-B_1(b))(v_1^\prime(b)-v_2^\prime(b))}{(v_1^\prime(b)-v_2^\prime(b))v_3(b)-(v_1(b)-v_2(b))v_3^\prime(b)}.
\end{align}
The functions $v_1(\cdot)$ and $v_2(\cdot)$ are the solutions to
$
\frac{\sigma^2(x)}{2}g^{\prime\prime}(x)+\mu(x)g^\prime(x)-\delta g(x)=0,
$ 
with the respective sets of initial conditions:
$ v_1(0)=1, v_1^\prime(0)=1,\mbox{and\ }  v_2(0)=1,  v_2^\prime(0)=-1.$ 
The function $u(x)$ is the unique bounded solution to
$
\frac{\sigma^2(x)}{2}g^{\prime\prime}(x)+ (\mu(x)-\bar{l})g^\prime(x)-\delta g(x)+\Lambda + (\gamma-\beta) \bar{l} = 0,
$
on $[0,\infty)$ with initial condition $g(0)=0$.
The function $v_3(\cdot)$ is the unique bounded solution to
$
\frac{\sigma^2(x)}{2}g^{\prime\prime}(x)+ (\mu(x)-\bar{l})g^\prime(x)-\delta g(x) = 0
$
on $(0,\infty)$ with initial condition $g(0)=1$.

Additionally, we have
\begin{align}
&B_1(x)=2\Lambda \int_0^x \frac{ v_1(x)v_2(y)-v_2(x)v_1(y)}{v_1(y)v_2^\prime(y)-v_2(y)v_1^\prime(y)}\dif y,
\end{align}
which is the solution to $\frac{\sigma^2(x)}{2}g^{\prime\prime}(x) +
\mu(x)g^\prime(x)-\delta g(x)+\Lambda=0$ under $B_1(0)=0$ and $B_1^\prime(0)=0$.

\end{Lemma}

We can show that the value function $V^E(x)$ has the following property.

\begin{Lemma}\label{upperbound}
The 
function $V^E(x)$ is nonnegative, increasing and has an upper bound $\frac{(\gamma-\beta)\bar{l}+\Lambda}{\delta}$.
\end{Lemma}

Moreover, we can derive an expression for the value function under exponential discounting as follows.

\begin{Theorem} \label{optsol-exp} 
The value function under exponential discounting is given by:
\begin{eqnarray*}
V^E(x)=\left\{\begin{array}{ll}
C_1\,(b^*_E)(v_1(x)-v_2(x))+B_1(x), & 0 \leq x < b^*_E, \\
C_3(b^*_E) v_3(x)+u(x), & x \geq b^*_E,
\end{array}\right.
\end{eqnarray*}
where $C_{1}(\cdot)$, $C_3(\cdot)$, $v_1(\cdot)$, $v_2(\cdot)$, $v_3(\cdot)$, $u(\cdot)$, and $B_1(\cdot)$ are defined in 
Lemma \ref{28525-1}, and 
\begin{align}
b^*_E=\inf\left\{b>0: C_3(b) v_3^\prime(b)+u^\prime(b)\le \gamma-\beta \right\}. \label{17225-6}
\end{align}
It can be shown that $b^*_E<\infty$, and that if $\mu(0)<0$, then $b^*_E\le b_0$, where
\begin{align}\label{b0}
b_0=\inf\left\{b\ge 0: C_{1}\,(b)+\frac{\Lambda}{2\mu(0)}>0\right\}.
\end{align}
Finally, an optimal admissible strategy that attains the best performance according to the performance functional $\mathcal{P}^E$ is $L^{b^*_E}:=\{\bar{l}\cdot I\{X_t\ge b^*_E\}; t\ge 0\}$, where $b^*_E$ is defined in \eqref{17225-6}.
    \end{Theorem}

 The above  theorem shows how to determine the value function and the optimal strategy.
In this regard, the key is to compute the functions $ v_1(\cdot) $, $ v_2(\cdot) $, $ v_3(\cdot) $, and $ u(\cdot) $. Based on these functions, we then compute $ C_{1}\,(\cdot) $, $ C_3(\cdot) $, and $ B_1(\cdot) $. Analytical solutions are available for some cases, while for others, numerical solutions are required. 
Determining $ v_1(\cdot) $ and $ v_2(\cdot) $ numerically involves solving two second-order ordinary differential equations (ODEs) numerically. 
Similarly, $B_1(x)$ is a particular solution to $\frac{\sigma^2(x)}{2}g^{\prime\prime}(x) +
\mu(x)g^\prime(x)-\delta g(x)+\Lambda=0$ with $B_1(0)=0$ and $B_1^\prime(0)=0$, which again can be solved using standard numerical procedures. However, determining $ v_3(\cdot) $ and $ u(\cdot) $ numerically from the ODEs is more challenging because it involves finding bounded solutions on infinite intervals. 
To overcome this, we convert the problem to a bounded interval and identify the boundary values at both ends. The following result (with the proof provided in  Appendix \ref{aB}) is key to identifying the boundary values mentioned.

\begin{Lemma}\label{lemmm}
\label{1325-1}For the functions, $v_3(\cdot)$ and $u(\cdot)$ defined in Theorem \ref{optsol-exp}, we have 
\begin{align}
&v_3(x)=\E\bigg[e^{-\delta \hat{T}^x}\bigg],
\mbox{ and } u(x)=\E\bigg[\int_0^{\hat{T}^x} e^{-\delta s}((\gamma-\beta) \bar{l}+\Lambda)\dif s\bigg],\label{1325-40000}
\end{align}
where $Y_t^x$ is a stochastic process and $\hat{T}^x$ is a stopping time  defined by
\begin{align}
&Y_t^x=x+\int_0^t(\mu(Y_s^x)-\bar{l})ds +\int_0^t \sigma(Y_s^x)ds,\quad s >0,\\
&\hat{T}^x=\inf\{t\ge 0: Y_t^x\le 0\}.
\end{align} Moreover,  the following limiting results hold:
\begin{align}
&\lim_{x\rightarrow\infty} v_3(x)=0,\quad
\lim_{x\rightarrow\infty} u(x)=\frac{(\gamma-\beta)\bar{l}+\Lambda}{\delta}.\label{4325-1-0}
\end{align}
\end{Lemma}

From Theorem \ref{optsol-exp} we know that 
$u$ is the unique bounded solution to $
 \frac{\sigma^2(x)}{2}g^{\prime\prime}(x) +
(\mu(x)-\bar{l})g^\prime(x)-\delta g(x)+\bar{l}(\gamma-\beta)+\Lambda=0$ on $(0,\infty)$ with the initial value $g(0)=0$. This, combined with \eqref{4325-1-0}, implies that $u(x)$ is the unique solution with 
$u(0)=0$ and $\lim_{x\rightarrow\infty} u(x)=\frac{(\gamma-\beta)\bar{l}+\Lambda}{\delta}
$. Hence, $u$ can be numerically determined by choosing a sufficiently large number, say $\bar{x}$, and then solving the second-order ODE  
$\frac{\sigma^2(x)}{2}g^{\prime\prime}(x) +
(\mu(x)-\bar{l})g^\prime(x)-\delta g(x)+\bar{l}(\gamma-\beta)+\Lambda=0$ with boundary conditions $g(0)=0$
and $g(\bar{x})=\frac{(\gamma-\beta)\bar{l}+\Lambda}{\delta}
$.  
Similarly, $v_3$ can be computed by selecting a sufficiently large number, say $\bar{y}$, and then solving the boundary value  second order ODE 
$\frac{\sigma^2(x)}{2}g^{\prime\prime}(x) +
(\mu(x)-\bar{l})g^\prime(x)-\delta g(x)=0$ with $g(0)=1$  and $
 g(\bar{y})=0
$.

\subsection{The Brownian motion model}\label{BWexp}
If $\overline{\mu}(x)\equiv\overline{\mu} $ and $\sigma(x)\equiv\sigma >0$ are constant, the expressions simplify. Indeed, applying the main theorems from above to 
\begin{equation}
    \mathrm{d} X_t^L=(\overline{\mu}-\underline{l}-l_t)\mathrm{d} t+ \sigma\mathrm{d} W_t,\ t\ge 0,\label{that}
\end{equation}
we obtain (see Appendix \ref{aE} for the derivations):
\begin{equation}\label{sol1-000}
V_{b}^{E}(x)=%
\begin{cases}
\left( e^{\theta _{1}x}-e^{-\theta _{2}x})I_{1}(b)-\frac{2}{\sigma ^{2}}%
\frac{e^{\theta _{1}x}-1}{\theta _{1}(\theta _{1}+\theta _{2})}+\frac{2}{%
\sigma ^{2}}\frac{1-e^{-\theta _{2}x}}{\theta _{2}(\theta _{1}+\theta _{2})}%
\right) \Lambda \\
\ +(e^{\theta _{1}x}-e^{-\theta _{2}x})I_{2}(b), & 0\leq x<b, \\
\left(e^{-\theta _{4}x} I_{3}(b)+\frac{2}{\sigma ^{2}}\frac{1}{\theta
_{3}(\theta _{3}+\theta _{4})}+\frac{2}{\sigma ^{2}}\frac{1-e^{-\theta _{4}x}%
}{\theta _{4}(\theta _{3}+\theta _{4})}\right) \Lambda\\
\  +e^{-\theta _{4}x}I_{4}(b)+\frac{2((\gamma-\beta) \bar{l})}{\sigma ^{2}}\frac{1}{\theta _{3}(\theta _{3}+\theta _{4})}+\frac{2((\gamma-\beta) \bar{l})}{\sigma ^{2}}\frac{1-e^{-\theta _{4}x}}{\theta _{4}(\theta_{3}+\theta _{4})}, & x\geq b,
\end{cases}%
\end{equation}
where 
\begin{align}
&\theta_1=\frac{-\mu+\sqrt{\mu^2+2\sigma^2\delta}}{\sigma^2},\quad \theta_2=\frac{\mu+\sqrt{\mu^2+2\sigma^2\delta}}{\sigma^2},\label{theta1}\\
&\theta_3=\frac{-(\mu -\bar{l})+\sqrt{\rule{0pt}{2ex}(\mu -\bar{l})^2+2\sigma^2\delta}}{\sigma^2},\quad \theta_4=\frac{(\mu -\bar{l})+\sqrt{\rule{0pt}{2ex}(\mu -\bar{l})^2+2\sigma^2\delta}}{\sigma^2},\label{theta3}\\
& I_{1}(b)=\frac{\frac{2\left(\theta _{2}(\theta _{1}+\theta _{4})e^{\theta
_{1}b}+\theta _{1}(\theta _{4}-\theta _{2})e^{-\theta _{2}b}-\theta
_{4}(\theta _{1}+\theta _{2})\right)}{\sigma ^{2}\theta _{1}\theta _{2}(\theta
_{1}+\theta _{2})}+\frac{2}{\sigma ^{2}\theta _{3}}}{(\theta _{1}+\theta
_{4})e^{\theta _{1}b}+(\theta _{2}-\theta _{4})e^{-\theta _{2}b}}, \label{I125}\\
& I_{2}(b)=\frac{\frac{2(\gamma-\beta) \bar{l}}{\sigma ^{2}\theta _{3}}}{(\theta
_{1}+\theta _{4})e^{\theta _{1}b}+(\theta _{2}-\theta _{4})e^{-\theta _{2}b}},\label{I225}\\
& I_{3}(b)=\frac{\frac{2(e^{\theta _{1}b}-e^{-\theta _{2}b})}{\sigma
^{2}(\theta _{1}+\theta _{2})}+\frac{2e^{-\theta _{4}b}}{\sigma ^{2}(\theta
_{3}+\theta _{4})}-(\theta _{1}e^{\theta _{1}b}+\theta _{2}e^{-\theta
_{2}b})I_{1}(b)}{\theta _{4}e^{-\theta _{4}b}} ,\label{I325}\\
& I_{4}(b)=\frac{\frac{2(\gamma-\beta) \bar{l}e^{-\theta _{4}b}}{\sigma ^{2}(\theta
_{3}+\theta _{4})}-(\theta _{1}e^{\theta _{1}b}+\theta _{2}e^{-\theta
_{2}b})I_{2}(b)}{\theta _{4}e^{-\theta _{4}b}}.\label{I425}
\end{align}
The optimal strategy under exponential discounting is $L^{b_E^*}$ with $b_E^*$ determined by 
$b^*_E=\inf\{b>0: C_3(b) v_3^\prime(b)+u^\prime(b)\le \gamma-\beta \}$, cf.\ \eqref{17225-6}.

\section{Equilibrium Solution under the Stochastic Quasi-Hyperbolic Discounting}\label{sol-hyber-sec}

The equilibrium policy $L^*=\{l_t^*;t\ge 0\}$ is subject to the carbon emission budget and is a function of the state variable $X_t$.
This is a game with many players (the selves) where each self's objective is to optimize the total future profits, composed by their own state and control as well as the ones of the future selves who value cash-flows in any specified period inconsistently due to present-bias. We look for a Markov equilibrium solution, which is the policy that achieves the best outcome for a self assuming that all the future selves taking the actions according to the same equilibrium policy (\cite{HarrisLaibson2013}).  We start with establishing an extended Hamilton-Jacobi-Bellman  equation (\cite{BjorkKhapkoMurgoci2017}) for the game-theoretical problem and then construct solutions to the equation.  

For any Markov strategies $L$ and $\tilde{L}$ it follows by the Markov property and the definition of the objective function under exponential discounting $\mathcal{P}^E$ in \eqref{exponential-objective} that
\begin{align}
\mathcal{P}(x;L,\tilde{L})=\E_x\left[\int_0^{\tau^{\pi^{L,\tilde{L}}}\wedge \eta_0}((\gamma-\beta) l_t+\Lambda)\dif t +I\{\tau^{\pi^{L,\tilde{L}}}> \eta_0\}\alpha e^{-\delta \eta_0}\mathcal{P}^E(X_{\eta_0}^{\pi^{L,\tilde{L}}},\tilde{L})\right].\label{opt-problem}
\end{align}

As defined in \eqref{31525-1}, a strategy that attains 
\begin{align}
    \mathcal{P}(x;L^*,L^*)=\sup_{L\in\Pi} \mathcal{P}(x;L,L^*).
\end{align} 
is an equilibrium policy.


If the strategy $L^*=\{l^*(X_t);t\ge 0\}$ is an equilibrium solution that satisfies \eqref{opt-problem}, and  $v$ is the corresponding value function and is sufficiently smooth, by a standard differential argument for continuous stochastic processes, we can derive the following equation:
\begin{align}
&\frac{\sigma^2(x)}{2}v^{\prime\prime}(x)+ (\mu(x)-l^*(x))v^\prime(x)-(\lambda+\delta)
v(x)+(\gamma-\beta)l^*(x)+\Lambda+\lambda \alpha \mathcal{P}^{E}(x;L^*)=0,\label{12225-1}
\end{align}
and look for 
\begin{align}
&l^*(x)=\argmax_{l\in[0,\bar{l}]}\left(\frac{\sigma^2
(x)}{2}(x)v^{\prime\prime}(x)+ 
(\mu(x)-lv^\prime(x)-(\lambda+\delta)
v(x)+(\gamma-\beta)l+\Lambda+\lambda \alpha \mathcal{P}^{E}(x;L^*)\right)\nonumber\\
&\quad \quad \quad =\argmax_{l\in[0,\bar{l}]}\left((\gamma -\beta-v^\prime(x))l\right).\label{12225-2}
\end{align}

\noindent Note that $\mathcal{P}^{E}(x;L^*)$ in \eqref{12225-1} refers to the objective function under exponential discounting.\\

Let $L^b$ denote the threshold strategy  defined in \eqref{Lb-25} and define 
\begin{align}
    V_{b}(x):=\mathcal{P}(x,L^b,L^{b}). \label{Vb-25}  
\end{align}


We can obtain  the following key results.
\begin{Theorem}
\label{optimality-h}
The threshold strategy  
$
L^{b^*} := \{l_t = l\cdot{I} \{X_t \geq b^*\}; \, t \geq 0\}
$
is a stationary MPE strategy, and the associated (equilibrium) value function is given by
\begin{equation}
V_{b^*}(x) =
\begin{cases}
\overline{C}_1({b^*})(\overline{v}_1(x) - \overline{v}_2(x)) + \overline{B}_1(x;b^*), & 0 \leq x < {b^*}, \\[8pt]
\overline{C}_3({b^*}) \overline{v}_3(x) + \overline{u}_{b^*}(x), & x \geq {b^*}.
\end{cases}
\nonumber
\end{equation}
Here, the functions $ \overline{v}_1(\cdot) $ and $ \overline{v}_2(\cdot) $ are solutions of the differential equation
$
\frac{\sigma^2(x)}{2} g^{\prime\prime}(x) + \mu(x) g^\prime(x) - (\lambda+\delta) g(x) = 0$ for $ x \in [0,\infty),
$ 
with the respective initial conditions:
$
\overline{v}_1(0) = 1$ and $ \overline{v}_1^\prime(0) = 1$, {and} $\overline{v}_2(0) = 1$ and $\overline{v}_2^\prime(0) = -1.
$ 
The function $ \overline{v}_3(\cdot) $ is the bounded solution to
$
\frac{\sigma^2(x)}{2} g^{\prime\prime}(x) + (\mu(x) - \bar{l}) g^\prime(x) - (\lambda+\delta) g(x) = 0$ for $x \in [0,\infty)$
with initial condition $ g(0) = 1 $. The function $ \overline{u}_b(x) $ (for any $ b > 0 $) is the bounded solution to 
$
\frac{\sigma^2(x)}{2} g^{\prime\prime}(x) + (\mu(x) - \bar{l}) g^\prime(x) - (\lambda+\delta) g(x)  + \lambda \alpha V_b^E(x) + \Lambda + (\gamma - \beta) \bar{l}= 0$ for $x \in [0,\infty)$ 
with initial condition $ g(0) = 0 $, and the coefficients $ \overline{C}_1(b^*) $ and $ \overline{C}_3(b^*) $ are given by
\begin{align*}
\overline{C}_1(b^*) &= \frac{\left(\overline{B}_1(b^*,b^*) - \overline{u}_{b^*}(b^*)\right) \overline{v}_3^\prime(b^*) - \left(\overline{B}_1^\prime(b^*,b^*) - \overline{u}_{b^*}^\prime(b^*)\right) \overline{v}_3(b^*)}
{\left(\overline{v}_1^\prime(b^*) - \overline{v}_2^\prime(b^*)\right) \overline{v}_3(b^*) - \left(\overline{v}_1(b^*) - \overline{v}_2(b^*)\right) \overline{v}_3^\prime(b^*)}, \\
\overline{C}_3(b^*) &= \frac{\left(\overline{u}_{b^*}^\prime(b^*) - \overline{B}_1^\prime(b^*,b^*)\right) (\overline{v}_1(b^*) - \overline{v}_2(b^*)) - \left(\overline{u}_{b^*}(b^*) - \overline{B}_1(b^*,b^*)\right) (\overline{v}_1^\prime(b^*) - \overline{v}_2^\prime(b^*))}
{(\overline{v}_1^\prime(b^*) - \overline{v}_2^\prime(b^*)) \overline{v}_3(b^*) - (\overline{v}_1(b^*) - \overline{v}_2(b^*)) \overline{v}_3^\prime(b^*)}.
\end{align*}
Furthermore, 
$
\overline{W}_1(x) = \overline{v}_1(x) \overline{v}_2^\prime(x) - \overline{v}_2(x) \overline{v}_1^\prime(x)$ and $
\overline{B}_1(x;b) = \overline{v}_1(x) \int_0^x
\frac{\overline{v}_2(y)}{\overline{W}_1(y)} \frac{2(\Lambda + \lambda \alpha V_{b^*}^E(x))}{\sigma^2(y)} \, dy - 
\overline{v}_2(x) \int_0^x \frac{\overline{v}_1(y)}{\overline{W}_1(y)}
\frac{2(\Lambda + \lambda \alpha V_{b^*}^E(x))}{\sigma^2(y)} \, dy.
$ 
Here $\overline{B}_1(x;b)$ is a particular solution to $\frac{\sigma^2(x)}{2}g^{\prime\prime}(x) +
\mu(x)g^\prime(x)-(\lambda+\delta) g(x)+\lambda \alpha V_b^E(x)+\Lambda=0$ with $\overline{B}_1(0)=0$ and $\overline{B}_1^\prime(0)=0$.

The threshold $ b^* $ is determined through
\begin{align}
b^* = \inf \left\{ b > 0 : \overline{C}_3(b) \overline{v}_3^\prime(b) + \overline{u}_b^\prime(b) \leq \gamma - \beta \right \}.\label{b*000}
\end{align}
The function $ V_{b^*}^E $ can be computed using \eqref{14225-5}, with $ b $ replaced by $ b^* $.
Finally, $ b^* \leq b^*_E<\infty $.
\end{Theorem}
Theorem \ref{optimality-h} establishes the existence of equilibrium strategies and defines a rigorous procedure to determine them together with the associated value function. For their derivation, we need to compute the solutions to the given ODEs (whose existence and uniqueness are verified). As illustrated in later sections, in some cases these solutions can be determined explicitly. In other situations, numerical methods are required.


In the spirit of Lemma \ref{lemmm}, 
the following alternative representations for $ \overline{v}_3(\cdot) $ and $ \overline{u}_b(\cdot) $ will be helpful later for numerical evaluations. 
\begin{Lemma}
\label{1325-1-00}For the functions, $\overline{v}_3(\cdot)$ and $\overline{u}_b(\cdot)$,  we have 
\begin{align}
&\overline{v}_3(x)=\E\bigg[e^{-(\lambda+\delta) \hat{T}^x}\bigg],\label{1325-3}\\
&\overline{u}_b(x)=\E\bigg[\int_0^{\hat{T}^x\wedge \eta_0} e^{-\delta s}((\gamma-\beta) \overline{l}+\Lambda)\dif s+I\{\hat{T}^x>\eta_0\}\;\alpha \;e^{-\delta \hat{T}^x}\;V_b^E(Y^x_{\hat{T}^x})\bigg],\label{1325-4}
\end{align}
where $Y_t^x$ is a stochastic process and $\hat{T}^x$ is a stopping time  defined by
$Y_t^x=x+\int_0^t(\mu(Y_s^x)-\bar{l})ds +\int_0^t \sigma(Y_s^x)ds$ for $\quad s >0$,  and $ \hat{T}^x=\inf\{t\ge 0: Y_t^x\le 0\}$, respectively.  Moreover,  the following limiting results hold:
\begin{align}
&\lim_{x\rightarrow\infty} \overline{v}_3(x)=0,\quad 
\lim_{x\rightarrow\infty} \overline{u}_b(x)=\frac{\lambda\alpha+\delta}{\lambda+\delta}\frac{((\gamma-\beta)\bar{l}+\Lambda)}{\delta}.\label{4325-1}
\end{align}
\end{Lemma}

Similar to the last section, the two functions $\bar{v}_3$ and $\bar{u}_b$ can be computed numerically by selecting sufficiently large $\bar{x}$ and $\bar{y}$ and solving the following two boundary value ODEs, respectively: 
$\frac{\sigma^2(x)}{2} g^{\prime\prime}(x) + (\mu(x) - \bar{l}) g^\prime(x) - (\lambda+\delta) g(x) = 0$ 
 with $g(0)=1$ and $ g(\bar{x})=0$,
and
$\frac{\sigma^2(x)}{2} g^{\prime\prime}(x) + (\mu(x) - \bar{l}) g^\prime(x) - (\lambda+\delta) g(x) + \Lambda + (\gamma - \beta) \bar{l} + \lambda \alpha V_b^E(x) = 0$ 
  with $g(0)=0$ and $g(\bar{y})=\frac{\lambda\alpha+\delta}{\lambda+\delta}\frac{((\gamma-\beta)\bar{l}+\Lambda)}{\delta}$.

\subsection{The Brownian motion model}\label{subsec:BM-nonexp}
For the case \eqref{that} with constant coefficients, we obtain (cf.\ Appendix \ref{appendix:BM-nonexp} for details)
\begin{equation*}
V_{b}(x)=%
\begin{cases}
N_{1}(b)(e^{\tilde{\theta} _{1}x}-e^{-\tilde{\theta} _{2}x})+P_{3}(x;b), & 0\leq x<b, \\
N_{4}(b)e^{-\tilde{\theta} _{4}x}+P_{5}(x;b), & x\geq b,
\end{cases}
\end{equation*}
where 
\begin{align}
    &\tilde{\theta}_1=\frac{-\mu+\sqrt{\mu^2+2\sigma^2(\lambda+\delta)}}{\sigma^2},\quad \quad \tilde{\theta}_2=\frac{\mu+\sqrt{\mu^2+2\sigma^2(\lambda+\delta)}}{\sigma^2},\nonumber\\
    &\tilde{\theta}_3=\frac{-(\mu -\bar{l})+\sqrt{(\mu -\bar{l})^2+2\sigma^2(\lambda+\delta)}}{\sigma^2},\quad \quad \tilde{\theta}_4=\frac{(\mu -\bar{l})+\sqrt{(\mu -\bar{l})^2+2\sigma^2(\lambda+\delta)}}{\sigma^2},\nonumber\\
&P_{3}(x;b)=-\frac{2\Lambda }{\sigma ^{2}}\frac{e^{\tilde{\theta} _{1}x}-1}{\tilde{\theta}_{1}(\tilde{\theta} _{1}+\tilde{\theta} _{2})}
+\frac{2\Lambda }{\sigma ^{2}}\frac{1-e^{-\tilde{\theta} _{2}x}}{\tilde{\theta} _{2}(\tilde{\theta} _{1}+\tilde{\theta} _{2})}
+\frac{2\lambda \alpha }{\sigma ^{2}\tilde{\theta}  _{1}\tilde{\theta}  _{2}}M_{3}(b)\nonumber \\
&-\frac{2\lambda \alpha }{\sigma ^{2}(\tilde{\theta} _{1}+\tilde{\theta} _{2})}\left[\left( \frac{M_{1}(b)}{\tilde{\theta} _{1}-\theta _{1}}+\frac{M_{2}(b)}{\tilde{\theta} _{1}+\theta _{2}}+\frac{M_{3}(b)}{\tilde{\theta} _{1}}\right) e^{\tilde{\theta}_{1}x}
+\left( \frac{M_{1}(b)}{\theta _{1}+\tilde{\theta} _{2}}+\frac{M_{2}(b)}{\tilde{\theta} _{2}-\theta _{2}}+\frac{M_{3}(b)}{\tilde{\theta}_{2}}\right) e^{-\tilde{\theta} _{2}x}\right]\nonumber \\
&+\frac{2\lambda \alpha }{\sigma ^{2}(\tilde{\theta} _{1}+\tilde{\theta} _{2})}\left[\left( \frac{M_{1}(b)}{\tilde{\theta} _{1}-\theta _{1}}+\frac{M_{1}(b)}{\theta _{1}+\tilde{\theta} _{2}}\right) e^{\theta_{1}x}
+\left( \frac{M_{2}(b)}{\tilde{\theta} _{1}+\theta _{2}}+\frac{M_{2}(b)}{\tilde{\theta} _{2}-\theta_{2}}\right) e^{-\theta_{2}x}\right], \label{P325} 
\end{align}
\begin{align}
&P_{5}(x;b)=\frac{2((\gamma-\beta) \bar{l}+\Lambda)}{\sigma ^{2}}\frac{1}{\tilde{\theta}_{3}(\tilde{\theta} _{3}+\tilde{\theta} _{4})}
+\frac{2((\gamma-\beta) \bar{l}+\Lambda)}{\sigma ^{2}}\frac{1-e^{-\tilde{\theta} _{4}x}}{\tilde{\theta} _{4}(\tilde{\theta} _{3}+\tilde{\theta} _{4})}
+\frac{2\lambda \alpha }{\sigma ^{2}\tilde{\theta}  _{3}\tilde{\theta}  _{4}}M_{5}(b)\nonumber \\
&-\frac{2\lambda \alpha }{\sigma ^{2}(\tilde{\theta} _{3}+\tilde{\theta} _{4})}\left( \frac{M_{4}(b)}{\tilde{\theta} _{4}-\theta_{4}}+\frac{M_{5}(b)}{\tilde{\theta}_{4}}\right) e^{-\tilde{\theta} _{4}x}
+\frac{2\lambda \alpha }{\sigma ^{2}(\tilde{\theta} _{3}+\tilde{\theta} _{4})}\left( \frac{M_{4}(b)}{\tilde{\theta} _{3}+\theta _{4}}+\frac{M_{4}(b)}{\tilde{\theta} _{4}-\theta _{4}}\right) e^{-\theta_{4}x},\label{P525}\\
&N_{1}(b)=\frac{\tilde{\theta}  _{4}(P_{5}(b;b)-P_{3}(b;b))+P_{5}^{\prime }(b;b)-P_{3}^{\prime}(b;b)}{(\tilde{\theta} _{1}+\tilde{\theta} _{4})e^{\tilde{\theta}_{1}b}+(\tilde{\theta}_{2}-\tilde{\theta} _{4})e^{-\tilde{\theta} _{2}b}},\label{N125} \\
&N_{4}(b)=\frac{N_{1}(b)(e^{\tilde{\theta} _{1}b}-e^{-\tilde{\theta} _{2}b})+P_{3}(b;b)-P_{5}(b;b)}{e^{-\tilde{\theta}_{4}b}},\label{N425}\\
&M_{1}(b)=K_{1}(b)-\frac{2\Lambda }{\sigma ^{2}}\frac{1}{\theta
_{1}(\theta _{1}+\theta _{2})},\quad 
M_{2}(b)=-K_{1}(b)-\frac{2\Lambda }{\sigma ^{2}}\frac{1}{\theta_{2}(\theta _{1}+\theta _{2})},\label{M125}\\
&M_{3}(b)=\frac{2\Lambda }{\sigma ^{2}}\frac{1}{\theta _{1}\theta _{2}}, \quad
M_{4}(b)=K_{4}(b)-\frac{2((\gamma-\beta) \bar{l}+\Lambda )}{\sigma ^{2}}\frac{1}{\theta _{4}(\theta _{3}+\theta _{4})},\label{M425}\\
&M_{5}(b)=\frac{2((\gamma-\beta) \bar{l}+\Lambda )}{\sigma ^{2}}\frac{1}{\theta_{3}\theta _{4}}.\label{M525}
\end{align}

Here $b^*$ is the solution of $-\tilde{\theta}_4 N_{4}(b)e^{-\tilde{\theta} _{4}b}+P_{5}^{\prime }(b;b)=\gamma-\beta$.

\section{Probability of Early Depletion}\label{expectation}
A further quantity of interest is the probability of early depletion when implementing the optimal threshold strategy with and without taking into consideration the present-biasedness of the decision makers. For any threshold strategy $L^{b}$ with (not necessarily optimal) threshold $b$ we define the time of depletion
\begin{equation}
  \tau^b=\inf \left \{t\geq 0: X_t^b  = 0 \right \},\label{ruin-time_barr}
\end{equation}
where $X_t^b$ follows the dynamics \eqref{dynamics-00s} for the threshold strategy $L={L_t^b}$. Note that $\pr(\tau^b<\infty)=1$, as soon as $\bar{l}\ge \mu(x)$ for all $x\ge 0$. Its Laplace transform $\widetilde{L^b}(x;s):=\E_x \left[e^{-s \tau^{b}}\right]$ is more amenable for analytical expressions (see e.g.\ \cite{gerber1998time,albrechercani}), and in the present case is given as follows (see Appendix \ref{aD} for the proof). 
\begin{Theorem}\label{tm:laplace}For any $b\ge 0$ we have
 \begin{eqnarray*}
\widetilde{L_b}(x;s)=\left\{\begin{array}{ll}
C_4(b;s)v_4(x;s)+v_5(x;s),&0\le x< b,\\
C_6(b;s) v_6(x;s)+u(x;s),&x\ge b,
\end{array}\right.
\end{eqnarray*}
where
for $v_4(\cdot; s)$ and $v_5(\cdot;s)$ are the unique solutions to 
  $    \frac{\sigma^2(x)}{2}g^{\prime\prime}(x)
+\mu(x)g^\prime(x)-s g(x)=0$ on $[0,\infty)$ 
with initial values, respectively, 
$v_4(0;s)=0$ and $v_4^\prime(0;s)=1$, and \\
    $v_5(0;s)=1$ and $ v_5^\prime(0;s)=1$.
Likewise, $v_6(x;s)$ and $u(x;s)$ are the unique bounded solutions to 
$\frac{\sigma^2(x)}{2}g^{\prime\prime}(x)+
(\mu(x)-\bar{l})g^\prime(x)-s g(x) =0$ with initial value $v_6(0;s)=0$ and $u(0;s)=1$, respectively, and 
\begin{align}
C_4(b;s) &= \frac{\left(u(b;s) - v_5(b;s)\right) v_6^\prime(b;s) - \left(u^\prime(b;s) - v_5^\prime(b;s)\right) v_6(b;s)}{v_4(b;s) v_6^\prime(b;s) - v_4^\prime(b;s) v_6(b;s)}, \label{C425}\\
C_6(b;s) &= \frac{\left(u^\prime(b;s) - v_5^\prime(b;s)\right) v_4(b;s) - \left (u(b;s) - v_5(b;s)\right) v_4^\prime(b;s)}{v_4(b;s) v_6^\prime(b;s) - v_4^\prime(b;s) v_6(b;s)}.\label{C625}
\end{align}

\end{Theorem}
Define the finite-time depletion probability
\begin{align}
    &\psi_b(x;t)=\pr_x\left(\tau^b\le t\right),\quad x,t\ge 0.    \end{align}
Clearly, $\psi_b(x;t)$ can be obtained as the inverse Laplace transform w.r.t.\ $s$ of $\widetilde{L_b}(x;s)/s$. Finally, the finite-time depletion probability of the optimal strategy under exponential discounting is denoted by $\psi^E(x;t):=\psi_{b^E}(x;t)$ and under stochastic quasi-hyperbolic discounting by $\psi^*(x;t):=\psi_{b^*}(x;t)$, where $b^E$ and $b^*$ are the optimal thresholds in the respective cases.

\section{Numerical Illustrations for the Brownian Motion Model}\label{brownian}
In this section, we present a numerical illustration that allows quantitative insight into the impact of present-bias on the optimal production and emission strategies identified in the previous sections. We focus here on the Brownian model (a model with surplus-dependent diffusion coefficients will be considered in Section \ref{surp}). We first need to choose numerical values for the involved parameters whose magnitudes are motivated by practical considerations, but naturally remain rough magnitudes. Suppose the total global remaining carbon budget until 2050 (as of $2025$) is $340\ \text{GtCO}_2$ (which is a rough estimate based on the \cite{GlobalCarbonProject2022}. 
If we consider the company under consideration to receive a share of $0.0001\%$ of that amount, we have $x_0=34$ (in units of $10^4$ $\text{tCO}_2$)
to be used over the next $25$ years. In terms of the drift parameter, we may assume that the expected annual increase in capacity due to Direct Air Capture (DAC) and other carbon removal technology advancement could be set to $\bar{\mu}=0.05$.\footnote{\cite{Lebling2025CCUS} estimate that the total global CCUS (Carbon Capture, Utilization and Sequestration) capacity will reach between $416$ and $520$ MtCO$_2$/yr. The above choice refers to the proportional share for the value $500$ MtCO$_2$/yr.} Furthermore we choose the volatility parameter to be $\sigma=2$.\footnote{One may justify such a magnitude as follows. The parameter $\sigma$ captures uncertainties due to both earth system dynamics and technology development. One may want to use the concept of interannual variability (IAV) as a basis here. 
In the context of the carbon cycle, IAV commonly describes annual variations in net ecosystem exchange (NEE), net primary production (NPP), or the carbon sink strength. 
According to \cite{MarcollaRodenbeckCescatti2017}, the average annual NEE globally is approximately $120~\text{gC}~\text{m}^{-2}~\text{yr}^{-1}$, which means terrestrial ecosystems absorb around $120$ grams of carbon per square meter each year. The reported IAV is about $15$--$20~\text{gC}~\text{m}^{-2}~\text{yr}^{-1}$, implying a relative fluctuation of $12\%$--$17\%$ of the mean capacity. 
For simplicity, we assume the IAV  to be $6\%$ of the initial capacity $x_0=34$ (since the emission capacity declines as the budget is gradually depleted): 
 leading to $\sigma=34\cdot 0.06=2$.}

\subsection{Impact of present-bias on the emission schedule }\label{sec7}

Recall that the parameter $\lambda$ (the arrival intensity of the future periods, and correspondingly the `disappearance intensity' of the present period) and the discounting weight $\alpha$ capture the impatience of the decision-maker. 
For any fixed $\alpha$, a larger $\lambda$ implies a higher intensity rate for eliminating the present period and transitioning to future periods. Since future periods are discounted further by the additional factor $\alpha<1$, this leads to greater impatience and stronger present-bias.
At the same time, for any fixed $\lambda > 0$, a smaller $\alpha$ places less weight on future cash flows, also indicating higher impatience. Note that either $\lambda = 0$ or $\alpha = 1$ remove the present-bias. \\


\noindent \textbf{Impact of $\lambda$.} In addition to 
$x_0=34$, $\bar{\mu}=0.05$ and $\sigma =2$, we set $\underline{l}=1$,  $\delta =0.1$, $\alpha =0.9$, $\gamma =0.9$, $\overline{\Lambda} =0.5$, $\overline{l}=3$,  $c_{ind}\,=0.1$ and $c_{tax}=0.05$. Then $\beta=c_{ind}\,\gamma +c_{tax} =0.086$ and  $\Lambda=\overline{\Lambda}+(\gamma-\beta)\underline{l}=1.344$. Note that with this choice $\underline{l}=1$ of the baseline emission rate, we have a negative net drift $\mu(0)=-0.95$ for the optimization problem.
The optimal threshold levels $b^*$, calculated using the formulas from Section \ref{sol-hyber-sec} together with the resulting early depletion probabilities $\psi^*(34;25)$ within the next 25 years, are given in the following tables for various combinations of $\lambda$ and $\alpha$. 

\begin{table}[H]
\centering
\scriptsize 
\setlength{\tabcolsep}{2.5pt} 
\renewcommand{\arraystretch}{1.1} 

\begin{minipage}[t]{0.47\textwidth}
\centering
\begin{tabular}{|c|c|c|c|c|c|c|}
    \hline
    \textbf{$\lambda$} & 0 & 0.1 & 0.25 & 1 & 4 & 12 \\
    \specialrule{.1em}{.05em}{.05em}
    \textbf{$b^*$} & \cellcolor{gray!30}5.51 & 5.08 & 4.68 & 3.86 & 3.06 & 2.58 \\
    \hline
    $\psi^*(34;25)$ & 0.9968 & 0.9985 & 0.9993 & 0.9997 & 1 & 1 \\
    \hline
\end{tabular}
\caption{Optimal $b^*$ and $\psi^*(34;25)$ for varying $\lambda$ ($\alpha=0.9$).}
\label{tab:lambda_optimal-sig=2}
\end{minipage}\hspace{0.04\textwidth}
\begin{minipage}[t]{0.47\textwidth}
\centering
\begin{tabular}{|c|c|c|c|c|c|c|}
    \hline
    \textbf{$\alpha$} & 0.5 & 0.7 & 0.8 & 0.9 & 0.95 & 1.0 \\
    \specialrule{.1em}{.05em}{.05em}
    \textbf{$b^*$} & 0.83 & 1.66 & 2.53 & 3.86 & 4.66 & \cellcolor{gray!30}5.51 \\
    \hline
    $\psi^*(34;25)$ & 1 & 1 & 1 & 0.9997 & 0.9994 & 0.9968 \\
    \hline
\end{tabular}
\caption{Optimal $b^*$ and $\psi^*(34;25)$ for varying $\alpha$ ($\lambda=1$).}
\label{tab:alpha_optimal-sig=2}
\end{minipage}
\end{table}

In Table \ref{tab:lambda_optimal-sig=2}, we fix $\alpha=0.9$ and vary $\lambda$ within the interval $[0, 12]$ (recall that $\lambda = 0$ refers to no present-bias, and larger values of $\lambda$ indicate a higher degree of present-bias). Specifically, $\lambda = 1$ implies that the expected duration of the “present” period is 1 year, $\lambda = 0.25$ corresponds to an expected present period of 4 years, and $\lambda = 12$ represents an extremely impatient case in which the present period lasts, on average, only 1 month. As expected, the optimal threshold for excess production/emission decreases with increasing present-bias from the exponential discounting case $\lambda=0$, implying earlier excess emissions and higher overall emission amounts. 
 This suggests that if policies are made under the assumption that decision-makers are not present-biased, while in reality they are, the  amount of resulting emissions is underestimated.
 
\begin{figure}[htbp]
    \centering
    \includegraphics[width=0.6\linewidth]{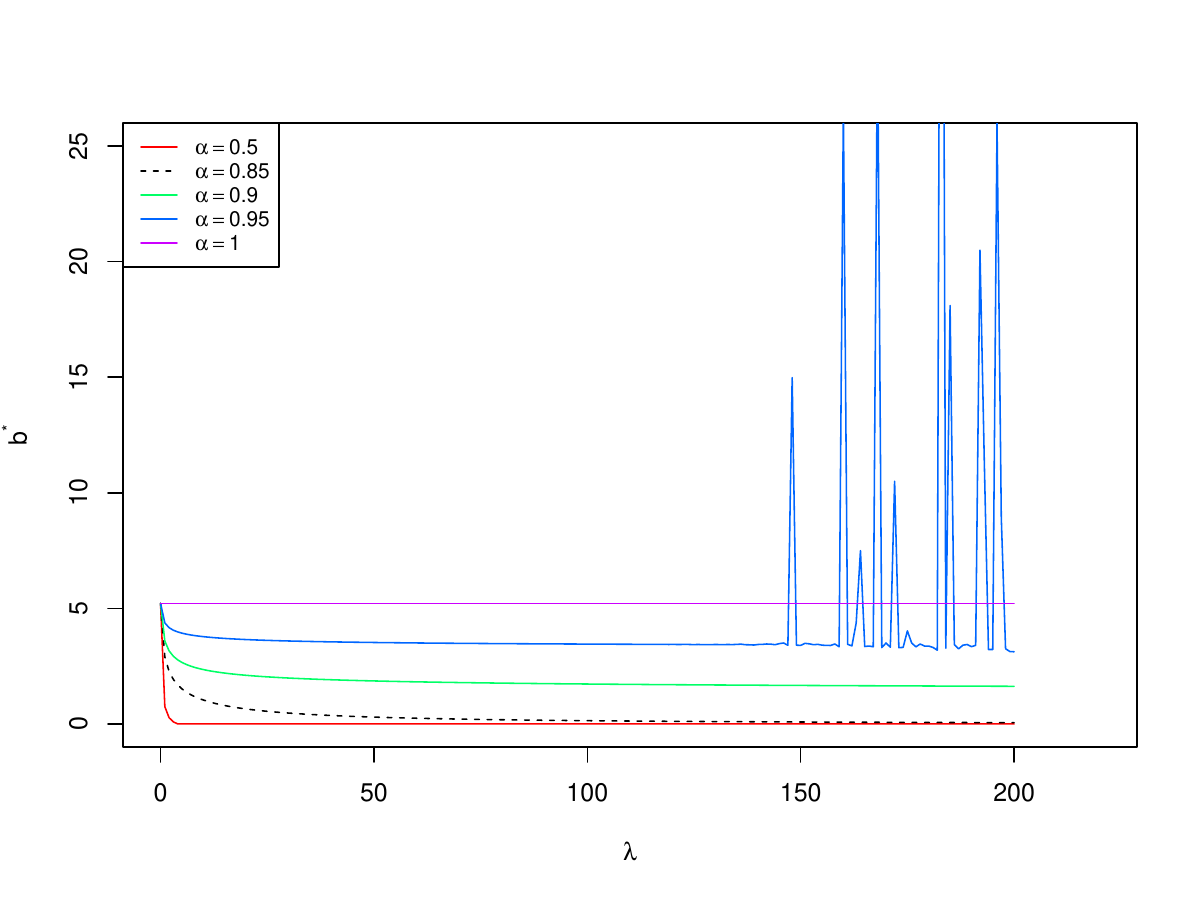}
    \caption{Optimal threshold levels $b^*$ as a function of $\lambda$ for various values of $\alpha$. }
    \label{fig:bstar_lambda_multi_alp_sig=2}
\end{figure}

 Although large values of $\lambda$ are not central to our analysis, 
 we highlight an interesting phenomenon that may be of mathematical interest, particularly in theoretical studies of hyperbolic discounting. When $\lambda$ is very large, the impact of increasing $\lambda$ on strategies may become non-monotonic, cf.\ Figure~\ref{fig:bstar_lambda_multi_alp_sig=2}. For $\alpha$-values close to 1, the interaction between parameters can lead to higher threshold levels (i.e., lower emissions) as $\lambda$ increases. This suggests that when the present period is extremely short but future profits are still significantly weighted, it may be optimal to prioritize maximizing total future benefits resulting in reduced present emissions, as reflected in a higher threshold. At the same time, for other large values of $\lambda$ the threshold is smaller again, making it more beneficial to prioritize immediate gains. Such a non-monotonicity is noteworthy, although it only appears for certain specific parameter ranges. 
\\
 
\noindent
\textbf{Impact of $\alpha$.} Table \ref{tab:alpha_optimal-sig=2} shows the results for $\lambda = 1$ and variable values of $\alpha$. One observes that the sensitivity of $b^*$ to changing values of $\alpha$ is more pronounced ($\alpha = 1$ again refers to exponential discounting without present-bias). 
 As $\alpha$ decreases from $1$, the present-bias is increased and the excess production threshold is lowered.\\

\begin{figure}[htbp]
    \centering
    \includegraphics[width=0.6\linewidth]{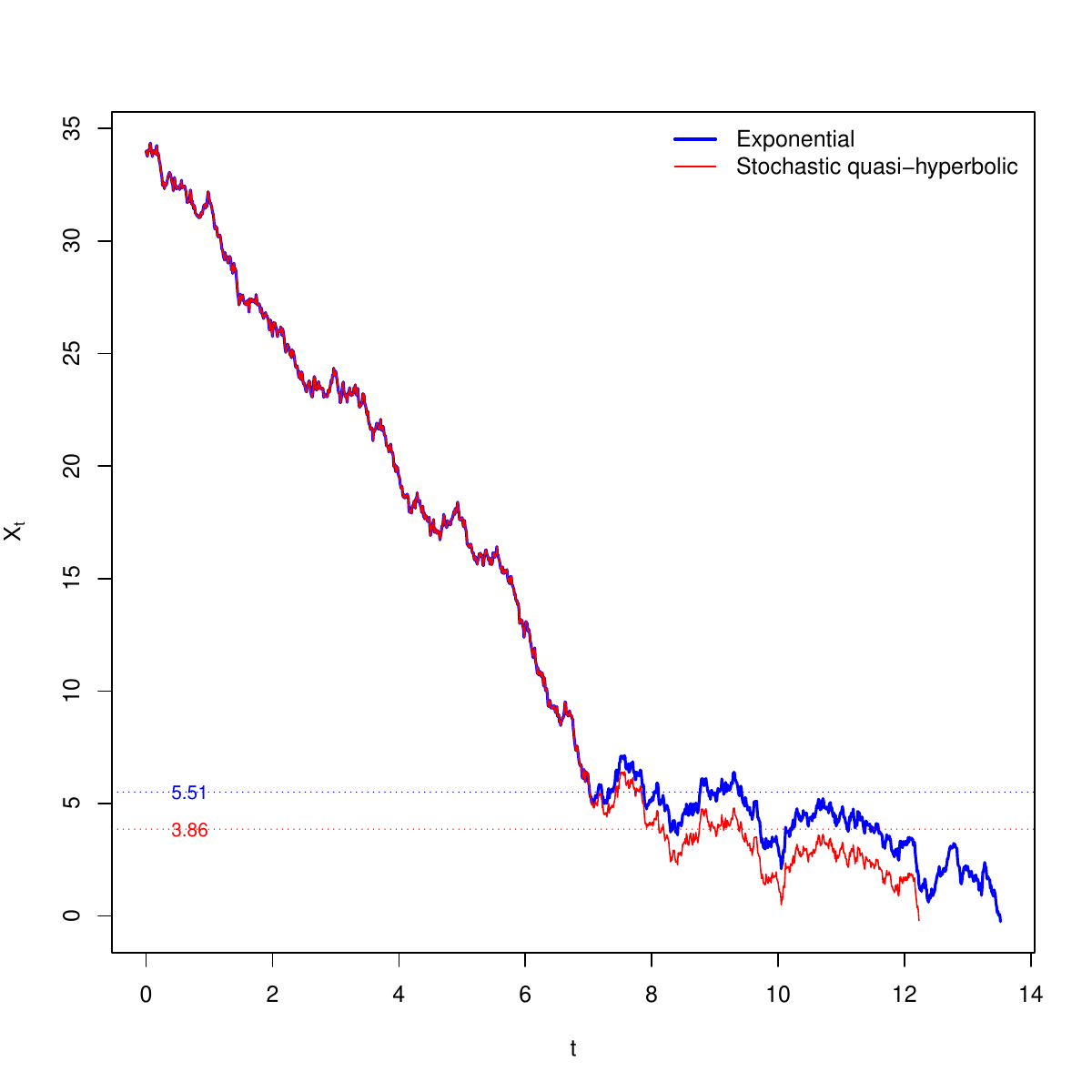}
       \caption{Comparison of two sample paths generated with the same random seed: optimal strategy under exponential discounting (with optimal threshold $b_E^* = 3.86$) versus stochastic quasi-hyperbolic discounting ($\lambda = 1$, $\alpha = 0.9$) with optimal threshold $b^* = 5.51$.}
          \label{fig:Paths_sig=2}
          \end{figure}

\noindent \textbf{Comparison to Exponential Discounting.} In line with Theorem \ref{optimality-h}, we observe that the excess production threshold is lower under present bias than under exponential discounting, which results in increased emissions and a reduced budget, as illustrated in Figure~\ref{fig:Paths_sig=2} with a comparison of two sample paths generated with the same random seed.  In both cases, the budget is fully depleted before time $T=25$. Note also that for the concrete choice of parameters, the strategies conincide for the first seven years. 

\begin{figure}[H]
    \centering
    \includegraphics[width=0.4\linewidth]{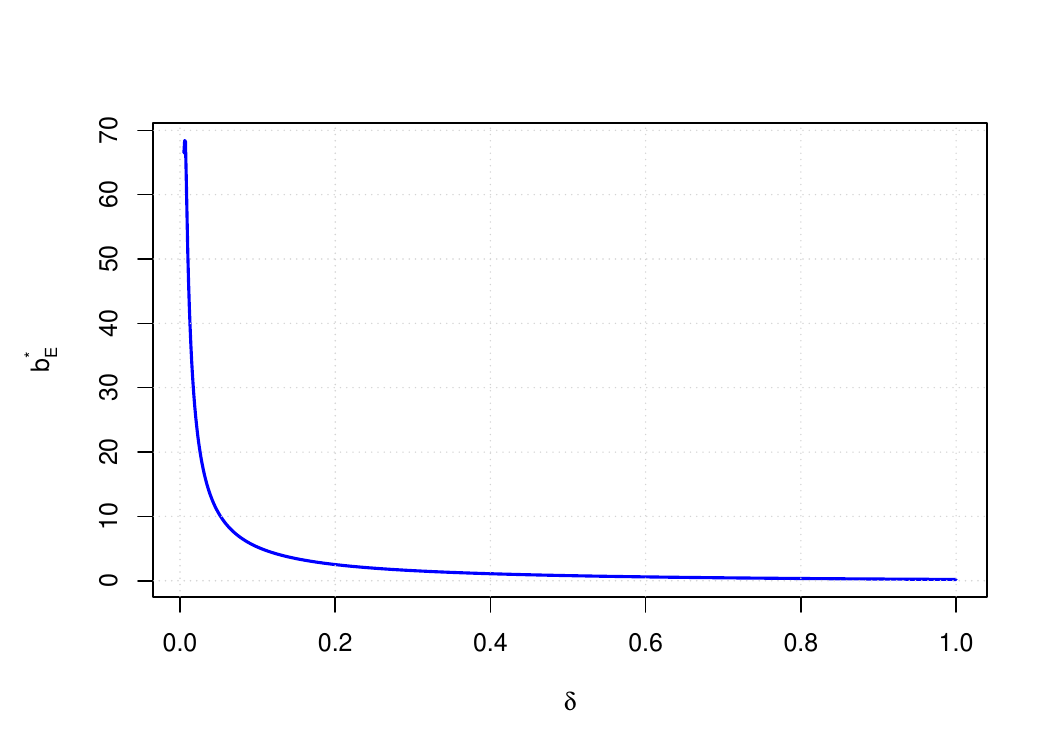}
    \caption{Optimal threshold $b_E^*$ as a function of $\delta$ for exponential discounting ($c_{tax}=0$).}
    \label{fig:bEvsDelta_c2=0_sig=2}
\end{figure}

\begin{figure}[H]
\centering
\begin{minipage}[t]{0.4\textwidth}
    \centering
    \includegraphics[width=\linewidth]{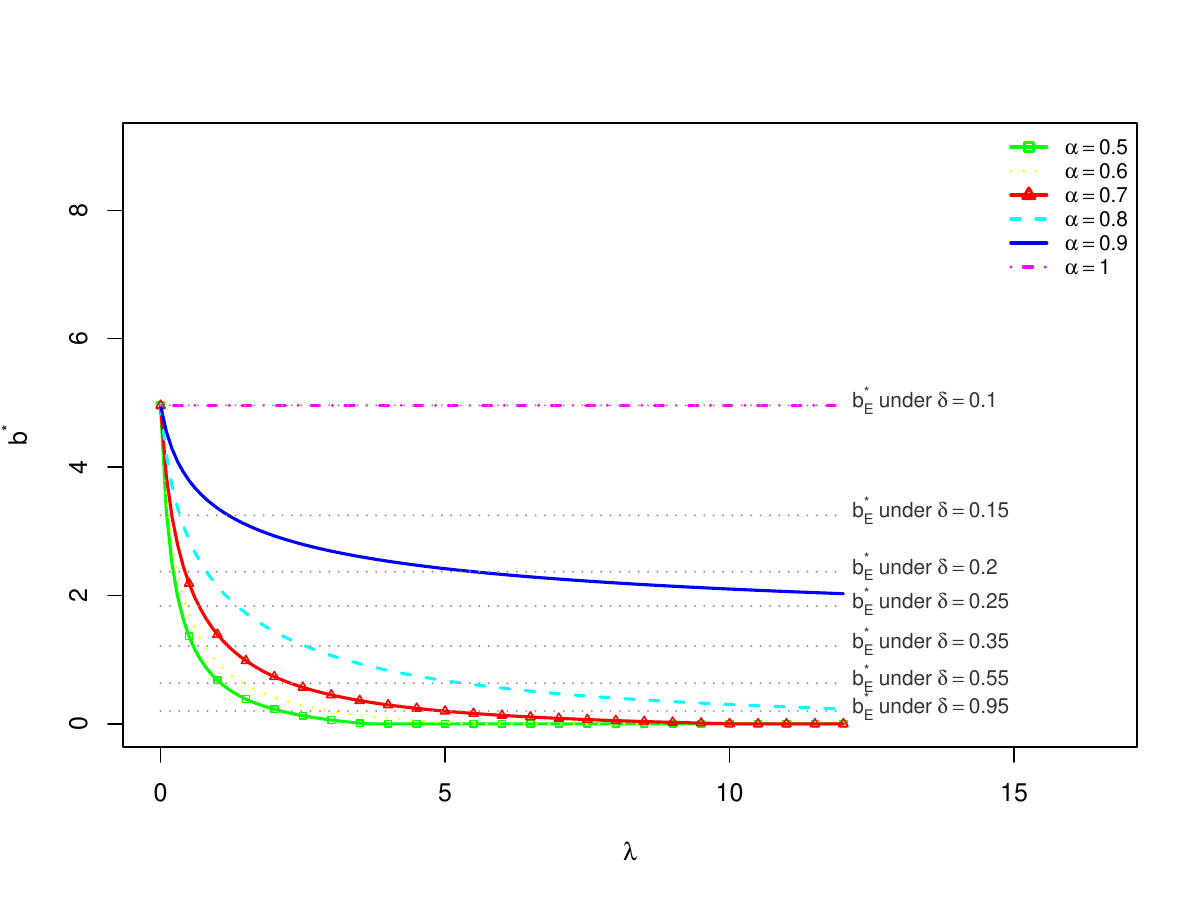}
    \caption{Comparison of optimal threshold levels $b^*$ for stochastic hyperbolic discounting (with $\delta=10\%$) and for exponential discounting (for various $\delta$-values, indicated by black dotted horizontal lines), as a function of $\lambda$.}
    \label{fig:exp-nonexp_vs_lambda_multi_alp_sig=2}
\end{minipage}
\hspace{0.04\textwidth} 
\begin{minipage}[t]{0.47\textwidth}
    \centering
    \includegraphics[width=\linewidth]{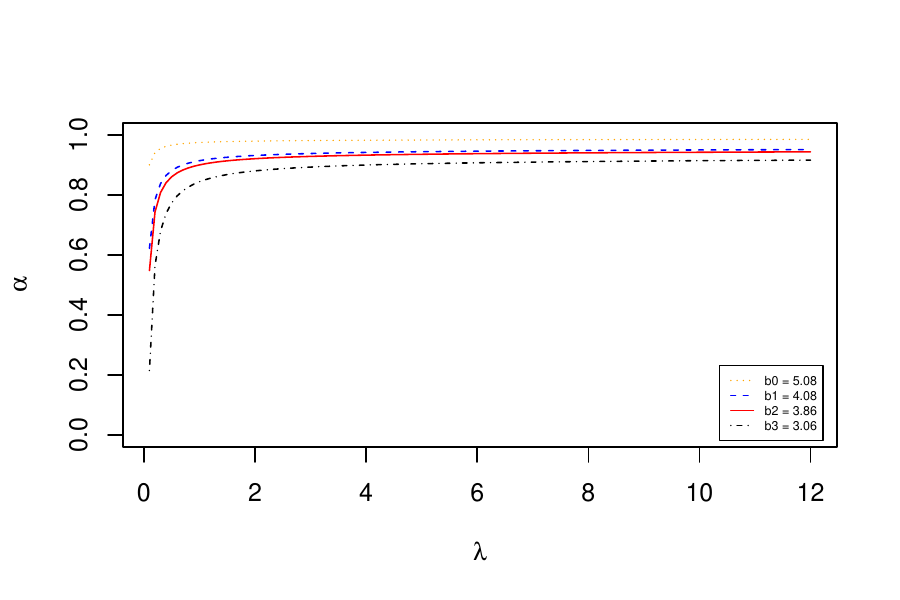}
    \caption{Pairs of $\lambda$ and $\alpha$ that yield the same equilibrium threshold $b^*$ for different target values $b^*= b_0, b_1, b_2$, or $b_3$.}
    \label{fig:alpha_lambda_given_b_star}
\end{minipage}
\end{figure}

In the exponential discounting case without present-bias, one should also expect a lower optimal threshold when increasing the discount rate $\delta$, which downgrades future contributions. Indeed, Figure  \ref{fig:bEvsDelta_c2=0_sig=2} depicts the optimal threshold level under exponential discounting for the above parameters, now as a function of $\delta$. 
It is of interest to compare which level of $\delta$ without present-bias leads to the same optimal decisions (threshold levels) as the effect of present-bias for a lower level of $\delta$. Figure \ref{fig:exp-nonexp_vs_lambda_multi_alp_sig=2} shows the optimal threshold for equilibrium strategies of various stochastic quasi-hyperbolic discounting settings $(\lambda, \alpha)$ for $\delta=0.1$. 
The dotted horizontal lines represent the optimal threshold levels $b_E^*$ for exponential discounting for various other $\delta$-levels, so that one can identify which parameters in each of the two discounting regimes lead to the same eventual optimal strategy. For instance, the equilibrium strategy for $\delta=0.1$, $\lambda = 2$ and $\alpha=0.9$ under stochastic quasi-hyperbolic discounting leads to the same threshold (and hence value function) as exponential discounting with a discount rate around $\delta=0.15$ (Figure \ref{fig:alpha_lambda_given_b_star} gives a more detailed account on matching levels). 
In other words, the effect of present bias in this case is comparable to increasing the exponential discount rate $\delta$ from 
 0.1 to 0.15. This raises the question of whether explicitly accounting for present bias could, in general, be replaced by using a higher discount rate within a standard exponential discounting model. The answer is no, and we will elaborate on this in Section \ref{carbon}  (Remark \ref{remm}).

\subsection{Impact of level of social responsibility on emission schedule }
Recall that the term $\overline{\Lambda}$ in the objective function rewards avoiding early depletion of the carbon budget. 
We can interpret it as a measure of how much the company values preserving its emission budget, which can, to some extent, reflect its social responsibility and sustainability awareness.  

\begin{figure}[H]
\begin{subfigure}[t]{0.45\textwidth}
    \centering
    \includegraphics[width=\textwidth]{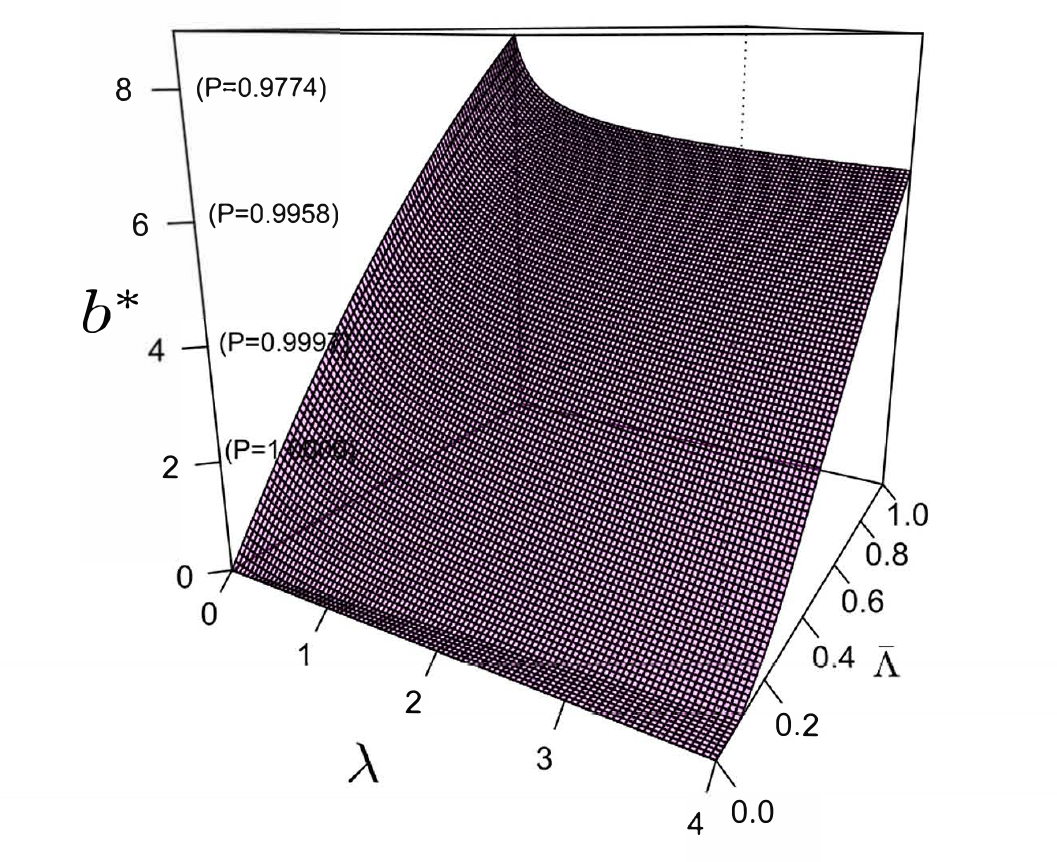}
\end{subfigure}
\begin{subfigure}[t]{0.45\textwidth}
    \centering
    \includegraphics[width=\textwidth]{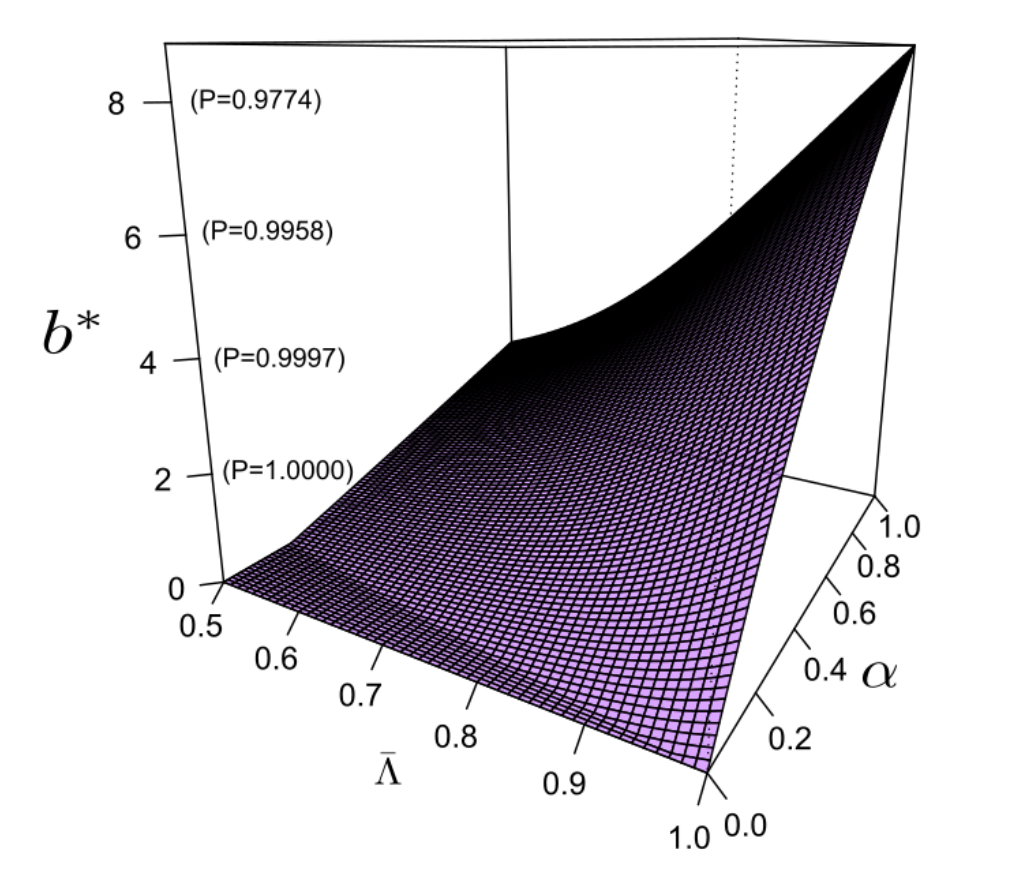}
\end{subfigure}
\caption{Optimal threshold $b^*$ (and corresponding early depletion probability $P=\psi^*(34;25)$) of the equilibrium strategy under different behavioral parameters with $\alpha = 0.9$: 
as a function of future arrival intensity $\lambda$ and $\bar{\Lambda}$ (left) and as a function of the additional future period discount factor $\alpha$ and and the social responsibility factor $\bar{\Lambda}$ (right). }
\label{fig:combined_bstar_plots_sig=2}
\end{figure}

We now examine how $\overline{\Lambda}$ affects decision-making by varying its value while keeping all other parameters fixed. Fixing again $\delta = 0.1$, $\alpha=0.9$, $\lambda = 1$, we now vary $\bar{\Lambda}$ (and correspondingly $\Lambda=\bar{\Lambda}+(\gamma-\beta)\underline{l}$). Table \ref{tab:CLambda_optimal_sig=2} shows how 
additional weight on sustainability increases the optimal excess production/emission threshold in  the present-biased case. As expected, higher 
sustainability awareness postpones emissions, resulting in lower overall emissions.
Figure \ref{fig:combined_bstar_plots_sig=2} gives a more detailed picture on how choices of present-bias parameters and the sustainability weight $\bar{\Lambda}$ affect the optimal emission schedule. It quantifies how $b^*$ changes as a function of intensity $\lambda$ of arrival for the future period, weight $\alpha$ for future profits and sustainability weight, $\bar{\Lambda}$, respectively. Along the vertical axis, the probability of early depletion resulting from implementing $b^*$ is also indicated.

\begin{table}[H]
    \centering
\begin{tabular}{|c|c|c|c|c|c|}
\hline
$\overline{\Lambda}$ & 0.0 & 0.1 & 0.2 & 0.5 & 0.8 \\
\hline
$b^*$ & 0 & 0.52 & 1.32 & 3.86 & 6.12 \\
\hline
\end{tabular}
    \caption{Optimal threshold levels for varying $\bar{\Lambda}$}
    \label{tab:CLambda_optimal_sig=2}
    \end{table}

\subsection{Impact of carbon tax on emission schedule}\label{carbon}
Finally, we want to examine how setting a carbon tax level $c_{tax}$ impacts decision-making by varying its value while keeping all other parameters fixed. Choosing again $c_{ind}\,=0.1$, $x_0=34$,  $\bar{\mu} = 0.05$, $\underline{l} = 1$, $\sigma = 2$, $\delta = 0.1$, $\gamma = 0.9$,  $\bar{\Lambda} =0.5$,  $\bar{l} = 3$  and $\alpha=0.9$, we now vary $c_{tax}$. Note that both  $\beta=c_{ind}\,\gamma +c_{tax} $ and  $\Lambda=\bar{\Lambda}+(\gamma-\beta)\underline{l}$ vary with $c_{tax}$.

\begin{figure}[H]
    \centering
    \includegraphics[width=0.6\linewidth]{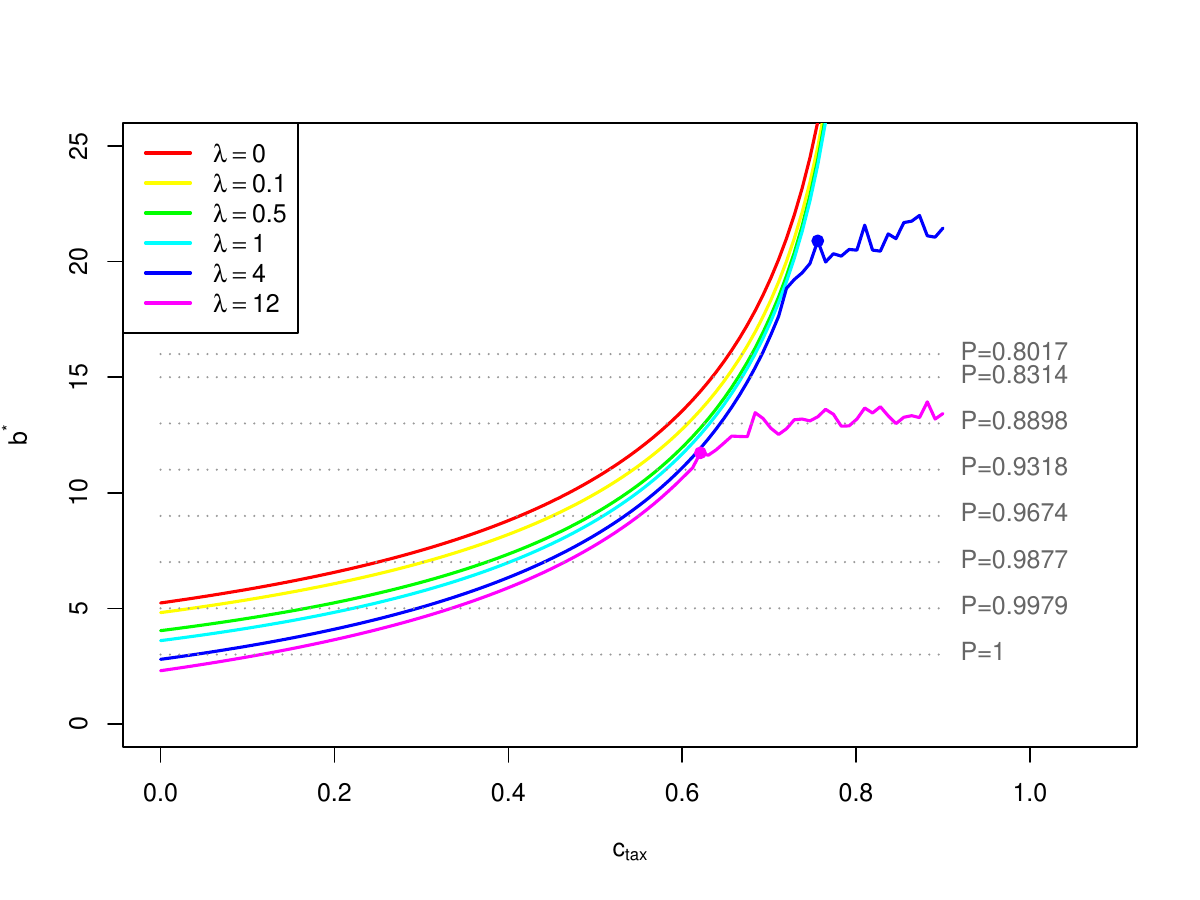}
    \caption{Optimal threshold $b^*$ as a function of tax rate $c_{tax}$ for various levels of $\lambda$ ($\alpha=0.9$).}
    \label{fig:bstar_carbonT_multi_lam_sig=2}
\end{figure}

 We calculate the optimal strategy for various levels of carbon tax and present the results for in Tables \ref{tab:carbon_tax_bstar_sig=2} and \ref{tab:carbon_tax_bstar_3_1_sig=2} for $\lambda=1$ and $\lambda=12$, respectively. Moreover, Figure \ref{fig:bstar_carbonT_multi_lam_sig=2} plots how  $b^*$ changes as the tax rate $c_{tax}$ increases, for various levels of present bias represented by different values of $\lambda$, with $\alpha$ fixed at $0.9$. The probability indicated at the end of each horizontal dashed line represents the likelihood of early depletion if the production strategy or emission policy uses the maximal excess production threshold at that level. For example, the second dashed line from the bottom indicates that if the production policy begins maximal excess production when the budget exceeds $5$, the probability of early depletion is $0.9979$.

\begin{table}\centering
\begin{tabular}{|c|c|c|c|c|c|c|c|c|c|c|c|c|}
\hline
\textbf{$c_{tax}$} & 0 & 0.05 & 0.1 & 0.3 & 0.5 & 0.7 & 0.8 & 0.802 & 0.803 & 0.804 & 0.805 & 0.809 \\
\hline
\textbf{$b^*$} & 3.36 & 3.58 & 3.83 & 5.21 & 7.72 & 13.71 & 22.47 & 22.79 & 22.95 & 23.12 & 23.29 & 23.99 \\
\hline
\end{tabular}
\caption{Optimal $b^*$ for varying carbon tax levels ($\lambda=1$)}
\label{tab:carbon_tax_bstar_sig=2}
\end{table}

\begin{table}
\centering
\resizebox{\textwidth}{!}{%
\begin{tabular}{|l|c|c|c|c|c|c|c|c|c|c|c|c|c|c|}
\hline
\textbf{ $c_{tax}$} &   0 
&   0.05 &   0.10 &   0.30 &   0.50 &  0.66&0.67&0.68&0.69& 0.70 &   0.72 &   0.73&   0.74 &   0.75  \\
\hline
$b^*$& 1.88 
& 2.14 & 2.42 & 3.92 & 6.52 &10.38&	\cellcolor{gray!30}24.00&	11.04&	\cellcolor{gray!30}19.00
 & 10.98 & \cellcolor{gray!30}11.16 & 10.98 &\cellcolor{gray!30} 38.00 & \cellcolor{gray!30}11.92  \\
\hline\hline
\textbf{$c_{tax}$}  &   0.76 &   0.77 &  0.78 &   0.79 &   0.8 & 0.801 & 0.802 & 0.803 & 0.804 & 0.805 & 0.806 & 0.807 & 0.808 & 0.809 \\
\hline
\textbf{$b^*$}& 11.49 & \cellcolor{gray!30}15.50& 10.88  & 10.99 & \cellcolor{gray!30}22.50  & 23.41 & 24.89 & 10.94 & 12.29 & \cellcolor{gray!30}22.48 & \cellcolor{gray!30}17.88 & 10.98 & \cellcolor{gray!30}31.00 & 10.99 \\
\hline
\end{tabular}
}
\caption{Optimal $b^*$ for varying carbon tax levels ($\lambda=12$)}
\label{tab:carbon_tax_bstar_3_1_sig=2}
\end{table}

One can observe that 
without carbon tax ($c_{tax} =0$) the excess production threshold $b^*$ is at its lowest, indicating a strong desire for early consumption of the budget and higher production, which results in higher emissions. As carbon tax increases, the incentive for production and consumption decreases (reflected in an increased threshold $b^*$), lowering carbon emissions. The curbing effect strengthens as the carbon tax increases, up to certain turning points that will be discussed below.

 Generally, for a higher present bias (larger $\lambda$), a larger tax rate is required to bring down the emission patterns to the same level as for lower present bias 
  (at least as long as the tax rate is not excessively high). 
  The concrete needed trade-off can be spotted in Figure \ref{fig:bstar_carbonT_multi_lam_sig=2}.
  Therefore, if carbon tax rates are designed ignoring present bias, they may fail to achieve their intended effect. For example, if a carbon policy is designed to restrict the probability of early depletion to around $96.74\%$, and present-bias is ignored,  $c_{tax}$ would be set around $0.4$ (see the red curve). 
  However, if there is some level of present-bias (e.g., $\lambda = 1$), to achieve that effect, the tax should have been set around $0.52$.

However, beyond a certain threshold (indicated by the dots on the curves in Figure \ref{fig:bstar_carbonT_multi_lam_sig=2} and the first cell highlighted in gray in Table \ref{tab:carbon_tax_bstar_3_1_sig=2} for the case $\lambda=12$) the impact of further increases in carbon tax  $c_{tax}$ becomes more variable. 
  This suggests that excessively high carbon tax rates may be suboptimal, particularly when combined with stronger present bias (larger $\lambda$), which aligns with findings  by \cite{MacKenzieOhndorf2012} that ``revenue-raising instruments, such as carbon taxes, are suboptimal” (see also \cite{BorissovBretschger2022}). One can observe in Figure \ref{fig:bstar_carbonT_multi_lam_sig=2} that this phenomenon is more pronounced (and occurs at lower $c_{tax}$ levels) for higher degrees of present-biasedness (higher $\lambda$). 

\begin{figure}[H]
    \centering
    \includegraphics[width=0.50\linewidth]{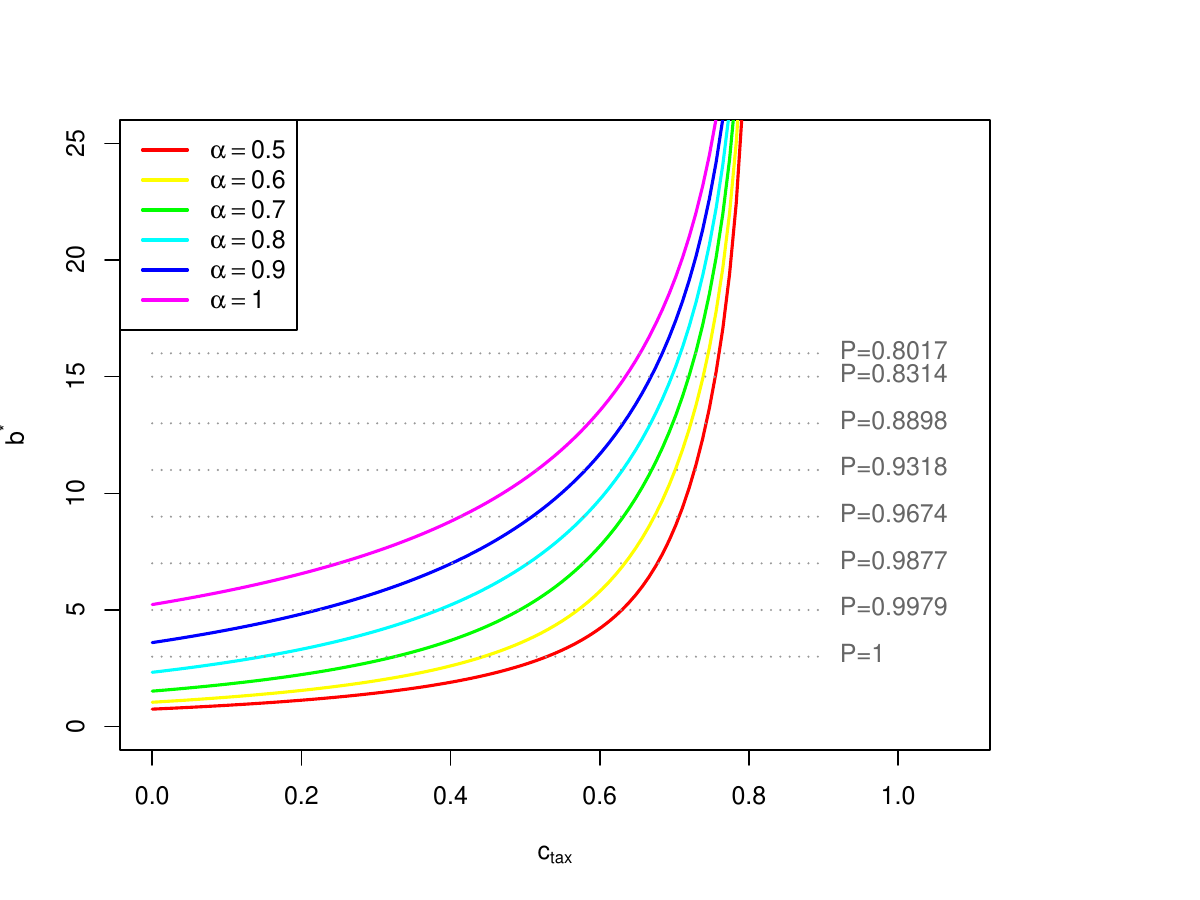}
    \includegraphics[width=0.49\linewidth]{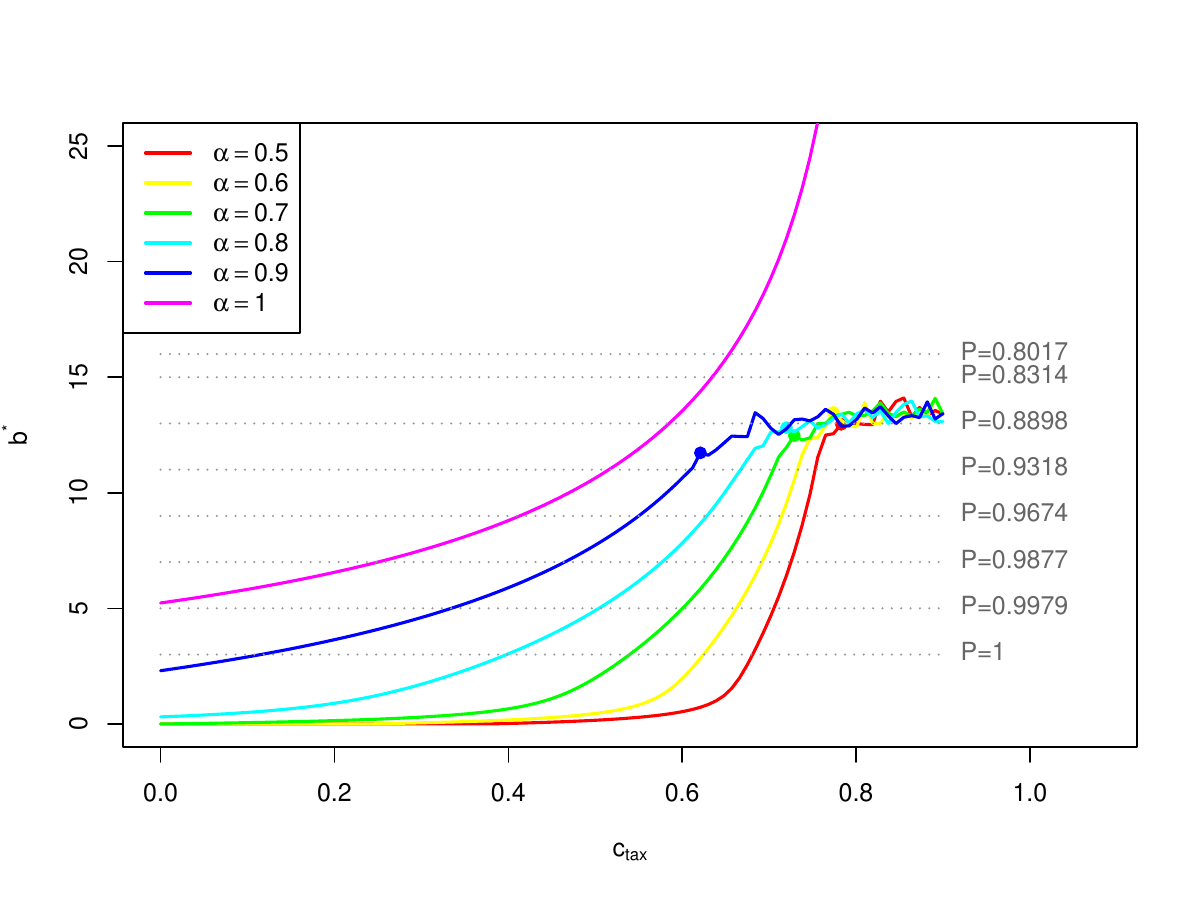}
\caption{Optimal threshold $b^*$ as a function of tax rate $c_{tax}$ for various levels of $\alpha$ ($\lambda=1$ (left) and $\lambda=12$ (right))}    
\label{fig123_sig=2}
\end{figure}

Figure \ref{fig123_sig=2} shows the effect of carbon tax for a fixed $\lambda > 0$ but varying $\alpha$ (which is another way to measure present-bias). It reveals similar patterns on the impact of carbon tax on production and emission strategies. If the tax rate is determined under the assumption that there is no present-bias, but in reality present-bias exists, then actual emissions will be higher than targeted. Specifically, if $c_{tax}$ is chosen using the curve corresponding to $\alpha = 1$ (no present bias) and based on a targeted probability of early depletion (e.g., $P = 80.17\%$, indicated by the first dashed line from the top), then the tax rate $c_{tax}$ would be approximately $0.64$. However, under present-biased preferences (e.g., $\alpha = 0.9$), the resulting production strategy under such a tax rate (around $0.64$) yields a lower threshold $b^*$, leading to a higher probability of early depletion—around $86\%$. This illustrates again that ignoring present-bias when setting policy negatively affects the achievement of emission targets set by social planners. Furthermore, present-bias may also undermine the effectiveness of carbon taxation, as higher tax rates do not necessarily lead to lower emissions— indicated at the dot on the curve for $\alpha=0.9$ on the right-hand panel in Figure \ref{fig123_sig=2}.

  In conclusion, the findings suggest that present-bias negatively impacts the effectiveness of carbon taxes, with stronger biases having a greater effect.   
  This highlights the importance for  social planners and governments to account for present-biased behavior when designing effective carbon tax policies.

\begin{Remark}\label{remm}\normalfont
    Finally, let us return to a question raised at the end of Section \ref{sec7}. As it was shown there, the equilibrium strategy under stochastic quasi-hyperbolic discounting can match that of an exponential discounting model with a higher discount rate (in the example given there by an increase from $\delta = 0.1$ to $0.15$).
However, this does not mean that the effect of present-biasedness can be equivalently replaced by using an exponential discounting model with a suitably higher discount rate. In that example, without carbon tax the probability of early depletion was about $99\%$ for both cases. 
Suppose we want to determine how much  carbon tax should be imposed in order to reduce the probability of early depletion to about $90\%$. If we ignore present-bias and instead adopt the exponential discounting model with the equivalent higher discount rate of $14.08\%$, we would need to set the carbon tax instead at $0.635$. Under this increased carbon tax, the optimal strategy in the exponential discounting case corresponds to a threshold of $9.06$.
However, if we apply the same carbon tax in the actual present-biased scenario, the resulting threshold becomes $8.19$, and the early depletion probability is reduced only to about $93\%$, missing the desired target.
This example illustrates that determining the carbon tax based on an exponential discounting model with an adjusted higher discount rate, calibrated to match the pre-tax equilibrium, results in a less effective policy when applied to agents exhibiting present-bias.

\end{Remark}

\section{Numerical Illustrations for More General Models}\label{surp} 
Our general diffusion setup in this paper in principle also allows to study more involved stochastic processes for the carbon budget. As an alternative model, let us here briefly consider an Ornstein–Uhlenbeck type process with state-dependent volatility for the cumulative carbon emission budget available to a company over time. For instance, one could assume that the target atmospheric CO$_2$ concentration in $2050$ is $450$ ppm (which translates to cumulative net emissions since pre-industrial times of approximately $1,330$ GtCO$_2$, see e.g.\ \cite{BennedsenHillebrandKoopman2023}). The aggregate carbon emission capacity available at any time $t$ is then linked to the difference between the target and the current concentration level, which evolves dynamically. Translating this into an individual target level $\theta$ of the company may then justify an adaptive budget available at time $t$ of the form 
\begin{align*}
dX_t^L = \kappa(\theta -  X_t^L)dt + (\sigma_0+\sigma_1 X_t^L) dW_t-(l_0+l_t)dt. 
\end{align*}
The volatility term $\sigma_0+\sigma_1 X_t^L$ may reflect policy uncertainty, technological change and estimation uncertainty, and $l_t$ is determined by the emission schedule $L$. The choice $\theta=35$, $\sigma_0=0.5$, $\sigma_1=0.11$ leads to similar initial values as before, and according to \cite{BennedsenHillebrandKoopman2023}, one may choose $\kappa =0.018$. The other parameters we choose again as $\underline{l}=3$,  $\delta =0.1$,  $\gamma =0.9$, $\overline{\Lambda} =0.5$, $\overline{l}=6$,  $c_{ind}\,=0.1$ and $c_{tax}=0.05$. \\

\noindent \textbf{Impact of $\lambda$ and $\alpha$.} We calculate 
the optimal threshold $b^*$ and the resulting probability of early depletion  for various combinations of $\lambda$ and $\alpha$.

\begin{table}[H]
\centering
\begin{tabular}{lcccccc}
\hline
 & $\lambda=0.1$ & $\lambda=0.25$ & $\lambda=0.5$ & $\lambda=1$ & $\lambda=4$ & $\lambda=12$ \\
\hline
$\alpha=0.5$  & 4.52 & 3.56 & 2.79 & 2.00 & 0.15 & 0.00 \\
$\alpha=0.6$  & 4.81 & 4.00 & 3.37 & 2.70 & 0.62 & 0.00 \\
$\alpha=0.7$  & 5.11 & 4.49 & 4.02 & 3.51 & 1.77 & 0.66 \\
$\alpha=0.8$  & 5.43 & 5.01 & 4.70 & 4.37 & 3.13 & 2.32 \\
$\alpha=0.9$  & 5.76 & 5.55 & 5.40 & 5.23 & 4.56 & 4.08 \\
$\alpha=0.95$ & 5.92 & 5.82 & 5.75 & 5.66 & 5.31 & 5.05 \\
$\alpha=1$    & 6.09 & 6.09 & 6.09 & 6.09 & 6.09 & 6.09 \\
\hline
\end{tabular}
\caption{Optimal threshold $b^*$ for various combinations of $\lambda$ and $\alpha$}
\end{table}

\noindent \textbf{Impact of level of social responsibility on emission schedule.}
Recall that the term $\overline{\Lambda}$ indicates how much the company values preserving its emission budget and reflect its social responsibility and sustainability awareness in some sense.  
We now examine how $\overline{\Lambda}$ affects decision-making by varying its value while keeping all other parameters fixed. 
Table \ref{tab:CLambda_optimal-e2} shows how 
additional weight on sustainability increases the optimal excess production/emission threshold and lowers the probability of early depletion in  the present-biased  case. As expected, higher 
sustainability awareness postpones emissions, resulting in lower overall emissions.

\begin{table}[H]
    \centering
    \begin{tabular}{|c|c|c|c|c|c|}
        \hline
        $\overline{\Lambda}$ & 0 & 0.1 & 0.2 & 0.5 & 0.8 \\
        \hline
        $b^*$ & 3.00 & 3.52 & 4.00 & 5.23 & 6.25 \\
            \hline
    \end{tabular}
    \caption{Optimal threshold levels and resulting early depletion probabilities for varying $\bar{\Lambda}$}
    \label{tab:CLambda_optimal-e2}
\end{table}

\begin{figure}[H]
\begin{subfigure}[t]{0.48\textwidth}
    \centering
    \includegraphics[width=0.9\textwidth]{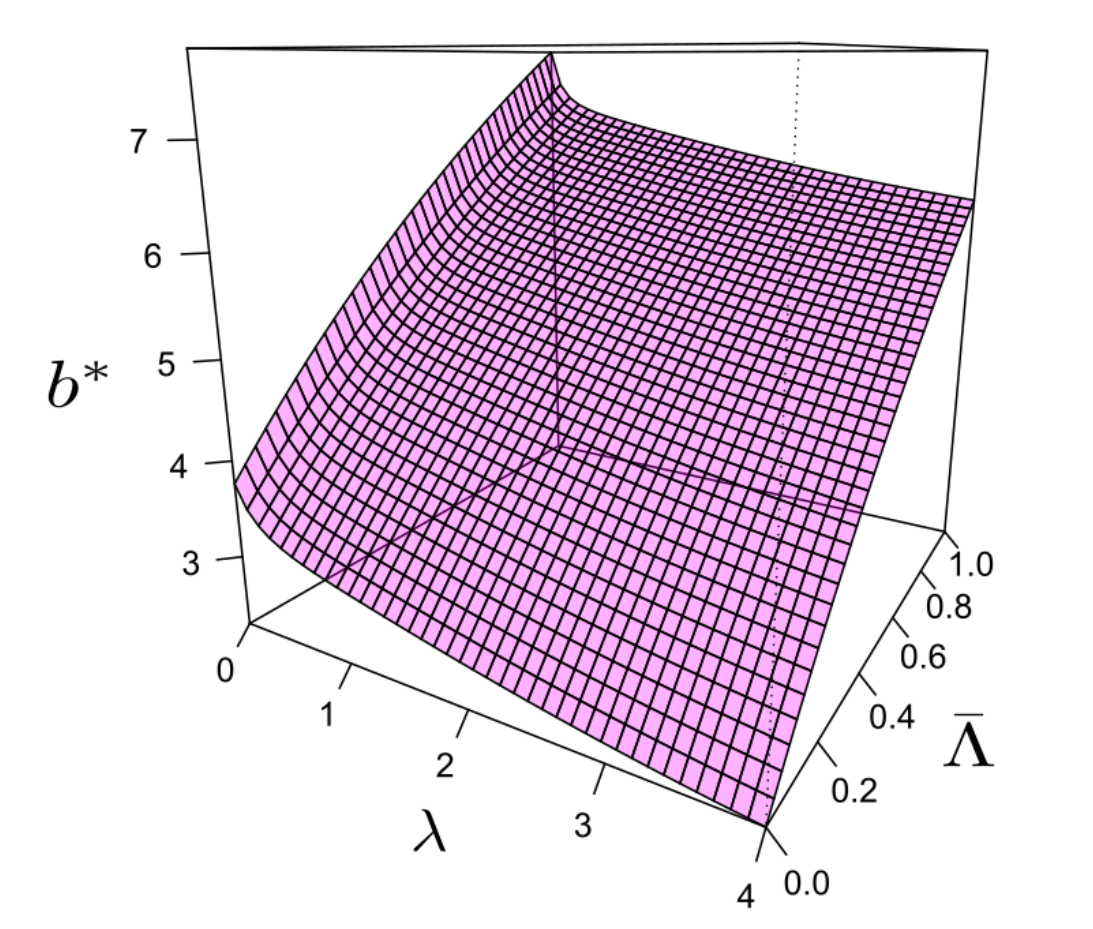}
     \end{subfigure}
\begin{subfigure}[t]{0.48\textwidth}
    \centering
    \includegraphics[width=0.9\textwidth]{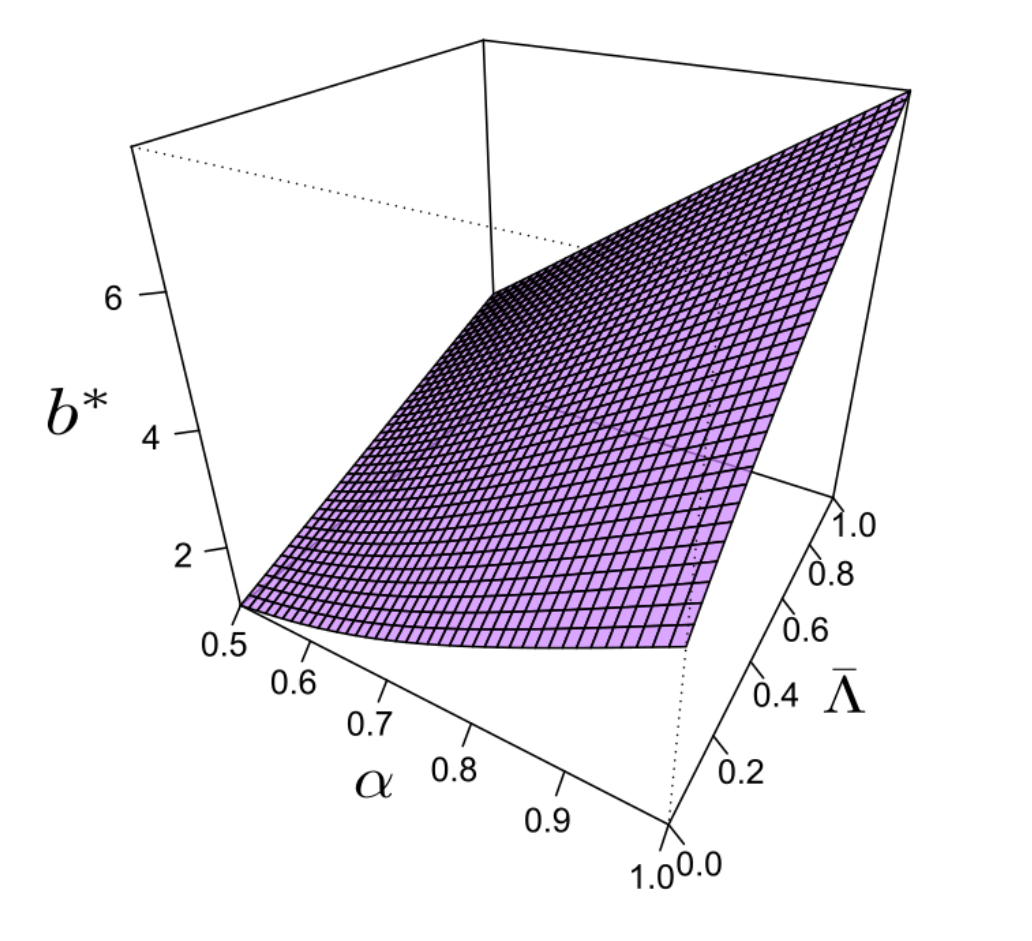}
   \end{subfigure}
\caption{Optimal threshold $b^*$ of the equilibrium strategy under different behavioral parameters: as a function of future arrival intensity $\lambda$ and social responsibility factor $\bar{\Lambda}$ for $\alpha = 0.9$ (left) and as a function of the additional future period discount factor $\alpha$ and $\bar{\Lambda}$ for $\lambda=1$ (right).}
\label{fig:combined_bstar_plots-e2a}
\end{figure}

Figure \ref{fig:combined_bstar_plots-e2a} gives a more detailed picture on how choices of present-bias parameters and the sustainability weight $\bar{\Lambda}$ affect the optimal emission schedule. It quantifies how $b^*$ changes as a function of intensity $\lambda$ of arrival for the future period, weight $\alpha$ for future profits and sustainability weight, $\bar{\Lambda}$, respectively. 

\section{Conclusion}\label{conclu}

In this paper, we provide a framework to study optimal carbon emission schedules for an agent aiming to maximize profit, while being subject to emission constraints and incorporating social responsibility awareness. In particular, we looked into the effects of present-biasedness on the optimal emission behavior. The problem was formulated as an intra-personal game, where the objective is to search for equilibrium solutions. We established the existence of these equilibrium solutions and provided detailed procedures for finding the equilibrium value function and equilibrium emission/production strategy in a general diffusion setup, under stochastic quasi-hyperbolic discounting.

In a detailed numerical illustration for the case of a diffusion setting with constant coefficients, we showed that present-bias leads companies to consume carbon emissions earlier and more aggressively. This behavior results in a higher probability of early depletion compared to the exponential discounting case (the case with no present-bias). Furthermore, the higher the degree of present-bias, the greater the impatience regarding the consumption of the emission budget. 
We also examined the impact of the level of sustainability preferences and showed in what way it has a positive effect on emission patterns and later depletion of the allocated carbon budget. We furthermore studied how carbon tax can provide incentives to an individual company's reduced emission behavior. As the tax increases, the effect becomes more significant. However, when the tax reaches a certain level, the effect begins to diminish. A further insight provided to the social planners is that if policies (carbon tax levels in particular) are set ignoring present-bias of companies, the desired effects may not materialize.

As indicated in the introduction, while the exposition of the paper was formulated for the case of a firm looking for optimal production decisions with implied carbon emission patterns, the results may also be interpreted for rational individuals who decide about carbon-intensive consumption patterns when facing a carbon budget constraint and potential taxes on carbon-intensive activities or consumption goods. 

A main purpose of this paper was to establish a link between the above questions and solution techniques developed in insurance risk theory, which enabled to determine the optimal production/consumption behavior, where the remaining surplus in a dividend-paying insurance company now took the role of the remaining carbon-budget to spend. We deliberately restricted the analysis to a simple diffusion dynamic, allowing a transparent view into the effects of some background parameters and the drivers of a certain optimal behavior. There are many ways in which this line of thinking can be extended to integrate further factors of real-life constraints and objectives into such a study. In particular, it can be interesting to also consider future-biased decision makers (i.e. $\alpha>1$), and situations where the relationship between emission and profit is more complex than the linear relationship applied in the present study. We leave such extensions to future research. 
\bibliographystyle{apalike}
\bibliography{BibPool1}

\begin{thebibliography}{}

\bibitem[Albrecher and Cani, 2017]{albrechercani}
Albrecher, H. and Cani, A. (2017).
\newblock Risk theory with affine dividend payment strategies.
\newblock In {\em Number Theory--Diophantine Problems, Uniform Distribution and
  Applications: Festschrift in Honour of Robert F. Tichy’s 60th Birthday},
  pages 25--60. Springer.

\bibitem[Albrecher and Thonhauser, 2009]{albrecher2009optimality}
Albrecher, H. and Thonhauser, S. (2009).
\newblock Optimality results for dividend problems in insurance.
\newblock {\em Revista de la Real Academia de Ciencias Exactas, Fisicas y
  Naturales}, 103(2):295.

\bibitem[Azcue and Muler, 2014]{azcue2014stochastic}
Azcue, P. and Muler, N. (2014).
\newblock {\em Stochastic optimization in insurance: a dynamic programming
  approach}.
\newblock Springer.

\bibitem[Bennedsen et~al., 2023]{BennedsenHillebrandKoopman2023}
Bennedsen, M., Hillebrand, E., and Koopman, S.~J. (2023).
\newblock A multivariate dynamic statistical model of the global carbon budget
  1959--2020.
\newblock {\em Journal of the Royal Statistical Society Series A: Statistics in
  Society}, 186:20--42.

\bibitem[Bj\"ork et~al., 2017]{BjorkKhapkoMurgoci2017}
Bj\"ork, T., Khapko, M., and Murgoci, A. (2017).
\newblock Inconsistent stochastic control in continuous time: Theory and
  examples.
\newblock {\em Finance and Stochastics}, 21:331--360.

\bibitem[Borissov and Bretschger, 2022]{BorissovBretschger2022}
Borissov, K. and Bretschger, L. (2022).
\newblock Optimal carbon policies in a dynamic heterogeneous world.
\newblock {\em European Economic Review}, 148:104253.

\bibitem[Bourgey et~al., 2024]{bourgey2024}
Bourgey, F., Gobet, E., and Jiao, Y. (2024).
\newblock Bridging socioeconomic pathways of {CO}2 emission and credit risk.
\newblock {\em Annals of Operations Research}, 336(1):1197--1218.

\bibitem[Chekriy et~al., 2025]{chekriy2025probabilistic}
Chekriy, K., Kiesel, R., and Stahl, G. (2025).
\newblock Probabilistic assessment of corporate net-zero transition.
\newblock {\em Available at SSRN 5255705}.

\bibitem[Chen et~al., 2014]{ChenLiZeng2014}
Chen, S., Li, Z., and Zeng, Y. (2014).
\newblock Optimal dividend strategies with time-inconsistent preferences.
\newblock {\em Journal of Economic Dynamics and Control}, 46:150 -- 172.

\bibitem[Chen et~al., 2016]{ChenWangDengZeng2016}
Chen, S., Wang, X., Deng, Y., and Zeng, Y. (2016).
\newblock Optimal dividend-financing strategies in a dual risk model with
  time-inconsistent preferences.
\newblock {\em Insurance: Mathematics and Economics}, 67:27 -- 37.

\bibitem[Colaneri et~al., 2024]{colaneri2024random}
Colaneri, K., Frey, R., and K\"ock, V. (2024).
\newblock Random carbon tax policy and investment into emission abatement
  technologies.
\newblock {\em arXiv preprint arXiv:2406.01088}.

\bibitem[Eisenberg, 2015]{eisenberg2015optimal}
Eisenberg, J. (2015).
\newblock Optimal dividends under a stochastic interest rate.
\newblock {\em Insurance: Mathematics and Economics}, 65:259--266.

\bibitem[Frederick et~al., 2002]{FrederickLoewensteinODonoghue2002}
Frederick, S., Loewenstein, G., and O'Donoghue, T. (2002).
\newblock Time discounting and time preference: A critical review.
\newblock {\em Journal of Economic Literature}, 40(2):351--401.

\bibitem[Fries and Quante, 2024]{FriesQuante2024}
Fries, C.~P. and Quante, L. (2024).
\newblock Intergenerational equitable climate change mitigation: Negative
  effects of stochastic interest rates; positive effects of financing.
\newblock {\em SSRN}.

\bibitem[Gerber and Shiu, 1998]{gerber1998time}
Gerber, H.~U. and Shiu, E.~S. (1998).
\newblock On the time value of ruin.
\newblock {\em North American Actuarial Journal}, 2(1):48--72.

\bibitem[Gikhman and Skorokhod, 1972]{GikhmanSkorokhod1972}
Gikhman, I.~I. and Skorokhod, A.~V. (1972).
\newblock {\em Stochastic differential equations}.
\newblock Springer-Verlag, New York.

\bibitem[{Global Carbon Project}, 2022]{GlobalCarbonProject2022}
{Global Carbon Project} (2022).
\newblock Global carbon budget 2022 highlights.
\newblock
  \url{https://www.globalcarbonproject.org/carbonbudget/22/highlights.htm}.
\newblock p. 1. Accessed on 20 June 2025.

\bibitem[Grenadier and Wang, 2007]{GrenadierWang2007}
Grenadier, S.~R. and Wang, N. (2007).
\newblock Investment under uncertainty and time-inconsistent preferences.
\newblock {\em Journal of Financial Economics}, 84:2--39.

\bibitem[Harris and Laibson, 2013]{HarrisLaibson2013}
Harris, C. and Laibson, D. (2013).
\newblock Instantaneous gratification.
\newblock {\em Quarterly Journal of Economics}, 128:205--248.

\bibitem[Hu and Zhou, 2025]{hu2025equilibrium}
Hu, S. and Zhou, Z. (2025).
\newblock Equilibrium policy on dividend and capital injection under
  time-inconsistent preferences.
\newblock {\em arXiv preprint arXiv:2505.23511}.

\bibitem[Ikeda and Watanabe, 1977]{IkedaWatanabe1977}
Ikeda, N. and Watanabe, S. (1977).
\newblock A comparison theorem for solutions of stochastic differential
  equations and its applications.
\newblock {\em Osaka Journal of Mathematics}, 14(3):619--633.

\bibitem[Iverson and Karp, 2021]{IversonKarp2021}
Iverson, T. and Karp, L. (2021).
\newblock Carbon taxes and climate commitment with non-constant time
  preference.
\newblock {\em The Review of Economic Studies}, 88(2):764--799.

\bibitem[Korn, 2025]{korn2025framework}
Korn, R. (2025).
\newblock A framework for optimal portfolios with sustainable assets and
  climate scenarios.
\newblock {\em European Actuarial Journal}, 15(1):1--13.

\bibitem[Korn and Nurkanovic, 2025]{korn20252}
Korn, R. and Nurkanovic, A. (2025).
\newblock Sustainable portfolio optimization and sustainable taxation.
\newblock {\em European Actuarial Journal}.
\newblock to appear.

\bibitem[Krylov, 1996]{Krylov1996}
Krylov, N.~V. (1996).
\newblock {\em Lectures on Elliptic and Parabolic Equations in Holder Spaces}.
\newblock The American Mathematical Society.

\bibitem[Laibson, 1998]{Laibson1998}
Laibson, D. (1998).
\newblock Life-cycle consumption and hyperbolic discount functions.
\newblock {\em European Economic Review}, 42(3):861 -- 871.

\bibitem[Lebling et~al., 2025]{Lebling2025CCUS}
Lebling, K., Gangotra, A., Hausker, K., and Byrum, Z. (2025).
\newblock 7 things to know about carbon capture, utilization and sequestration.
\newblock World Resources Institute.

\bibitem[Li et~al., 2016]{LiLiZeng2015}
Li, Y., Li, Z., and Zeng, Y. (2016).
\newblock Equilibrium dividend strategy with non-exponential discounting in a
  dual model.
\newblock {\em Journal of Optimization Theory and Applications},
  168(2):699--722.

\bibitem[Li et~al., 2015]{LiChenZeng2015}
Li, Z., Chen, S., and Zeng, Y. (2015).
\newblock Optimal dividend strategy for a diffusion model with
  time-inconsistent preferences.
\newblock {\em Systems Engineering - Theory \& Practice}, 35(7):1633.

\bibitem[MacKenzie and Ohndorf, 2012]{MacKenzieOhndorf2012}
MacKenzie, I.~A. and Ohndorf, M. (2012).
\newblock Cap-and-trade, taxes, and distributional conflict.
\newblock {\em Journal of Environmental Economics and Management},
  63(1):51--65.

\bibitem[Marcolla et~al., 2017]{MarcollaRodenbeckCescatti2017}
Marcolla, B., R{\"o}denbeck, C., and Cescatti, A. (2017).
\newblock Patterns and controls of inter-annual variability in the terrestrial
  carbon budget.
\newblock {\em Biogeosciences}, 14:3815--3829.

\bibitem[Maskin and Tirole, 2001]{MaskinTirole2001}
Maskin, E. and Tirole, J. (2001).
\newblock Markov perfect equilibrium.
\newblock {\em Journal of Economic Theory}, 100:191--219.

\bibitem[Nordhaus, 2018]{Nordhaus2018}
Nordhaus, W. (2018).
\newblock Evolution of modeling of the economics of global warming: changes in
  the dice model, 1992-2017.
\newblock {\em Climatic Change}, 148:623--640.

\bibitem[Palacios-Huerta and P\`erez-Kakabadse,
  2011]{Palacios-HuertaPerez-Kakabadse2011}
Palacios-Huerta, I. and P\`erez-Kakabadse, A. (2011).
\newblock Consumption and portfolio rules with stochastic quasi-hyperbolic
  discounting.
\newblock {\em Working Paper}.

\bibitem[Pao, 1992]{Pao1992}
Pao, C.~V. (1992).
\newblock {\em Nonlinear parabolic and elliptic equations}.
\newblock Plenum Press, New York.

\bibitem[Phelps and Pollak, 1968]{PhelpsPollak1968}
Phelps, E.~S. and Pollak, R.~A. (1968).
\newblock {On Second-Best National Saving and Game-Equilibrium Growth1}.
\newblock {\em The Review of Economic Studies}, 35(2):185--199.

\bibitem[Popovski, 2018]{popovski2018implementation}
Popovski, V. (2018).
\newblock {\em The implementation of the Paris agreement on climate change}.
\newblock Routledge.

\bibitem[Reppen et~al., 2020]{reppen2020optimal}
Reppen, A.~M., Rochet, J.-C., and Soner, H.~M. (2020).
\newblock Optimal dividend policies with random profitability.
\newblock {\em Mathematical Finance}, 30(1):228--259.

\bibitem[Saleh et~al., 2025]{saleh2025estimating}
Saleh, H., Battiston, S., Monasterolo, I., Barreau, T., and Tankov, P. (2025).
\newblock Estimating firms’ emissions from asset level data helps revealing
  (mis) alignment to net zero targets.
\newblock {\em Available at SSRN 4661050}.

\bibitem[Schmidli, 2007]{schmidli2007stochastic}
Schmidli, H. (2007).
\newblock {\em Stochastic control in insurance}.
\newblock Springer, Heidelberg.

\bibitem[Shreve et~al., 1984]{ShreveLehoczkyGaver1984}
Shreve, S.~E., Lehoczky, J.~P., and Gaver, D.~P. (1984).
\newblock Optimal consumption for general diffusions with absorbing and
  reflecting barriers.
\newblock {\em SIAM Journal of Control and Optimization}, 22(1):55--75.

\bibitem[Stern, 2006]{stern2006review}
Stern, N. (2006).
\newblock The {S}tern {R}eview: The economics of climate change.
\newblock Technical report, HM Treasury, Government of the United Kingdom,
  London.

\bibitem[Strini and Thonhauser, 2023]{strini2023time}
Strini, J.~A. and Thonhauser, S. (2023).
\newblock Time-inconsistent view on a dividend problem with penalty.
\newblock {\em Scandinavian Actuarial Journal}, 2023(8):811--833.

\bibitem[Strotz, 1956]{Strotz1956}
Strotz, R.~H. (1956).
\newblock Myopia and inconsistency in dynamic utility maximization.
\newblock {\em The Review of Economic Studies}, 23(3):165--180.

\bibitem[Thonhauser and Albrecher, 2007]{ThAl07}
Thonhauser, S. and Albrecher, H. (2007).
\newblock Dividend maximization under consideration of the time value of ruin.
\newblock {\em Insurance: Mathematics and Economics}, 41(1):163--184.

\bibitem[Zhao et~al., 2014]{ZhaoWeiWang2014}
Zhao, Q., Wei, J., and Wang, R. (2014).
\newblock On dividend strategies with non-exponential discounting.
\newblock {\em Insurance: Mathematics and Economics}, 58:1--13.

\bibitem[Zhu, 2015]{Zhu2015a}
Zhu, J. (2015).
\newblock Dividend optimization for general diffusions with restricted dividend
  payment rates.
\newblock {\em Scandinavian Actuarial Journal}, 2015(7):592--615.

\bibitem[Zhu et~al., 2020]{ZhuSiuYang2020}
Zhu, J., Siu, T., and Yang, H. (2020).
\newblock Singular dividend optimization for a linear diffusion model with
  time-inconsistent preferences.
\newblock {\em European Journal of Operational Research}, 285(1):66--80.

\end{thebibliography}

\appendix

\section{Proofs of Section \ref{exponential}}\label{aB}

\textbf{Proof of Lemma \ref{28525-1}}  (i) We begin by proving the existence and uniqueness of a bounded solution that is continuously differentiable on $(0,\infty)$ and twice continuously differentiable on $(0,b)\cup(b,\infty)$ to Equations \eqref{1819-1}-\eqref{251025-1}, through an explicit construction.
 Let $v_1(\cdot)$ and $v_2(\cdot)$ be solutions to the initial value problems as defined in Lemma \ref{28525-1} and $B_1(x)$ as defined in the same. 
    The existence and uniqueness of $v_1$ and $v_2$ are guaranteed by  Theorem 5.4.2. of \cite{Krylov1996}. It is clear that $v_1$ and $v_2$ are linearly independent. Denote their  Wronskian by
$W_{v_1,v_2}(x)=v_1(x)v_2^\prime(x)-v_2(x)v_1^\prime(x).$ Then, $B_1(x)$ can be expressed as  
$B_1(x)=v_1(x)\int_0^x\frac{v_2(y)}{W_{v_1,v_2}(y)}\frac{2\Lambda}{\sigma^2(y)}\dif y-v_2(x)\int_0^x\frac{v_1(y)}{W_{v_1,v_2}(y)}\frac{2\Lambda}{\sigma^2(y)}\dif y, 
$ which implies that  $B_1(x)$ is a particular solution to the differential equation $\frac{\sigma^2(x)}{2}g^{\prime\prime}(x) +
\mu(x)g^\prime(x)-\delta g(x)+\Lambda=0$ with initial value conditions $B_1(0)=0$ and $B_1^\prime(0)=0$.
Let $u$ and $v_3$ be defined  as in Lemma \ref{28525-1}.  The existence of $v_3$ and $u$ can be  established   by extending  the differential equation to the domain $(-\infty,-1)\cup (0,\infty)$, imposing  the boundary condition $g(-1)=1$, and applying Corollary 8.1 of  \cite{Pao1992}.

Recall that $v_1$ and $v_2$  form a pair of independent solutions to  $\frac{\sigma^2(x)}{2}g^{\prime\prime}(x)
+\mu(x)g^\prime(x)-\delta g(x)=0$ on $[0,\infty)$.
Then all the solutions to  \eqref{1819-1} can be expressed in the following general form: $C_{1}\,v_1(x)+C_{2}v_2(x)+B_1(x)$ where $C_1$ and $C_2$ are constants.
 Recall $v_3(\cdot)$ and $u(\cdot)$ are both bounded solutions to
$\frac{\sigma^2(x)}{2}g^{\prime\prime}(x)+
(\mu(x)-\bar{l})g^\prime(x)-\delta g(x)=0$ on $(0,\infty)$.
 Then, for any constant $C_3$, the function $C_3v_3(x)+u(x)$ is a solution to \eqref{1819-2}. For $b\ge 0$, define a new function
\begin{eqnarray}
g_b(x)=\left\{\begin{array}{ll}
C_{1}\,(b)v_1(x)+C_{2}(b)v_2(x)+B_1(x)&0\le x< b,\\
C_3(b) v_3(x)+u(x)&x\ge b,
\end{array}\right.\label{013819-1}
\end{eqnarray}
 where $C_{1}\,(b)$ and $C_3(b)$ are constants (depending on $b$ only) that satisfy the following:
\begin{eqnarray}
&g_b(0)=0, \mbox{ i.e.,}  &
C_{1}\,(b)v_1(0)+C_{2}(b)v_2(0)+B_1(0)=0\label{130225-1}\\
&g_b(b-)=g_b(b+), \mbox{ i.e.,} &C_{1}\,(b)v_1(b)+C_{2}(b)v_2(b)+B_1(b)=C_3(b)v_3(b)+u(b),\ \ b>0,\label{0300413-2}\\
&g_b^\prime(b-)=g_b^\prime(b+), \mbox{ i.e.,}  &C_{1}\,(b)v_1^\prime(b)+C_{2}(b)v_2^\prime(b)+B_1^\prime(b)
=C_3(b)v_3^\prime(b)+u^\prime(b),\ b>0.\label{0300413-03}
\end{eqnarray}
We can see that $C_{1}\,(b)$, $C_{2}(b)$ and $C_3(b)$ can be uniquely determined with $C_{2}(b)=-C_{1}(b)$. Taking $\lim_{b\downarrow 0}$ on both sides of \eqref{0300413-2},  using $v_1(0)=v_2(0)=v_3(0)=1$ and $u(0)=0$, and noting $B_1(0)=0$, we obtain $\lim_{b\downarrow 0} C_3(b)=0=C_3(0)$.  The function $g_b(x)$ satisfies \eqref{1819-1} and \eqref{1819-2}, and is bounded due to the boundedness of $v_3(x)$ and $u(x)$. From the structure of $g_b$ and noting \eqref{130225-1}-\eqref{0300413-03} and $C_3(0)=0$, we can find that $g_b(0)=0$, and that when $b>0$, $g_b(b-)=g_b(b+)$ and $g_b(x)$ is continuously differentiable in $[0,\infty)$. We can also see that $g_b(x)$ is twice continuously differentiable except for $x=b$. So $g_b(x)$ is the desired unique solution.
Since $v_1$, $v_2$, $v_3$ and $u$ are continuously differentiable functions, from \eqref{130225-1} - \eqref{0300413-03} we can observe that  $C_1\,(b)$, $C_2(b)$ and $C_3(b)$ are continuous functions. 

\noindent (ii)   We now proceed to prove that the above solution is unique and coincides with $V_b^E(x)$.  Let $g$ be any bounded solution that meets all the requirements in (i).
It follows by \cite[Lemma A.1]{ZhuSiuYang2020} that
  \begin{align}
 &\E_x\left[e^{-\delta (\tau^b\wedge
\tau_n\wedge t)}g(X^{b}_{\tau^b\wedge\tau_n\wedge t})\right]\nonumber\\
=&g(x)+\E_x\left[\int^{\tau^b \wedge
\tau_n\wedge t}_0e^{-\delta s}\left(\frac{1}{2}\sigma^2(X^b_{s})g^{\prime\prime}(X^b_{s})
+(\mu(X^b_{s})-l_s^b)g^{\prime}(X^b_{s})-\delta g(X^b_{s})\right)\dif s\right],\nonumber
\end{align}
where $\{\tau_n\}$ is a sequence of stopping times converging to $\infty$.
Note that $l_s^b=\bar{l}I\{X^b_{s}\ge b\}$ and that $g$ satisfies \eqref{1819-1} and \eqref{1819-2}, and so we have
 \begin{align*}
 \frac{1}{2}\sigma^2(X^b_{s})g^{\prime\prime}(X^b_{s})
+(\mu(X^b_{s})-l_s^b)g^{\prime}(X^b_{s})-\delta g(X^b_{s})=-\Lambda-\bar{l}(\gamma-\beta) I\{X^b_{s}\ge b\}.
 \end{align*}
 Consequently,
 \begin{align}
g(x)
=&\E_x\left[e^{-\delta (\tau^b\wedge
\tau_n\wedge t)}g(X^{b}_{\tau^b\wedge\tau_n\wedge t})\right]+\E_x\left[\int_0^{\tau^b\wedge
\tau_n\wedge t}(\Lambda+\bar{l}(\gamma-\beta) I\{X^b_{s}\ge b\})\dif s\right],\ \ x\ge 0.\label{06}
\end{align}
Since the function $g(\cdot)$ is bounded, using the dominated convergence twice we can obtain
\begin{eqnarray}
\lim_{t\rightarrow\infty}\lim_{n\rightarrow\infty}\E_x\left[e^{-\delta (\tau^b\wedge
\tau_n\wedge t)}g(X^{b}_{\tau^b\wedge\tau_n\wedge t})\right]=\E_x\left[
e^{-\delta \tau^b}g(X^{b}_{\tau^b})\right]=0,\label{08}
\end{eqnarray}
where the last equality follows by noticing $X^{b}_{\tau^b}=0$ and $g(0)=0$.
By using the monotone convergence twice we have
\begin{align}
&\lim_{t\rightarrow\infty}\lim_{n\rightarrow\infty}\E_x\left[\int^{\tau^b \wedge
\tau_n\wedge t}_0e^{-\delta s}(\Lambda+\bar{l}I\{X^{b}_{s}\ge b\})\dif s\right]
=\E_x\left[\int^{\tau^b}_0e^{-\delta s}(\Lambda+(\gamma-\beta)\bar{l}I\{X^{b}_{s}\ge b\})\dif s\right]\nonumber\\
=&\E_x\left[\int^{\tau^b}_0e^{-\delta s}(\gamma-\beta) l^b_s\dif s+\int^{\tau^b}_0e^{-\delta s}\Lambda\dif s\right]
=V_b^E(x).\label{0811}
\end{align}
By letting $t\rightarrow\infty$ and $n\rightarrow\infty$ on both sides of \eqref{06}, and then using \eqref{08} and \eqref{0811}, we can obtain
$
g(x)=V_b^E(x)$ for $x\ge 0$.

\noindent (iii)
It follows immediately from the above derivations that
 \begin{eqnarray*}
V_b^E(x)=\left\{\begin{array}{ll}
C_1\,(b)v_1(x)-C_1\,(b)v_2(x)+B_1(x)&0\le x< b,\\
C_3(b) v_3(x)+u(x)&x\ge b,
\end{array}\right.
\end{eqnarray*}
where $C_1\,(b)$ and $C_3(b)$ are determined by solving \eqref{130225-1}-\eqref{0300413-03}. \hfill $\square$\\

\noindent \textbf{Proof of Lemma \ref{upperbound}} 
  The non-negativity of $V^E(x)$ is obvious from its definition in \eqref{Value-Exponential}. 
By noting that the excess emission rate for any admissible strategy is bounded by $\bar{l}$, it follows that\\
$
  V^E(x)
=\sup_{L\in \Pi}\E\left[\int_0^{\tau^L} e^{-\delta s}((\gamma-\beta) l_s+\Lambda)\dif s|X_0=x\right]\nonumber\le \int_0^{\infty}e^{-\delta s}((\gamma-\beta)\bar{l}+\Lambda)\dif s
=\frac{(\gamma-\beta)\bar{l}+\Lambda}{\delta}$. 

 For any $x>0$, let $X_t^{x,b}$  represent the controlled stochastic process $\dif X_t^{x,b}= (\mu(X_{t}^{x,b})-\bar{l}I\{X_t^{x,b}\ge b\})\dif t+\sigma(X_{t-}^{x,b})\dif W_t$ with $X_{0-}^{x,b}=x$. By adapting the comparison theorem (Theorem 1.1 in \cite{IkedaWatanabe1977}),  we can show that
  with probability $1$, $X_t^{x+h,b}\ge X_{t}^{x ,b}$ for all $t\ge 0$. This, along with the fact that, under $ L^b $, excess emissions  only occur when the controlled stochastic process is above $ b $, implies that when there are excess emissions (at rate $ \bar{l} $) at time $ t $ under the process $ X_{t}^{x+h,b} $, there may or may not be excess emissions under $ X_{t}^{x,b} $. However, when there are excess emissions at time $ t $ under $ X_{t}^{x,b} $, there will also be excess emissions at the same rate $ \bar{l} $ under $ X_{t}^{x+h,b} $ with probability 1. 
 As a result,
$
V_{b}^E(x)\le V_b^E(x+h)$ for $h>0$, and so $V_b^E(x)$ is non-decreasing.
 \hfill $\square$\\

\begin{Lemma}\label{st30-iii}
The function $h^E(b):={V_b^E}^\prime(b)$ is continuous on $[0,\infty)$ and the following holds: $\lim_{b\downarrow 0}{V_b^E}^{\prime\prime}(0+)={V_0^E}^{\prime\prime}(0+)$.
\end{Lemma}
\proof 
From the proof of Lemma \ref{28525-1} (i) and (ii) we know that $V_b^E(x)=g_b(x)$ for $x\ge 0$ and $b\ge 0$. As mentioned there,  $C_1\,(b)$ and $C_3(b)$ are continuous functions of $b$ for $b\ge 0$.
Consequently, the function $
h^E(b):={V_b^E}^\prime(b) =g_b^\prime(b) =C_3(b)v_3^\prime(b)+u^\prime(b)
$ is continuous in $b$ for  $b\in[0,\infty)$.
It follows  by \eqref{013819-1} that
\begin{align}\lim_{b\downarrow 0}{V_b^E}^{\prime\prime}(0+)&=\lim_{b\downarrow 0}g_b^{\prime\prime}(0+)=\lim_{b\downarrow 0}C_1\,(b)(v_1^{\prime\prime}(0)-v_2^{\prime\prime}(0)) = \lim_{b\downarrow 0}\left(C_1\,(b)(v_1^{\prime\prime}(b)-v_2^{\prime\prime}(b))\right)\nonumber\\
&= \lim_{b\downarrow 0} \left(C_3(b)v_3^{\prime\prime}(b)+u^{\prime\prime}(b)\right)=\lim_{b\downarrow 0} C_3(b)v_3^{\prime\prime}(0)+u^{\prime\prime}(0)\label{20819-1}\\
&= C_3(0)v_3^{\prime\prime}(0)+u^{\prime\prime}(0)={V_0^E}^{\prime\prime}(0+),\nonumber
\end{align}
where the first equality in  \eqref{20819-1} is due to \eqref{0300413-03}.
 \hfill $\square$

\begin{Lemma} \label{VEconcavity} (i) For $b\ge 0$,  ${V_b^E}^{\prime\prime}(x)\le 0$ for $x\in(b,\infty)$. 
(ii) For $b\ge 0$, if ${V_b^E}^{\prime}(b)\ge \gamma-\beta$,  then ${V_b^E}^{\prime\prime}(x)\le 0$ for $x\in(0,b)$. 
(iii) For $b\ge 0$, if $\mu(0)<0$ and  {$C_1\,(b)>\frac{\Lambda}{-2\mu(0)}$}, 
then ${V_b^E}^\prime(b)<\gamma-\beta$ and furthermore, ${V_b^E}^\prime(x)<\gamma-\beta$ for $x>b$. (iv)  Moreover,
$
     C_1\,(b)\ge 0\mbox{\ for $b\ge 0$}.
    $ 
\end{Lemma}

\proof  (i) We proceed with an indirect proof. Suppose the statement in (i) is not true. Then ${V_b^E}^{\prime\prime}(y_0)>0$ for some $y_0>b$. Since $V_b^E$ is bounded and  increasing, eventually ${V_b^E}^{\prime\prime}(x)<0$ for sufficiently large $x$. Let $y_1$ represent the first point after $y_0$ such that the function, ${V_b^E}^{\prime\prime}(x)$,  becomes concave. Then,
\begin{align}
{V_b^E}^{\prime\prime}(y_1)=0,\ \mbox{ and } {V_b^E}^{\prime\prime}(x)>0 \mbox{ for $x\in(y_0,y_1)$.}\label{convexity}
\end{align} Thus,
\begin{align}
&\mu(x){V_b^E}^\prime(x)-\delta {V_b^E}(x)+\bar{l}((\gamma-\beta)-{V_b^E}^\prime(x)) +\Lambda =-\frac{\sigma^2(x)}{2}{V_b^E}^{\prime\prime}(x)<0\ \mbox{ for $x\in(y_0,y_1)$},\label{5719-1}\\
&\mu(y_1){V_b^E}^\prime(y_1)-\delta {V_b^E}(y_1)+\bar{l}((\gamma-\beta)-{V_b^E}^\prime(y_1))+\Lambda=-\frac{\sigma^2(y_1)}{2}{V_b^E}^{\prime\prime}(y_1)=0.\label{5719-2}
 \end{align}
Also, $(\mu(y_1){V_b^E}^\prime(y_1)-\mu(x){V_b^E}^\prime(x))-\delta ({V_b^E}(y_1)-{V_b^E}(x))-\bar{l}({V_b^E}^\prime(y_1)-{V_b^E}^\prime(x))>0$ for $x\in(y_0,y_1)$. As a result, by dividing both sides by $y_1-x$,  then taking the limit $\lim_{x\uparrow y_1}$ and using ${V_b^E}^{\prime\prime}(y_1)=0$ we obtain $(\mu^\prime(y_1)-\delta ){V_b^E}^\prime(y_1) \ge 0$. On the other hand, since $\mu^\prime(y_1)<\delta $ and ${V_b^E}^\prime(y_1)>0$ (by the increasing property of $V_b^E$ and \eqref{convexity}), we have $(\mu^\prime(y_1)-\delta ){V_b^E}^\prime(y_1) <0$, which is a contradiction.

\noindent (ii) Recall from Lemma \ref{28525-1} we know that $V_b^E$ satisfies \eqref{1819-1} and hence, 
\begin{align}
{V_b^E}^{\prime\prime}(b-)&=-\frac{2(\mu(b){V_b^E}^\prime(b)-\delta {V_b^E}(b) +\Lambda)} {\sigma^2(b)}\nonumber\\
&=-\frac{2(\mu(b){V_b^E}^\prime(b)-\delta {V_b^E}(b)+\bar{l}((\gamma-\beta)-{V_b^E}^\prime(b)) +\Lambda-\bar{l}((\gamma-\beta)-{V_b^E}^\prime(b))} {\sigma^2(b)}\nonumber\\
&={V_b^E}^{\prime\prime}(b+)+\frac{2\bar{l}((\gamma-\beta)-{V_b^E}^\prime(b))}{\sigma^2(b)}\le {V_b^E}^{\prime\prime}(b+)\le 0,\label{16335-1}
\end{align}
where the second to the last inequality follows by noting ${V_b^E}^\prime(b)\ge \gamma-\beta$ and the last equality follows by the result in (i).\\
 We use proof by contradiction again. Suppose there exists some $y_0\in(0,b)$ such that  ${V_b^E}^{\prime\prime}(y_0)>0$. By noting ${V_b^E}^{\prime\prime}(b-)\le 0$ (\eqref{16335-1}) and the continuity of ${V_b^E}^{\prime\prime}(x)$ on $(0,b)$,  we know there exists a $y_1\in(y_0,b)$ such that ${V_b^E}^{\prime\prime}(y_1)=0$ and ${V_b^E}^{\prime\prime}(x)>0$ for $x\in(y_0,y_1)$.
 Following the same lines starting from \eqref{5719-1} until the end of proof for (i)  we can obtain a contradiction. 

 \noindent
 (iii) We now consider the situation where $\mu(0)<0$, and  $C_1\,(b)>\frac{\Lambda}{-2\mu(0)}$.  By using the expression for $V_b^E$ in Lemma \ref{28525-1} we can obtain  ${V_b^E}^\prime(0+)=2C_1\,(b)>\frac{\Lambda}{-\mu(0)}$. 
  Suppose the statement is not true, that is,   ${V_b^E}^\prime(b)\ge \gamma-\beta$. Then it follows by (ii) that \begin{align}
     {V_b^E}^{\prime\prime}(0+)\le 0.\label{17225-2}     \end{align}
     On the other hand, however,
 \begin{align}
    -\frac{\sigma^2(0)}{2}{V_b^E}^{\prime\prime}(0+)&= \mu(0){V_b^E}^\prime(0+)-\delta {V_b^E}(0)+\Lambda=\mu(0){V_b^E}^\prime(0+)+\Lambda\nonumber\\
    &<\mu(0)\frac{\Lambda}{-\mu(0)}+\Lambda=0, 
    \label{17225-1}
 \end{align}
 where the last inequality follows by noting  $\mu(0)<0$, and  ${V_b^E}^\prime(0+)>\frac{\Lambda}{-\mu(0)}$. The inequality \eqref{17225-1} implies that ${V_b^E}^{\prime\prime}(0+)>0$, which is a contradiction to \eqref{17225-2}. Hence, ${V_b^E}^\prime(b)<\gamma-\beta$.
From (i) we know ${V_b^E}^{\prime\prime}(x)\le 0$ for $x>b$. Therefore,  Hence, ${V_b^E}^\prime(x)\le {V_b^E}^\prime(b)<\gamma-\beta$. 
 
 \noindent (iv) By using the expression for $V_b^E$ in Lemma \ref{28525-1} we can obtain  ${V_b^E}^\prime(0+)=2C_1\,(b)$. Further note from Lemma \ref{upperbound} that ${V_b^E}^\prime(0+)\ge 0$. From this we can conclude $C_1\,(b)\ge 0$, which completes the proof.
 \hfill $\square$\\

The following is an immediate consequence of Lemma \ref{VEconcavity}. 
\begin{Corollary}\label{VbE+}
If $0<b^*_E<\infty$, then ${V_{b^*_E}^E}^\prime(b^*_E)=\gamma-\beta$, ${V_{b^*_E}^E}^{\prime\prime}(x)\le 0$ for $x\ge 0$, ${V_{b^*_E}^E}^\prime(x)\ge \gamma-\beta$ for $0\le x<b^*_E$ and ${V_{b^*_E}^E}^\prime(x)\le \gamma-\beta$ for $ x\ge b^*_E$. If $b^*_E=\infty$, then for any $b\ge 0$, ${V_{b}^E}^\prime(b)>\gamma-\beta$ and ${V_{b}^E}^{\prime\prime}(x)\le 0$ for $x\in [0,b)$. \end{Corollary}

\begin{Lemma} \label{VbE0} If $b_E^*=0$, then ${V_{b^*_E}^E}^\prime(x)\le \gamma-\beta$ for $x\ge 0$.
\end{Lemma}
\vspace{-0.35cm}
\proof 
 From  \eqref{17225-6} 
 we know that if $b_E^*=0$, ${V_{0}^E}^\prime(0)\le \gamma-\beta$. Since from Lemma \ref{VEconcavity} (i)  we know ${V_0^E}^{\prime\prime}(x)\le 0$ for $x>0$, we can obtain
${V_{b^*_E}^E}^\prime(x)={V_{0}^E}^\prime(x)\le {V_{0}^E}^\prime(0)\le \gamma-\beta$  for $x\ge 0$.
\hfill $\square$\\

\noindent \textbf{Proof of Theorem \ref{optsol-exp} }
(i)  We first use proof by contradiction to show $b_E^*<\infty$.
Suppose $b_E^*=\infty$. Then, from  \eqref{17225-6} 
we have
${V_b^E}^\prime(b)>\gamma-\beta$ for all $b>0$, and thus by  Lemma \ref{VEconcavity}(ii) we can obtain ${V_b^E}^{\prime\prime}(x)\le 0$ for $0<x<b$ and all $b\ge 0$. This implies ${V_b^E}^\prime(x)\ge {V_b^E}^\prime(b)>\gamma-\beta$ for all  $b>0$ and $0<x<b$.  As a result, for all $b\ge 0$, $V_b^E(x) \ge (\gamma-\beta) x$ for $0\le x \le b$. Thus, for any $x\ge 0$, $\limsup_{b\uparrow \infty}V_b^E(x) \ge (\gamma-\beta) x$. However, according to Lemma \ref{upperbound} we know $V_b^E(x)\le \sup_{L\in \Pi}V^E_L(x)=V^E(x)\le \frac{(\gamma-\beta)\bar{l}+\Lambda}{\delta}$. This is a contradiction and completes the proof. 

\noindent (ii)  We now consider the case $\mu(0)<0$ and  proceed to prove $b_E^*\le b_0$. If $b_0=\infty$, then this is obviously true. So we only need to consider the case $b_0<\infty$. Recall the definition $b_0=\inf\{b\ge 0: C_{1}\,(b)+\frac{\Lambda}{2\mu(0)}>0\}$.
Then $C_{1}\,(b)\le -\frac{\Lambda}{2\mu(0)}$ for $0\le b\le b_0$, 
$C_{1}\,(b_0)= -\frac{\Lambda}{2\mu(0)}$ and there exists a sequence $b_n\downarrow b_0 $ such that  
$
C_{1}\,(b_n)> -\frac{\Lambda}{2\mu(0)}.
  $
 It then follows from Lemma \ref{VEconcavity} (iii)) that ${V_{b_n}^E}^\prime (b_n)<\gamma-\beta$.
 Note that 
 ${V_b^E}$ is continuously differentiable, and ${V_b^E}^\prime(b)=C_3(b)V_E^\prime(b)+u^\prime(b)$. Hence,
 $C_3(b_n){V^E_{b_n}}^\prime(b_n)+u^\prime(b_n)<\gamma-\beta$ for all $n\ge 1$.
 This combined with the definition for $b^*$ (see \eqref{17225-6}),  $b^*_E=\inf\{b>0: C_3(b) v_3^\prime(b)+u^\prime(b)\le \gamma-\beta \}$, and the fact that $b_n\downarrow b_0$ implies $b^*_E\le b_0$.
 
\noindent (iii)   Note we have already shown in (i)  that $b_E^*$ is finite, and in Lemma   \ref{28525-1} that $V^E_{b^*_E}$ is continuously differentiable on $[0,\infty)$ and twice continuously differentiable on $(0,\infty)$. From \cite[Lemma A.1]{ZhuSiuYang2020} with $g$ there being set to $V^E_{b^*_E}$, we obtain that for any $L\in\pi$ and some positive sequence $\{\tau_n\}$ increasing to $\infty$,
 \begin{align}
 &\E_x\left[e^{-\delta (\tau^L\wedge
\tau_n\wedge t)}V^E_{b^*_E}(X^{L}_{\tau^L\wedge\tau_n\wedge t})\right]\nonumber\\
=&V^E_{b^*_E}(x)+\E_x\left[\int^{\tau^L \wedge
\tau_n\wedge t}_0e^{-\delta s}\left(\frac{1}{2}\sigma^2(X^L_{s}){V^E_{b^*_E}}^{\prime\prime}(X^L_{s})
+(\mu(X^L_{s})-{l_s}){V^E_{b^*_E}}^{\prime}(X^L_{s})-\delta V^E_{b^*_E}(X^L_{s})\right)\dif s\right]. \label{5819-2}
\end{align}
Since  $V^E_{b^*_E}(x)$ is a solution to \eqref{1819-1} and \eqref{1819-2}, we have
\begin{align}
&\frac{\sigma^2(X^L_{s}}{2}){V^E_{b^*_E}}^{\prime\prime}(X^L_{s})
+(\mu(X^L_{s})-{l_s}){V^E_{b^*_E}}^{\prime}(X^L_{s})-\delta V^E_{b^*_E}(X^L_{s})\nonumber\\
=&\bigg(\frac{\sigma^2(X^L_{s})}{2}{V^E_{b^*_E}}^{\prime\prime}(X^L_{s})
+(\mu(X^L_{s})-\bar{l}I\{X^L_{s}\ge b^*_E\}){V^E_{b^*_E}}^{\prime}(X^L_{s})-\delta V^E_{b^*_E}(X^L_{s})+\bar{l}(\gamma-\beta) I\{X^L_{s}\ge b^*_E\}\nonumber\\
&+\Lambda\bigg)+(\bar{l}I\{X^L_{s}\ge b^*_E\}-l_s){V^E_{b^*_E}}^{\prime}(X^L_{s})-\bar{l}(\gamma-\beta) I\{X^L_{s}\ge b^*_E\})-\Lambda\nonumber\\
=&0+(\bar{l}I\{X^L_{s}\ge b^*_E\}-l_s){V^E_{b^*_E}}^{\prime}(X^L_{s})-\bar{l}(\gamma-\beta) I\{X^L_{s}\ge b^*_E\})-\Lambda\nonumber\\
=&(\bar{l}-l_s){V^E_{b^*_E}}^{\prime}(X^L_{s})I\{X^L_{s}\ge b^*_E\}-l_s{V^E_{b^*_E}}^{\prime}(X^L_{s})I\{X^L_{s}< b^*_E\}-\bar{l}(\gamma-\beta) I\{X^L_{s}\ge b^*_E\})-\Lambda\nonumber\\
\le & (\bar{l}-l_s)(\gamma-\beta) I\{X^L_{s}\ge b^*_E\}-l_s(\gamma-\beta) I\{X^L_{s}< b^*_E\}-\bar{l}(\gamma-\beta) I\{X^L_{s}\ge b^*_E\})-\Lambda\nonumber\\
=& -(\gamma-\beta) l_s-\Lambda,\label{5819-1}
\end{align}
where the last inequality follows by noting $0\le l_s\le \bar{l}$,  ${V^E_{b^*_E}}^{\prime}(x)\le \gamma-\beta$ for $x\ge {b^*_E}$ and ${V^E_{b^*_E}}^{\prime}(x)\ge \gamma-\beta$ for $0\le x< {b^*_E}$ (see Corollary \ref{VbE+} and Lemma \ref{VbE0}).
Combining \eqref{5819-2} and \eqref{5819-1} yields:
\begin{align}
 V^E_{b^*_E}(x)\ge&\E_x\left[e^{-\delta (\tau^L\wedge
\tau_n\wedge t)}V^E_{b^*_E}(X^{L}_{\tau^L\wedge\tau_n\wedge t})+\int^{\tau^L \wedge
\tau_n\wedge t}_0e^{-\delta s}((\gamma-\beta) l_s+\Lambda)\dif s\right],\ \ x\ge 0.\label{5819-3}
\end{align}
Note that $V^E_{b^*_E}$ is a bounded function and $V^E_{b^*_E}(X^{L}_{\tau^L})=V^E_{b^*_E}(0)=0$. By letting $n\rightarrow \infty$ and $t\rightarrow \infty$ on \eqref{5819-3} and applying dominated convergence and monotone convergence, we arrive at
\begin{align}
 V^E_{b^*_E}(x)\ge&\;\;\E_x\left[\int^{\tau^L}_0e^{-\delta s}((\gamma-\beta) l_s+\Lambda)\dif s\right]=\mathcal{P}^E(x;L),\ \ x\ge 0.
\end{align}
As the above inequality holds for all admissible strategies, $V^E_{b^*_E}(x)\ge \sup_{L\in\Pi}\mathcal{P}^E(x;L)=V^E(x)$ for $x\ge 0$.
On the other hand, $L^{b^*_E}$ is an admissible strategy and so
$V^E_{b^*_E}(x)=\mathcal{P}^E(x;L^{b^*_E})\le \sup_{L\in\Pi}\mathcal{P}^E(x;L)=V^E(x)$ for $x\ge 0$. This implies $L^{b^*_E}$ is an optimal strategy. \hfill $\square$\\

\noindent \textbf{Proof  to Lemma \ref{1325-1}}.  
It follows by
\cite[Lemma A.1]{ZhuSiuYang2020} that for any finite $t>0$,
  \begin{align}
 &\E\left[e^{-\delta (\hat{T}^x\wedge \tau_n\wedge t)}v_3(Y_{\hat{T}^x\wedge  \tau_n\wedge  t}^x)\right]\nonumber\\
=&v_3(x)+\E\left[\int^{\hat{T}^x \wedge \tau_n\wedge t}_0e^{-\delta s}\left(\frac{1}{2}\sigma^2(Y_{s}^x)
+(\mu(Y_{s}^x)-\bar{l})v_3^{\prime}(Y_{s}^x)-\delta v_3(Y_{s}^x)\right)\dif s\right].\nonumber
\end{align}
 Note that we have
 $
 \frac{1}{2}\sigma^2(Y_{s}^x)v_3^{\prime\prime}(Y_{s}^x)
+(\mu(Y_{s}^x)-\bar{l})v_3^{\prime}(Y_{s}^x)-\delta v_3(Y_{s}^x)=0.
$ 
 Therefore,\\
$
v_3(x)
=\E\left[ e^{-\delta (\hat{T}^x\wedge\tau_n\wedge t)}v_3(Y_{\hat{T}^x\wedge\tau_n\wedge t}^x)\right]$ for $x\ge 0$. 
 Since the function $v_3(\cdot)$ is bounded, by using the dominated convergence twice we can obtain
\begin{eqnarray}
\lim_{n\rightarrow \infty}\lim_{t\rightarrow\infty}\E\left[e^{-\delta (\hat{T}^x\wedge
\tau_n\wedge t)}v_3(Y_{\hat{T}^x\wedge\tau_n\wedge t}^x)\right]=\E\left[
e^{-\delta \hat{T}^x}v_3(Y_{\hat{T}^x}^x)\right]=\E\left[
 e^{-\delta \hat{T}^x}\right],\quad x\ge 0\label{08-25}
\end{eqnarray}
It then follows by \eqref{1325-40000} that 
$
\lim_{x\rightarrow \infty}v_3(x)=\lim_{x\rightarrow \infty}E\bigg[e^{-\delta \hat{T}^x}\bigg]=E\bigg[\lim_{x\rightarrow \infty}e^{-\delta \hat{T}^x}\bigg]=0,$ 
where the second-to-last equality follows from the dominated convergence theorem, and the last equality follows by noting
 $\hat{T}^x \rightarrow \infty$ as $x$ goes to $\infty$.

Similarly, 
we know that for any finite $t>0$,
  \begin{align}
 &\E\left[e^{-\delta (\hat{T}^x\wedge \tau_n\wedge t)}u(Y_{\hat{T}^x\wedge \tau_n\wedge t}^x)\right]\nonumber\\
=&\; u(x)+\E\left[\int^{\hat{T}^x \wedge \tau_n\wedge  t}_0e^{-\delta s}\left(\frac{1}{2}\sigma^2(Y_{s}^x)u^{\prime\prime}(Y_{s}^x)
+(\mu(Y_{s}^x)-\bar{l})u^{\prime}(Y_{s}^x)-\delta u(Y_{s}^x)\right)\dif s\right].\nonumber
\end{align}
 Note we have
$
 \frac{1}{2}\sigma^2(Y_{s}^x)u^{\prime\prime}(Y_{s}^x)
+(\mu(Y_{s}^x)-\bar{l})u^{\prime}(Y_{s}^x)-\delta u(Y_{s}^x)+(\bar{l}(\gamma-\beta) +\Lambda)=0.
$
Hence,
 \begin{align}
u(x)
=&\E\left[e^{-\delta (\hat{T}^x\wedge \tau_n\wedge t)}u(Y_{\hat{T}^x\wedge \tau_n\wedge t}^x)\right]+\E\left[\int^{\hat{T}^x \wedge \tau_n\wedge t}_0e^{-\delta s}\left(\bar{l}(\gamma-\beta) +\Lambda 
  \right)\dif s\right],\ \ x\ge 0.\label{06-2500}
\end{align}
Since the function $u(\cdot)$ is bounded, by using the dominated convergence  we can obtain
\begin{eqnarray}
\lim_{n\rightarrow \infty}\lim_{t\rightarrow\infty}\E\left[ e^{-\delta (\hat{T}^x\wedge \tau_n\wedge t)}u(Y_{\hat{T}^x\wedge \tau_n\wedge t}^x)\right]=\E\left[
e^{-\delta \hat{T}^x}u(Y_{\hat{T}^x}^x)\right]=0,\quad x\ge 0,\label{08-2500}
\end{eqnarray}
where the last equality follows by noticing $Y_{\hat{T}^x}^x=0$ and $u(0)=0$.
Employing monotone convergence, we get
\begin{align}
&\lim_{n\rightarrow \infty}\lim_{t\rightarrow\infty}\E\left[\int^{\hat{T}^x \wedge \tau_n\wedge t}_0e^{-\delta s}(\Lambda+(\gamma-\beta)\bar{l})\dif s\right]=\E\left[\int^{\hat{T}^x}_0e^{-\delta s}(\Lambda+(\gamma-\beta)\bar{l})\dif s\right],\quad x\ge 0.\label{1325-6}
\end{align}
Combining \eqref{06-2500}, \eqref{08-2500} and  \eqref{1325-6} yields \eqref{1325-40000}.
Then from \eqref{1325-40000}  we know that
\begin{align*}
\lim_{x\rightarrow \infty}u(x)&=\lim_{x\rightarrow \infty}\E\left[\int^{\hat{T}^x}_0 e^{-\delta s}(\Lambda+(\gamma-\beta)\bar{l})\dif s\right]\\
&=\E\left[\lim_{x\rightarrow \infty}\int^{\hat{T}^x}_0 e^{-\delta s}(\Lambda+(\gamma-\beta)\bar{l})\dif s\right]=\frac{\Lambda+(\gamma-\beta)\bar{l}}{\delta},
\end{align*} where the second to the last equality follows from the dominated convergence theorem, and the last equality follows by noting
 $\hat{T}^x \rightarrow \infty$ as $x$ goes to $\infty$. 
\hfill $\square$

\section{Derivations for Subsection \ref{BWexp}}
\label{aE}
Recall that  $V^E_b(\cdot)$ is the bounded and continuously differentiable
solution to the following equations:
$\frac{\sigma^2}{2}g^{\prime\prime}(x) +\mu g^\prime(x)-\delta
g(x)+\Lambda=0$ for $0< x< b,$ and $\frac{\sigma^2}{2}g^{\prime\prime}(x)+ (\mu-\bar{l})g^\prime(x)-\delta
g(x)+(\gamma-\beta)\bar{l}+\Lambda=0$ for  $x> b$ with $g(0)=0$. Here $e^{\theta_1 x}$ and $e^{-\theta_2 x}$ form a set of linearly
independent solutions to $\frac{\sigma^2}{2}g^{\prime\prime}(x) +\mu
g^\prime(x)-\delta g(x)=0$ and have a Wronskian $-(\theta_1+\theta_2)e^{(%
\theta_1-\theta_2) x}$, and 
$e^{\theta _{3}x}$ and $e^{-\theta _{4}x}$ form a set of
linearly independent solutions to $\frac{\sigma ^{2}}{2}g^{\prime \prime
}(x)+(\mu -\bar{l})g^{\prime }(x)-\delta g(x)=0$ and have a Wronskian $%
-(\theta _{3}+\theta _{4})e^{(\theta _{3}-\theta _{4})x}$. 
By  using the variation of parameters method, we obtain
\begin{equation*}
V_{b}^{E}(x)=%
\begin{cases}
K_{1}(b)e^{\theta _{1}x}-K_{1}(b)e^{-\theta _{2}x}-\frac{2\Lambda }{\sigma
^{2}}\frac{e^{\theta _{1}x}-1}{\theta _{1}(\theta _{1}+\theta _{2})}+\frac{%
2\Lambda }{\sigma ^{2}}\frac{1-e^{-\theta _{2}x}}{\theta _{2}(\theta
_{1}+\theta _{2})},& 0\leq x<b, \\
K_{4}(b)e^{-\theta _{4}x}-\frac{2((\gamma-\beta) \bar{l}+\Lambda )}{\sigma ^{2}}\frac{%
-1}{\theta _{3}(\theta _{3}+\theta _{4})}+\frac{2((\gamma-\beta) \bar{l}+\Lambda )}{%
\sigma ^{2}}\frac{1-e^{-\theta _{4}x}}{\theta _{4}(\theta _{3}+\theta _{4})},& x\geq b,%
\end{cases}%
\end{equation*}%
where $K_{1}(b)$ and $K_{4}(b)$ are determined by ${V_{b}^{E}}(b-)={%
V_{b}^{E}}(b+)$, ${V_{b}^{E}}^{\prime }(b-)={V_{b}^{E}}^{\prime }(b+)$:
\begin{align*}
K_{1}(b)& =\frac{\frac{2\Lambda \left( \theta _{2}(\theta _{1}+\theta
_{4})e^{\theta _{1}b}+\theta _{1}(\theta _{4}-\theta _{2})e^{-\theta
_{2}b}-\theta _{4}(\theta _{1}+\theta _{2})\right)}{\sigma ^{2}\theta _{1}\theta
_{2}(\theta _{1}+\theta _{2})}+\frac{2((\gamma-\beta) \bar{l}+\Lambda )}{\sigma
^{2}\theta _{3}}}{(\theta _{1}+\theta _{4})e^{\theta _{1}b}+(\theta
_{2}-\theta _{4})e^{-\theta _{2}b}} \triangleq I_{1}(b)\Lambda +I_{2}(b),\\
K_{4}(b)& =\frac{\frac{2\Lambda (e^{\theta _{1}b}-e^{-\theta _{2}b})}{\sigma
^{2}(\theta _{1}+\theta _{2})}+\frac{2((\gamma-\beta) \bar{l}+\Lambda )e^{-\theta
_{4}b}}{\sigma ^{2}(\theta _{3}+\theta _{4})}-(\theta _{1}e^{\theta
_{1}b}+\theta _{2}e^{-\theta _{2}b}) K_{1}(b)}{\theta _{4}e^{-\theta
_{4}b}}\triangleq I_{3}(b)\Lambda +I_{4}(b),
\end{align*}%
where $I_1(b)$–$I_4(b)$ are given in \eqref{I125}-\eqref{I425}.
This leads to \eqref{sol1-000}.

\section{Proofs of Section \ref{sol-hyber-sec}}\label{ac}

\begin{Lemma}\label{bound}
For any strategy, the associated objective function is nonnegative and has an initial value $0$ (the value when the initial state is $0$), and has an upper bound $\frac{((\gamma-\beta)\bar{l}+\Lambda)}{\delta}\frac{\lambda \alpha+\delta}{\lambda+\delta}$.
 \end{Lemma}

\proof From its definition, we can see  that for any $L,\tilde{L}\in \Pi$, $\mathcal{P}_0(x;L,\tilde{L})\ge 0$ for $x\ge 0$.
Note that for any admissible strategy, the excess emission rate has an upper bound $ \bar{l}$. So for any $L,\tilde{L}\in \Pi$,
 \begin{align*}
  \mathcal{P}(x;L,\tilde{L})&\le \E_{x}\left[\int_0^{\eta_0} e^{-\delta t}(  (\gamma-\beta) \bar{l}+\Lambda)\dif t+\int_{\eta_0}^\infty \alpha e^{-\delta t}(  (\gamma-\beta) \bar{l}+\Lambda)\dif t
    \right]\\
    &=\frac{(\gamma-\beta)\bar{l}+\Lambda}{\delta}\E_x[1-(1-\alpha)e^{-\delta \eta_0}]=\frac{(\gamma-\beta)\bar{l}+\Lambda}{\delta}\frac{\lambda \alpha+\delta}{\lambda+\delta},\ \ x\ge 0.\hspace{1cm} \hspace{1cm} \hspace{1cm}\square
 \end{align*}

\begin{Lemma}\label{Vb-ODE}
(i)  For any $b\ge 0$,
there is a bounded function  that is  continuously differentiable on $[0,\infty)$ and twice continuously differentiable on $[0,b)\cup (b,\infty)$, and satisfies the following equations: \begin{align}
&\frac{\sigma^2(x)}{2}g^{\prime\prime}(x) +
\mu(x)g^\prime(x)-(\lambda +\delta)g(x)+\lambda \alpha V^E_{b}(x)+\Lambda=0 \mbox{ for $0< x< b$},\label{13819-11}\\
& g(0)=0,\ \ \\
&\frac{\sigma^2(x)}{2}g^{\prime\prime}(x) +
(\mu(x)-\bar{l})g^\prime(x)-(\lambda +\delta)g(x)+\lambda \alpha V^E_{b}(x)+\Lambda+\bar{l}(\gamma-\beta)=0 \mbox{ for $ x> b$},\label{13819-12}
\end{align}
\noindent
(ii) The above solution is unique and equals $V_b(x)$.  (iii)  The function $\bar{h}(b):=V_b^\prime(b)$ as a function of $b$ is continuous on $[0,\infty)$, and $\lim_{b\downarrow 0}V_b^{\prime\prime}(0+)=V_0^{\prime\prime}(0+)$.
\end{Lemma}
\vspace{-0.35cm}
\proof (i) This  can be proven by employing arguments similar to the proof for Lemma \ref{28525-1}. 
 Define
 $\overline{v}_1(\cdot)$ and $\overline{v}_2(\cdot)$ to be the solutions  to
  $\frac{\sigma^2(x)}{2}g^{\prime\prime}(x)
+\mu(x)g^\prime(x)-(\lambda+\delta) g(x)=0$ on $[0,\infty)$ with the following two sets of initial values respectively, $\overline{v}_1(0)=1$ and $\overline{v}_1^\prime(0)=1$, and $\overline{v}_2(0)=1$ and $\overline{v}_2^\prime(0)=-1$. The existence and uniqueness of $\overline{v}_1$ and $\overline{v}_2$ are guaranteed by Theorem 5.4.2. of \cite{Krylov1996}.
We can see that $\overline{v}_1$ and $\overline{v}_2$ are linearly independent.
Define
\begin{align*}
&\overline{W}_1(x)=\overline{v}_1(x)\overline{v}_2^\prime(x)
-\overline{v}_2(x)\overline{v}_1^\prime(x),\\
&\overline{B}_1(x)=\overline{v}_1(x)\int_0^x
\frac{\overline{v}_2(y)}{\overline{W}_1(y)}\frac{2(\Lambda+\lambda \alpha V_b^E(x))}{\sigma^2(y)}\dif y-\overline{v}_2(x)\int_0^x\frac{\overline{v}_1(y)}{\overline{W}_1(y)}
\frac{2(\Lambda+\lambda \alpha V_b^E(x))}{\sigma^2(y)}\dif y.
\end{align*}
Then all the solutions to  \eqref{13819-11} have the following general form: $c_1\,\overline{v}_1(x)+c_2\overline{v}_2(x)+\overline{B}_1(x)$, where $c_{1}\,$ and $c_{2}$ are constants.
Here $\overline{B}_1(x)$ is a particular solution to $\frac{\sigma^2(x)}{2}g^{\prime\prime}(x) +
\mu(x)g^\prime(x)-(\lambda+\delta) g(x)+\lambda \alpha V_b^E(x)+\Lambda+\bar{l}(\gamma-\beta)=0$ with $\overline{B}_1(0)=0$ and $\overline{B}_1^\prime(0)=0$.

 Let $\overline{v}_3(\cdot)$ be a bounded solution to
$\frac{\sigma^2(x)}{2}g^{\prime\prime}(x)+
(\mu(x)-\bar{l})g^\prime(x)-(\lambda+\delta) g(x)=0$ on $[0,\infty)$ with initial value $g(0)=1$.  Let $\overline{u}_b(x)$ be a bounded solution to $\frac{\sigma^2(x)}{2}g^{\prime\prime}(x)+
(\mu(x)-\bar{l})g^\prime(x)-(\lambda+\delta) g(x)+\lambda\alpha V_b^E(x)+\Lambda +(\gamma-\beta) \bar{l}=0$ on $[0,\infty)$ with initial value $g(0)=0$. Note $V_b^E(x)$ is bounded on $[0,\infty)$ (see Lemma \ref{upperbound}) and $V_b^E(x)=0$ for $x<0$.   By extending  the differential equation to $(-\infty,-1)\cup (0,\infty)$ and adding the boundary condition $g(-1)=1$, and then using Corollary 8.1 of  \cite{Pao1992} we can show  $\overline{v}_3$ and $\overline{u}_b$ exist. Then, for any constant $C_3$, the function $C_3\overline{v}_3(x)+\overline{u}_b(x)$ is a solution to \eqref{13819-12}. For $b\ge 0$, define a new function
\begin{eqnarray}
\overline{g}_b(x)=\left\{\begin{array}{ll}
\overline{C}_1(b)\overline{v}_1(x)-\overline{C}_1(b)\overline{v}_2(x)+\overline{B}_1(x)&0\le x< b\\
\overline{C}_3(b) \overline{v}_3(x)+\overline{u}_b(x)&x\ge b,
\end{array}\right.\nonumber
\end{eqnarray}
 where $\overline{C}_1(b)$ and $\overline{C}_3(b)$  satisfy the following: 
 \begin{eqnarray}
&\mbox{when $b>0$,\quad }&\overline{C}_1(b)\overline{v}_1(b)-\overline{C}_1(b)\overline{v}_2(b)+\overline{B}_1(b)=\overline{C}_3(b)\overline{v}_3(b)+\overline{u}_b(b),\label{300413-2}\\
&&\overline{C}_1(b)\overline{v}_1^\prime(b)-\overline{C}_1(b)\overline{v}_2^\prime(b)+\overline{B}_1^\prime(b)
=\overline{C}_3(b)\overline{v}_3^\prime(b)+\overline{u}_b^\prime(b),\label{300413-03}\\
&\mbox{and when $b=0$, } &\overline{C}_3(0)=0.
\end{eqnarray}
All the arguments  following \eqref{0300413-03} in the proof of Lemma \ref{28525-1} (i) 
can be adapted here after replacing $v_i(x)$ by $\overline{v}_i(x)$, $B_1(x)$ by $\overline{B}_1(x)$,  $C_i(b)$ by $\overline{C}_i(b)$ $i=1,2,3$, and $g_b(x)$ by $\overline{g}_b(x)$, respectively.

\noindent (ii) Consider any fixed $b\ge 0$. It is  sufficient to
 show  that  any solution in (i) coincides with $V_b(x)$ for $x\ge 0$. Let $g$ be any bounded solution that meets all the requirements in (i).
It follows by
\cite[Lemma A.1]{ZhuSiuYang2020} that for $x\ge0$,
  \begin{align}
 &\E_x\left[e^{-(\lambda+\delta) (\tau^b\wedge
\tau_n\wedge t)}g(X^{b}_{\tau^b\wedge\tau_n\wedge t })\right]\nonumber\\
=&\; g(x)+\E_x\left[\int^{\tau^b \wedge
\tau_n\wedge t }_0e^{-(\lambda+\delta) s}\left(\frac{1}{2}\sigma^2(X^b_{s})g^{\prime\prime}(X^b_{s})
+(\mu(X^b_{s})-l_s^b)g^{\prime}(X^b_{s})-(\lambda+\delta) g(X^b_{s})\right)\dif s\right].\nonumber
\end{align}
Note that $l_s^b=\bar{l}I\{X^b_{s}\ge b\}$ and that $g$ satisfies \eqref{13819-11} and \eqref{13819-12}, and so we have
 \begin{align*}
 \frac{1}{2}\sigma^2(X^b_{s})g^{\prime\prime}(X^b_{s})
+(\mu(X^b_{s})-l_s^b)g^{\prime}(X^b_{s})-(\lambda+\delta) g(X^b_{s})=-\Lambda-\bar{l}(\gamma-\beta) I\{X^b_{s}\ge b\}-\lambda\alpha V_b^E(x) .
 \end{align*}
 Consequently,
 \begin{align}
g(x)
=&\E_x\left[e^{-(\lambda+\delta) (\tau^b\wedge
\tau_n\wedge t)}g(X^{b}_{\tau^b\wedge\tau_n\wedge t})\right]\nonumber\\
&+\E_x\left[\int_0^{\tau^b\wedge
\tau_n\wedge t}e^{-(\lambda+\delta) s}(\Lambda+\bar{l}(\gamma-\beta) I\{X^b_{s}\ge b\}+\lambda\alpha V_b^E(x))\dif s\right].\label{6}
\end{align}
Since the function $g(\cdot)$ is bounded, using the dominated convergence twice we can obtain
\begin{eqnarray}
\lim_{t\rightarrow\infty}\lim_{n\rightarrow\infty}\E_x\left[e^{-(\lambda+\delta) (\tau^b\wedge
\tau_n\wedge t )}g(X^{b}_{\tau^b\wedge\tau_n\wedge t })\right]=\E_x\left[
e^{-(\lambda+\delta) \tau^b}g(X^{b}_{\tau^b})\right]=0,\label{8}
\end{eqnarray}
where the last equality follows by noticing $X^{b}_{\tau^b}=0$ and $g(0)=0$.
By using the monotone convergence twice we have
\begin{align}
&\lim_{t\rightarrow\infty}\lim_{n\rightarrow\infty}\E_x\left[\int^{\tau^b \wedge
\tau_n\wedge t}_0e^{-(\lambda+\delta) s}(\Lambda+(\gamma-\beta) \bar{l}I\{X^{b}_{s}\ge b\})\dif s\right]\nonumber\\
=&\E_x\left[\int^{\tau^b}_0e^{-(\lambda+\delta) s}(\Lambda+(\gamma-\beta)\bar{l}I\{X^{b}_{s}\ge b\})\dif s\right]\nonumber\\
=&\E_x\left[\int^{\tau^b\wedge \eta_0}_0e^{-\delta s}(\Lambda+(\gamma-\beta)\bar{l}I\{X^{b}_{s}\ge b\})\dif s\right],\label{811}
\end{align}
where the last equality follows by using \cite[Eq.(A.2)]{ZhuSiuYang2020}. We then have
\begin{align}
\E_x\left[\int^{\tau^b}_0\lambda\alpha e^{-(\lambda+\delta)s} V_b^E(X^b_{s})\dif s\right]=\E_x\left[\alpha e^{-\delta \eta_0} V_b^E(X^b_{\eta_0})I\{\eta_0\le \tau^b\}\right]. \label{19819-1}
\end{align}
By letting $t\rightarrow\infty$ and $n\rightarrow\infty$ on both sides of \eqref{6}, and then using \eqref{8} -\eqref{19819-1}, we conclude
$$
g(x)=\E_x\left[\int^{\tau^b\wedge \eta_0}_0e^{-\delta s}(\Lambda+(\gamma-\beta)\bar{l}I\{X^{b}_{s}\ge b\})\dif s+\alpha e^{-\delta \eta_0} V_b^E(X^b_{\eta_0})I\{\eta_0\le \tau^b\}\right]=V_b(x),  \ \ x\ge 0,$$
where the last equality follows  using the definition of $V_b$ in \eqref{Vb-25}.

\noindent (iii) This can be proven by following the same lines as in the proof of  Lemma \ref{28525-1} (iii).
\hfill $\square$

\begin{Remark}  \label{11216-1} 
From the last lemma it follows immediately  that $V_b(x)$ is continuously differentiable on $[0,\infty)$ and twice continuously differentiable on $[0,b)\cup(b,\infty)$; additionally,
\begin{align}
&\frac{\sigma^2(x)}{2}V_b^{\prime\prime}(x) +
\mu(x)V_b^\prime(x)-(\lambda +\delta)V_b(x)+\lambda \alpha V^E_{b}(x)+\Lambda=0, \quad 0< x< b, \label{Vb-ODE-1}
\\
 &\frac{\sigma^2(x)}{2}V_b^{\prime\prime}(x) +
(\mu(x)-\bar{l})V_b^\prime(x)-(\lambda +\delta)V_b(x)+\lambda \alpha V^E_{b}(x)+\bar{l}(\gamma-\beta)+\Lambda=0,\quad x> b,\label{Vb-ODE-2}\\
&
 V_b(0)=0.\label{Vb(0)}
 \end{align}
 Furthermore,
  \begin{eqnarray}
V_b(x)=\left\{\begin{array}{ll}
\overline{C}_1(b)\overline{v}_1(x)-\overline{C}_1(b)\overline{v}_2(x)+\overline{B}_1(x)&0\le x< b\\
\overline{C}_3(b) \overline{v}_3(x)+\overline{u}_b(x)&x\ge b,
\end{array}\right.\nonumber
\end{eqnarray}
 where $\overline{C}_1(b)$ and $\overline{C}_3(b)$ are constants  satisfying the following: 
\begin{eqnarray}
&&\overline{C}_1(b)\overline{v}_1(b)-\overline{C}_1(b)\overline{v}_2(b)+B_1(b)=\overline{C}_3(b)\overline{v}_3(b)+u_b(b),\\
&&\overline{C}_1(b)\overline{v}_1^\prime(b)-\overline{C}_1(b)\overline{v}_2^\prime(b)+\overline{B}_1^\prime(b)
=\overline{C}_3(b)\overline{v}_3^\prime(b)+\overline{u}_b^\prime(b), 
\end{eqnarray}
that is,
 \begin{align}
&\overline{C}_1(b)=\frac{(\overline{B}_1(b)-\overline{u}_{b}(b)\overline{v}_3^\prime(b)-(\overline{B}_1^\prime(b)-\overline{u}_{b}^\prime(b)\overline{v}_3(b)}{(\overline{v}_1^\prime(b)-\overline{v}_2^\prime(b)\overline{v}_3(b)-(\overline{v}_1(b)-\overline{v}_2(b) \overline{v}_3^\prime(b)},\label{C1b25}\\
&\overline{C}_3(b)=\frac{(\overline{u}_{b}^\prime(b)-\overline{B}_1^\prime(b)(\overline{v}_1(b)-\overline{v}_2(b)-(\overline{u}_{b}(b)-\overline{B}_1(b)(\overline{v}_1^\prime(b)-\overline{v}_2^\prime(b)}{ (\overline{v}_1^\prime(b)-\overline{v}_2^\prime(b)\overline{v}_3(b)-(\overline{v}_1(b)-\overline{v}_2(b)\overline{v}_3^\prime(b)}.
\end{align}

 If $V_b^\prime(b)=\gamma-\beta$, then $V_b(x)$ is  twice continuously differentiable on $[0,\infty)$.
 \end{Remark}

Recall that $V_b^E$ is the expected profit function associated with the threshold strategy, $L^b$, under the exponential discounting.
We can derive the following relationship between $V_b^E$ and $V_b$.

\begin{Lemma} \label{Deritive-Comparison}For any $b\ge 0$, 
$\alpha V_b^E(x)\le V_b(x)\le V_b^E(x)$ and $0\le V_b^\prime(x)\le {V_b^E}^\prime(x)$ for $x\ge 0$.
\end{Lemma}
\proof
It follows from the definitions of $V_b^E$  in \eqref{VbE-25} and  $V_b$ in \eqref{Vb-25} that when $0<\alpha<1$,
\begin{align}
V_b(x)&=\E_{x+h}\bigg[\int_0^{\eta_0\wedge \tau^b} e^{-\delta t}((\gamma-\beta) \bar{l}I\{X_t^b\ge b\}+\Lambda)\dif
t\nonumber\\
&\quad\quad\quad\quad\quad \quad  +I\{\eta_0<\tau^b\}
\int_{\eta_0}^{\tau^b}\alpha e^{-\delta t}((\gamma-\beta) \bar{l}I\{X_t^b\ge b\}+\Lambda)\dif t
\bigg]\nonumber\\
&\ge \alpha\left(\E_{x+h}\bigg[\int_0^{\eta_0\wedge \tau^b}  e^{-\delta t}((\gamma-\beta) \bar{l}I\{X_t^b\ge b\}+\Lambda)\dif
t\right.
\nonumber\\&
\quad\quad\quad\quad\quad \quad +
\left.\int_{\eta_0\wedge \tau^b}^{\tau^b} e^{-\delta t}((\gamma-\beta) \bar{l}I\{X_t^b\ge b\}+\Lambda)\dif t
\bigg]\right)=\alpha V_b^E(x),\ \ x\ge 0,\label{14225-1}
\end{align}
\begin{align}
V_b(x)
=&\;\E_{x+h}\bigg[\int_0^{\eta_0\wedge \tau^b} e^{-\delta t}((\gamma-\beta) \bar{l}I\{X_t^b\ge b\}+\Lambda)\dif
t\nonumber\\& \quad\quad\quad\quad
+
I\{\eta_0<\tau^b\}\int_{\eta_0}^{\tau^b}\alpha e^{-\delta t}((\gamma-\beta) \bar{l}I\{X_t^b\ge b\}+\Lambda)\dif t
\bigg]\nonumber\\
\le& \;\E_{x+h}\bigg[\int_0^{\eta_0\wedge \tau^b}  e^{-\delta t}((\gamma-\beta) \bar{l}I\{X_t^b\ge b\}+\Lambda)\dif
t\nonumber\\&\quad\quad\quad\quad
+I\{\eta_0<\tau^b\}
\int_{\eta_0}^{\tau^b} e^{-\delta t}((\gamma-\beta) \bar{l}I\{X_t^b\ge b\}+\Lambda)\dif t
\bigg]=V_b^E(x),\ \ x\ge 0,\label{14225-2}
\end{align}
where the last inequalities in \eqref{14225-1} and \eqref{14225-2} both follow by noting $\alpha\le 1$. 
%

For any $x>0$, let $X_t^{x,b}$  represent the controlled stochastic process $\dif X_t^{x,b}= (\mu(X_{t}^{x,b})-\bar{l}I\{X_t^{x,b}\ge b\})\dif t+\sigma(X_{t-}^{x,b})\dif W_t$ with $X_{0-}^{x,b}=x$. Now consider $X_t^{x,b}$ and $X_t^{x+h,b}$ with $h>0$. By adapting the comparison theorem (Theorem 1.1 in \cite{IkedaWatanabe1977})  we can show that with probability $1$, $X_t^{x+h,b}\ge X_{t}^{x ,b}$ for all $t\ge 0$, and therefore, when $ X_{t}^{x+h,b}$ produces excess emissions at the maximal rate $\bar{l}$, $ X_{t}^{x,b}$ may or may not produce excess emissions, and when it does, $ X_{t}^{x+h,b}$ also generates excess emissiona at the same rate with probability $1$.
 As a result, by noting the expression of $V_b$ in terms of excess emission rates, we can observe $
V_{b}(x)\le V_b(x+h)$ for $h>0$ and thus,  $V_b^\prime(x)\ge 0$ for $x\ge 0$.
It follows from \eqref{VbE-25} and  \eqref{Vb-25} that
\begin{align}
&\left(V_b^E(x+h)-V_b(x+h)\right)-\left(V_b^E(x)-V_b(x)\right)\nonumber\\
=&\;\Bigg(\E_{x+h}\bigg[\int_0^{ \tau^b} e^{-\delta t}((\gamma-\beta) \bar{l}I\{X_t^b\ge b\}+\Lambda)\dif
t
\bigg]\nonumber\\
&-\E_{x+h}\bigg[\int_0^{\eta_0\wedge \tau^b} e^{-\delta t}((\gamma-\beta) \bar{l}I\{X_t^b\ge b\}+\Lambda)\dif
t+
\int_{\eta_0\wedge \tau^b}^{\tau^b}\alpha e^{-\delta t}((\gamma-\beta) \bar{l}I\{X_t^b\ge b\}+\Lambda)\dif t
\bigg]\Bigg)\nonumber\\
&-\Bigg(\E_{x}\bigg[\int_0^{ \tau^b} e^{-\delta t}((\gamma-\beta) \bar{l}I\{X_t^b\ge b\}+\Lambda)\dif
t
\bigg]\nonumber\\
&-\E_{x}\bigg[\int_0^{\eta_0\wedge \tau^b} e^{-\delta t}((\gamma-\beta) \bar{l}I\{X_t^b\ge b\}\dif
t+\Lambda)+
\int_{\eta_0\wedge \tau^b}^{\tau^b}\alpha e^{-\delta t}((\gamma-\beta) \bar{l}I\{X_t^b\ge b\}+\Lambda)\dif t
\bigg]\Bigg)\nonumber\\
=&\;\E_{x+h}\bigg[
\int_{\eta_0\wedge \tau^b}^{\tau^b}(1-\alpha) e^{-\delta t}((\gamma-\beta) \bar{l}I\{X_t^b\ge b\}+\Lambda)\dif t
\bigg]\nonumber\\
&-\E_{x}\bigg[
\int_{\eta_0\wedge \tau^b}^{\tau^b}(1-\alpha) e^{-\delta t}((\gamma-\beta) \bar{l}I\{X_t^b\ge b\}+\Lambda)\dif t
\bigg]\nonumber\\
=&\;(1-\alpha)\left(\E_{x+h}\bigg[e^{-\delta \eta_0 }I\{\eta_0<\tau^b\}
V_b^E(X_{\eta_0}^b) \dif t
\bigg]-\E_{x}\bigg[ e^{-\delta \eta_0 }I\{\eta_0<\tau^b\}
V_b^E(X_{\eta_0}^b)\dif t
\bigg]\right),\label{2189-1}
\end{align}
where the second to the last equality  follows by first calculating  the integrals into two mutually exclusive scenarios $\eta_0\ge \tau^b$ (which makes the integral $0$) and $\eta_0< \tau^b$, and then  applying the Markov property and using the definition for $V_b^E$ in  \eqref{VbE-25}.
 
Let $X_t^{x,b}$  and $X_t^{x+h,b}$ be defined as before. We use $\tau^{x,b}$ and $\tau^{x+h,b}$ to represent the corresponding depletion times, respectively.  We can observe that
\begin{align}
&\E_{x+h}\bigg[e^{-\delta \eta_0 }I\{\eta_0<\tau^b\}
V_b^E(X_{\eta_0}^b) \dif t
\bigg]
=\E\bigg[ e^{-\delta \eta_0 }I\{\eta_0<\tau^{x+h,b}\}
V_b^E(X_{\eta_0}^{x+h,b}) \dif t
\bigg],\label{2189-2}\\
&\E_{x}\bigg[ e^{-\delta \eta_0 }I\{\eta_0<\tau^b\}
V_b^E(X_{\eta_0}^{b})\dif t
\bigg]
=\E\bigg[ e^{-\delta \eta_0 }I\{\eta_0<\tau^{x,b}\}
V_b^E(X_{\eta_0}^{x,b})\dif t
\bigg]. \label{2189-3}
\end{align}
Using the same stochastic comparison argument as above, we know that
  with probability $1$, $X_t^{x+h,b}\ge X_{t}^{x ,b}$ for all $t\ge 0$, and thus $\tau^{x+h,b}\ge \tau^{x,b}$ with probability $1$.  Note that $\eta_0$ is independent of  $\{X_t^{x+h,b}\}$ and $\{X_{t}^{x ,b}\}$ and thus, also independent of $\tau^{x+h,b}$ and $\tau^{x,b}$. Further note that the function $V_b^E$ is increasing. We can conclude that with probability $1$, $e^{-\delta \eta_0 }I\{\eta_0<\tau^{x+h,b}\}
V_b^E(X_{\eta_0}^{x+h,b})\ge e^{-\delta \eta_0 }I\{\eta_0<\tau^{x,b}\}
V_b^E(X_{\eta_0}^{x,b})$. As a result,
 \begin{align}
&\E\bigg[ e^{-\delta \eta_0 }I\{\eta_0<\tau^{x+h,b}\}
V_b^E(X_{\eta_0}^{x+h,b}) \dif t
\bigg]
\ge \E\bigg[ e^{-\delta \eta_0 }I\{\eta_0<\tau^{x,b}\}
V_b^E(X_{\eta_0}^{x,b})\dif t
\bigg].\label{14225-3}
\end{align}
  This, along with \eqref{2189-1}-\eqref{2189-3}, implies that
  \begin{align}
\left(V_b^E(x+h)-V_b(x+h)\right)-\left(V_b^E(x)-V_b(x)\right)\ge 0, \ \  x\ge 0,\ h>     
0,
\label{14225-4}\end{align}
which further implies ${V_b^E}^\prime(x)-V_b^\prime(x)\ge 
0$ for $x\ge 0$.
\hfill $\square$


\begin{Lemma} \label{VbConcavity}Let $C_1(b)$ be the same as that defined in \eqref{C1b25}.
  For $b\ge 0$, suppose 
either (a) $\mu(0)\ge 0$, or {(b) $\mu(0)<0$ and $C_1\,(b)\le  \frac{\Lambda}{-2\mu(0)}$}. (i) The function $V_0(x)$ is concave on $[0,\infty)$. (ii) If  $V_{b}^\prime(b)= \gamma-\beta$, then $V_{b}(x)$ is concave on $[0,\infty)$ and thus $V_{b}^\prime(x)\ge \gamma-\beta $ for $x\in[0,b]$ and $V_{b}^\prime(x)\le  \gamma-\beta$ for $x\in(b,\infty)$.
\end{Lemma}
\proof
    Note that $V_b(0)=0$,  and $V_b(x)$  is increasing and  bounded.
We first show that there exists a positive sequence $\{x_n\}$ with $\lim_{n\rightarrow \infty}x_n=\infty$ such that \begin{eqnarray}
V^{\prime\prime}_b(x_n)\le 0.\label{23513-1}
 \end{eqnarray}  We follow the same idea as in Theorem 3.4 of \cite{Zhu2015a}, and use a proof by contradiction to prove \eqref{23513-1}.
 Suppose the contrary, that is, for some $M>0$, $V_b^{\prime\prime}(x)>0$ for all  $x\ge M$. This implies $V_b^\prime(x)>V_b^\prime(M)$ for $x>M$ and consequently, $
 V_b(x)>V_b(M)+V_b^\prime(M)(x-b)$  for $x>M$. By noting $V_b^\prime(M)>0$ (due the strictly increasing property of $V_b$), we conclude that $V_b(x)$ converges to infinity when $x$ is infinitely large, which is a contradiction to the boundedness of $V_b$.

Recall from \eqref{Vb-ODE-1} and \eqref{Vb-ODE-2}   that for $x\ge 0$, \begin{eqnarray}
\frac{\sigma^2(x)}{2}V_b^{\prime\prime}(x)+
\mu(x)V_b^\prime(x)-(\lambda+\delta) V_b(x)+\bar{l}((\gamma-\beta)-V_b^\prime(x))I\{x\ge b\}+\lambda \alpha V^E_b(x)+\Lambda=0.\label{13-2-4}
 \end{eqnarray}
 Letting $x\downarrow 0$  and noticing
$V_b(0)=0$ (see \eqref{Vb(0)}) and $V_b^E(0)=0$ (see Lemma \ref{28525-1}),
 we can obtain that for $b\ge 0$, \begin{align}
    \frac{\sigma^2(0)}{2}V_b^{\prime\prime}(0+)&=
-\mu(0)V_b^\prime(0+)-\Lambda\nonumber\\
&\le \begin{cases}
0 & \mbox{ if } \mu(0)\ge 0,\\
-\mu(0){V_b^E}^\prime(0+)-\Lambda\le 0 & \mbox{ if } \mu(0)<0, c_{ind}\,(b)\le \frac{\Lambda}{-2\mu(0)},\label{17225-3}
\end{cases} \end{align}
where the inequality in the first case above follows by the non-negativity of $V_b^\prime(x)$ due to the fact that $V_b$ is increasing, and in the second case, the second-to-last inequality  follows by noting $-\mu(0)>0$ and $V_b^\prime(x)\le{V_b^E}^\prime(x)$ for $x>0$ (Lemma \ref{Deritive-Comparison}), and the last inequality follows by noticing ${V_b^E}^\prime(0+)=2C_1\,(b)\le \frac{\Lambda}{-\mu(0)}$ (which can be obtained from  the expression for $V_b^E$ in Lemma  \ref{28525-1}).  
Thus,  \begin{align}
  V^{\prime\prime}_b(0+)\le 0,\ \ b\ge 0.\label{13-2-1}
\end{align}

 Since $V_b^\prime(b)=\gamma-\beta$ for $b>0$, by Remark \ref{11216-1} we know that for any $b>0$,
$V_b(x)$  is twice continuously differentiable on $[0,\infty)$. For $b=0$, we already know that $V_0(x)$ is twice continuously differentiable.

We now use proof by contradiction to show that for $b\ge 0$, $V_b(x)$ is concave.
Suppose that the statement is not true. That is, for some $b\ge 0$, we can find $y_0>     0$ such that $V^{\prime\prime}_b(y_0)>0$. Let $\{x_n\}$ be the sequence defined in the same way as before: $\lim_{n\rightarrow \infty}x_n=\infty$ and $V^{\prime\prime}_b(x_n)\le 0$. We can find a positive integer  $N$ such that $x_N>y_0$.  By  noting that $V^{\prime\prime}_b(x_N)\le 0$, $V^{\prime\prime}_b(0+)\le 0$ (from \eqref{13-2-1}), and $V_b^{\prime\prime}(y_0)>0$ (the assumption made at the beginning of this proof by contradiction), and the continuity of $V_b^{\prime\prime}$, we conclude that there exists  $y_1, y_2$ with  $0\le y_1<y_0<y_2\le x_N$ such that \begin{eqnarray}
V^{\prime\prime}_b(y_1)=0,\ \ V^{\prime\prime}_b(y_2)=0,\ \ \mbox{and }\ V^{\prime\prime}_b(x)>0\ \mbox{ for $x\in (y_1,y_2)$.} \label{13-2-5}
 \end{eqnarray}
 Hence,
\begin{eqnarray}
V^{\prime}_b(y_2)> V^{\prime}_b(y_1).\label{13-2-2}
\end{eqnarray}
It follows by \eqref{13-2-4} that for $x\ge0$, \begin{eqnarray*}
\frac{\sigma^2(x)}2V_b^{\prime\prime}(x)=\left(
(\lambda+\delta) V_b(x)-\mu(x)V_b^\prime(x)-\bar{l}((\gamma-\beta)-V_b^\prime(x))I\{x\ge b\}\right)-\lambda \alpha V_b^E(x) -\Lambda. 
 \end{eqnarray*}
 As $V_b^\prime(b)=\gamma-\beta$, we have $\bar{l}((\gamma-\beta)-V_b^\prime(x))I\{x\ge b\}=\bar{l}((\gamma-\beta)-V_b^\prime(x))I\{x> b\}$. Hence,
 for $x\ge 0$, \begin{align}
\frac{\sigma^2(x)}2V_b^{\prime\prime}(x)=\left(
(\lambda+\delta) V_b(x)-\mu(x)V_b^\prime(x)-\bar{l}((\gamma-\beta)-V_b^\prime(x))I\{x> b\}\right)-\lambda \alpha V_b^E(x)-\Lambda.\label{13-2-6}
 \end{align}
By combining \eqref{13-2-5} and \eqref{13-2-6}, we can obtain that for $i=1,2$, and $x\in(y_1,y_2)$,
\begin{align}
&
(\lambda+\delta) V_b(y_i)-\mu(y_i)V_b^\prime(y_i)-\bar{l}((\gamma-\beta)-V_b^\prime(y_i))I\{y_i> b\}-\lambda \alpha V_b^E(y_i)-\Lambda\nonumber\\
=&
\frac12\sigma^2(y_i)V_b^{\prime\prime}(y_i)
=0<\frac12\sigma^2(x)V_b^{\prime\prime}(x),\nonumber\\
=&
(\lambda+\delta) V_b(x)-\mu(x)V_b^\prime(x)-\bar{l}((\gamma-\beta)-V_b^\prime(x))I\{x> b\}-\lambda \alpha V_b^E(x) -\Lambda, \label{13-2-7}
 \end{align}
 which implies that for $i=1,2$, and  for $x\in(y_1,y_2)$, 
 \begin{align}
&(\lambda+\delta) V_b(y_i)-\mu(y_i)V_b^\prime(y_i)-\bar{l}((\gamma-\beta)-V_b^\prime(y_i))I\{y_i> b\}-\lambda \alpha V_b^E(y_i)\nonumber\\
<&
(\lambda+\delta) V_b(x)-\mu(x)V_b^\prime(x)-\bar{l}((\gamma-\beta)-V_b^\prime(x))I\{x> b\}-\lambda \alpha V_b^E(x).
\label{13-2-8}
 \end{align}
By dividing \eqref{13-2-8}  by $x-y_i$ and rearranging the terms, it follows that for $x\in(y_1,y_2)$,
\begin{eqnarray*}
&&(\lambda+\delta) \frac{V_b(x)- V_b(y_1)}{x-y_1}+\frac{-\mu(x)V_b^\prime(x)+\mu(y_1)V_b^\prime(y_1)}{x-y_1}\nonumber\\
&&+\bar{l}\frac{(V_b^\prime(x)-(\gamma-\beta))I\{x> b\}-(V_b^\prime(y_1)-(\gamma-\beta))I\{y_1> b\}}{x-y_1}-\lambda \alpha \frac{V_b^E(x)- V_b^E(y_1)}{x-y_1}>0,\label{13-2-10}\\
 && \delta \frac{V_b(x)-V_b(y_2)}{x-y_2}+\frac{-\mu(x)V_b^\prime(x)+\mu(y_2)V_b^\prime(y_2)}{x-y_2}\nonumber\\
 &&+\bar{l}\frac{(V_b^\prime(x)-(\gamma-\beta))I\{x> b\}-(V_b^\prime(y_2)-(\gamma-\beta))I\{y_2> b\}}{x-y_2}-\lambda \alpha \frac{V_b^E(x)- V_b^E(y_2)}{x-y_2}<0.
 \label{13-2-11}
 \end{eqnarray*}
 By letting $x\downarrow y_1$ and $x\uparrow y_2$ in the above two equations respectively, we can obtain
 \begin{eqnarray*}
&&(\lambda+\delta) V_b^\prime(y_1)-\mu(y_1)V_b^{\prime\prime}(y_1)
-\mu^\prime(y_1)V_b^{\prime}(y_1)+\bar{l}V_b^{\prime\prime}(y_1)I\{y_1> b\}-\lambda \alpha {V_b^E}^\prime(y_1)\ge 0,\label{13-2-12}\\
  &&(\lambda+\delta) V_b^\prime(y_2)-\mu(y_2)V_b^{\prime\prime}(y_2)
  -\mu^\prime(y_2)V_b^{\prime}(y_2)+\bar{l}V_b^{\prime\prime}(y_2)I\{y_2> b\}-\lambda \alpha {V_b^E}^\prime(y_2)\le 0.
 \label{13-2-13}
 \end{eqnarray*}
Therefore, by noting $V_b^{\prime\prime}(y_1)=0=V_b^{\prime\prime}(y_2)$ (see \eqref{13-2-5}) we have
\begin{eqnarray}
(\lambda+\delta-\mu^\prime(y_1))V_b^{\prime}(y_1)-\lambda \alpha {V_b^E}^\prime(y_1)\ge 0 \ge
 (\lambda+\delta -\mu^\prime(y_2))V_b^{\prime}(y_2)-\lambda \alpha {V_b^E}^\prime(y_2).
 \label{13-2-14}
 \end{eqnarray}
Note that the  increasing property of $V_b$ and \eqref{13-2-2} imply \begin{align}
0\le V_b^{\prime}(y_1)<V_b^{\prime}(y_2).\label{21819-6}
\end{align}

Recall that we are under the assumption that ${V_b^E}^\prime(b)=\gamma-\beta$ for $b>0$. It follows from Lemma \ref{Deritive-Comparison} that ${V_b^E}^\prime(b)\ge V_b^\prime(b)=\gamma-\beta$ when $b>0$, and then from  Lemma \ref{VEconcavity}(ii) that when $b>0$, ${V_b^E}^{\prime\prime}(x)\le 0$ for $x>0$. For $b=0$, we know from  Lemma \ref{VEconcavity}(i) that ${V_b^E}^{\prime\prime}(x)=V_0^E(x)\le 0$ for $x>0$. These imply that for $b\ge 0$, 
\begin{align}
{V_b^E}^\prime(y_1)\ge {V_b^E}^\prime(y_2).\label{5525-1}
\end{align}


 Since $\mu^\prime(x)\le \delta$  and $\mu$ is concave,  by using \eqref{5525-1} and \eqref{21819-6}, we can obtain  $(\lambda+\delta-\mu^\prime(y_1))V_b^{\prime}(y_1)-\lambda \alpha {V_b^E}^\prime(y_1)<
 (\lambda+\delta -\mu^\prime(y_2))V_b^{\prime}(y_2)-\lambda \alpha {V_b^E}^\prime(y_2)$, which contradicts \eqref{13-2-14}. This completes the proof of concavity of $V_b(x)$ in (i) and (ii).

 For $b>0$, since  $V_{b}^\prime(b)= \gamma-\beta$ and  $V_{b}(x)$ has been shown to be concave  on $[0,\infty)$, we get  $V_{b}^\prime(x)\ge V_{b}^\prime(b)=\gamma-\beta$ for $x\in[0,b]$ and $V_{b}^\prime(x)\le V_{b}^\prime(b)=\gamma-\beta$ for $x\in(b,\infty)$.
     \hfill $\square$

\begin{Lemma} \label{Vb*D}  The following holds:
(i) $0\le b^*\le b^*_E<\infty$, 
(ii) If $b^*>0$, then $V_{b^*}^\prime(b^*)=\gamma-\beta$, ${V_{b^*}^E}^\prime(x)\ge \gamma-\beta$ for $0\le x<b^*$, ${V_{b^*}^E}^\prime(x)\le \gamma-\beta$ for $x>b^*$; and 
  (iii) If $b^*=0$, then $V_{b^*}^\prime(x)\le \gamma-\beta$ for $x\ge 0$.
\end{Lemma}
\proof (i) It is obvious from its definition that $b^*\ge 0$. From Lemma \ref{Deritive-Comparison} we know  that  $V_b^\prime(x)\le {V_b^E}^\prime(x)$ for $x\ge 0$. Hence, $V_b^\prime(b)\le {V_b^E}^\prime(b)$ for $b\ge 0$. Note that we have shown $b_E^*<\infty$ in Lemma  \ref{optsol-exp} 
and so by the continuity of ${V_b^E}^\prime (b)$ with respect to $b$ (see Lemma \ref{st30-iii}) 
and the definition of $b_E^*$ in \eqref{17225-6} we can obtain ${V_{b_E^*}^E}^\prime (b_E^*)\le \gamma-\beta$. Therefore, $V_{b_E^*}^\prime (b_E^*)\le {V_{b_E^*}^E}^\prime(b_E^*)\le \gamma-\beta$, which along with the definition of $b^*$ in  \eqref{b*000} implies $b^*\le b_E^*$.

\noindent (ii) 
Since $b^*>0$, it follows by its definition in  \eqref{b*000}  and the continuity  of $V_b^\prime(b)$ with respect to $b$ (see Lemma \ref{Vb-ODE}(iii)) that $V_{b^*}^\prime(b^*)=\gamma-\beta$. 

Note we have shown in Lemma \ref{optsol-exp} 
that if $\mu(0)<0$, then $b_E^*\le b_0$ and thus, in this case, $b^*\le b_E^*\le b_0$. Recall the definition of $b_0$ in \eqref{b0}:
$b_0=\inf\{b\ge 0: C_1\,(b)+\frac{\Lambda}{2\mu(0)}>0\}.
$ Hence, 
$C_1\,(b^*)+\frac{\Lambda}{2\mu(0)}\le 0$, that is, $C_1\,(b^*)\le -\frac{\Lambda}{2\mu(0)}$. In conclusion, in the case $b^*>0$, the following is guaranteed: either $\mu(0)\ge 0$ or $\mu(0)<0$ and $C_1\,(b^*)\le -\frac{\Lambda}{2\mu(0)}$. 
Then by applying Lemma \ref{VbConcavity} (ii) we obtain  $V_{b^*}^\prime(x)\ge \gamma-\beta $ for $x\in[0,b^*]$ and $V_{b^*}^\prime(x)< \gamma-\beta$ for $x\in[b^*,\infty]$.

\noindent (iii) Since $b^*=0$, it follows by  its definition in \eqref{b*000}  that  $V_{b^*}^\prime(b^*)=V_0^\prime(0)\le \gamma-\beta$. Note from Lemma \ref{VbConcavity}(i) we know that $V_{b^*}(x)=V_0(x)$ is concave on $[0,\infty)$. Therefore,  $V_{b^*}^\prime(x)\le V_{b^*}^\prime(0)\le \gamma-\beta$ for $x\ge 0$. \hfill $\square$\\

\noindent \textbf{Proof of Theorem \ref{optimality-h}}
We have shown in {Lemma} \ref{Vb*D} that  $0\le b^*\le b^*_E<\infty$.
 Note $\mathcal{P} (x;L^{b^*},L^{b^*})\le\sup_{L\in\Pi}\mathcal{P} (x;L,L^{b^*})$.  According to the definition of a MPE strategy, we can see that it is sufficient to show that
$\mathcal{P} (x;L^{b^*},L^{b^*})(x)\ge \sup_{L\in\Pi}\mathcal{P} (x;L,L^{b^*}), \ x\ge 0.$

  If $b^*>0$, then $V_{b^*}^\prime(b^*)=\gamma-\beta$ and thus by Remark \ref{11216-1} we know that
$V_{b^*}(x)$  is twice continuously differentiable on $[0,\infty)$, and it follows by   Lemma \ref{Vb*D} that
$
 V_{b^*}^\prime(x)\ge \gamma-\beta $ for $x< b^*$ and $    V_{b^*}^\prime(x)\le \gamma-\beta$  for $x\ge b^*$.
 If $b^*=0$, then $V_{b^*}(x)=V_0(x)$  is twice continuously differentiable. Moreover,  by combining  Lemma \ref{Vb*D}  we have 
$V_{0}^\prime(x)\le \gamma-\beta$ for $x\ge 0$. In summary,  $V_{b^*}(x)$ is twice continuously differentiable on $[0,\infty)$, and for $ b^*\ge 0$,
\begin{align}
 V_{b^*}^\prime(x)\ge \gamma-\beta \mbox{ for $x< b^*$},\quad \text{and\ \ }
    V_{b^*}^\prime(x)\le \gamma-\beta \mbox{ for $0\le x\ge b^*$}.\label{1725-5}
    \end{align}

Let $L$  be any admissible strategy  and define ${\bar{\pi}^{\eta_0,L,L^{b^*}}_s}$ by $\dif{\bar{\pi}^{\eta_0,L,L^{b^*}}_s}=l_sI\{s<\eta_0\}\dif s+l^{b^*}_sI\{s\ge \eta_0\}\dif s$. For convenience, we use ${\bar{\pi}}$ to represent this strategy and use $\bar{\pi}_s$ to represent the excess emission rate throughout this proof. We can see that  ${\bar{\pi}}$ is also  admissible.
  By applying It\^o's formula 
  we can obtain that for any $t>0$,
  \begin{align}
&\E_{x}\left[e^{-(\lambda+\delta)  ({\tau^{{\bar{\pi}}}\wedge\tau_n \wedge
t})}V_{b^*}(X^{{\bar{\pi}}}_{\tau^{{\bar{\pi}}} \wedge \tau_n\wedge
t})\right]\nonumber\\
=&V_{b^*}(x)+\E_{x}\bigg[\int^{\tau^{{{\bar{\pi}}}} \wedge \tau_n\wedge
t}_0e^{-(\lambda+\delta)  s}\bigg(\frac12\sigma^2(X^{{\bar{\pi}}}_{s}) V_{b}^{\prime\prime}(X^{{\bar{\pi}}}_{s})+\left(\mu(X^{{\bar{\pi}}}_{s})
-\bar{\pi}_s\right) V_{b^*}^{\prime}(X^{{\bar{\pi}}}_{s})\nonumber\\
&-(\lambda+\delta) V_{b^*}(X^{{\bar{\pi}}}_{s})\bigg)\dif s \bigg].\label{000st15-1}
\end{align}
Since the function $V_{b^*}$ satisfies \eqref{Vb-ODE-1} and \eqref{Vb-ODE-2}, we have
\begin{align}
&\frac12\sigma^2(X^{{\bar{\pi}}}_{s}) V_{b^*}^{\prime\prime}(X^{{\bar{\pi}}}_{s})+\left(\mu(X^{{\bar{\pi}}}_{s})
-\bar{\pi}_s\right) V_{b^*}^{\prime}(X^{{\bar{\pi}}}_{s})-(\lambda+\delta) V_{b^*}(X^{{\bar{\pi}}}_{s})\nonumber\\
=&\bar{l}I\{X^{{\bar{\pi}}}_{s}\ge b^*\}V_{b^*}^{\prime}(X^{{\bar{\pi}}}_{s})-
\bar{\pi}_sV_{b^*}^{\prime}(X^{{\bar{\pi}}}_{s})-(\gamma-\beta) \bar{l}I\{X^{{\bar{\pi}}}_{s}\ge b^*\}-\lambda \alpha V_{b^*}^E(X^{{\bar{\pi}}}_{s})-\Lambda\nonumber\\
=&(\bar{l}-\bar{\pi}_s)I\{X^{{\bar{\pi}}}_{s}\ge b^*\}V_{b^*}^{\prime}(X^{{\bar{\pi}}}_{s})-
\bar{\pi}_sV_{b^*}^{\prime}(X^{{\bar{\pi}}}_{s})I\{X^{{\bar{\pi}}}_{s}< b^*\}-(\gamma-\beta) \bar{l}I\{X^{{\bar{\pi}}}_{s}\ge b^*\}-\lambda \alpha V_{b^*}^E(X^{{\bar{\pi}}}_{s})-\Lambda\nonumber\\
\le & (\gamma-\beta)(\bar{l}-\bar{\pi}_s)I\{X^{{\bar{\pi}}}_{s}\ge b^*\}-
(\gamma-\beta) \bar{\pi}_sI\{X^{{\bar{\pi}}}_{s}< b^*\}
-(\gamma-\beta) \bar{l}I\{X^{{\bar{\pi}}}_{s}\ge b^*\}-\lambda \alpha V_{b^*}^E(X^{{\bar{\pi}}}_{s})-\Lambda\nonumber\\
=&-(\gamma-\beta) \bar{\pi}_s-\lambda \alpha V_{b^*}^E(X^{{\bar{\pi}}}_{s})-\Lambda,
\end{align}
where the last inequality follows by noting $\bar{l}-\bar{\pi}_s\ge 0$, $V_{b^*}^\prime(x)\le
(\gamma-\beta)$ for $x\ge b^*$ and  $V_{b^*}^\prime(x)\ge
(\gamma-\beta)$ for $0\le x< b^*$ (see \eqref{1725-5}). Hence,
\begin{align}
V_{b^*}(x)\ge& \;\E_{x}\left[e^{-(\lambda+\delta)  ({\tau^{{\bar{\pi}}}\wedge\tau_n \wedge
t})}V_{b^*}(X^{{\bar{\pi}}}_{\tau^{{\bar{\pi}}} \wedge \tau_n\wedge
t})\right]\nonumber\\
&+\E_{x}\bigg[\int^{\tau^{{{\bar{\pi}}}} \wedge \tau_n\wedge
t}_0e^{-(\lambda+\delta)  s}\bigg(
(\gamma-\beta) \bar{\pi}_s+\lambda \alpha V_{b^*}^E(X^{{\bar{\pi}}}_{s})+\Lambda\bigg)\dif s\bigg].\label{22819-1}
\end{align}
Since the function $V_{b^*}(\cdot)$ is bounded, using dominated convergence twice we obtain
\begin{eqnarray}
\lim_{t\rightarrow\infty}\lim_{n\rightarrow\infty}\E_x\left[e^{-(\lambda+\delta) (\tau^{\bar{\pi}}\wedge
\tau_n\wedge t )}V_{b^*}(X^{\bar{\pi}}_{\tau^{\bar{\pi}}\wedge\tau_n\wedge t })\right]=\E_x\left[
e^{-(\lambda+\delta) \tau^{\bar{\pi}}}V_{b^*}(X^{\bar{\pi}}_{\tau^{\bar{\pi}}})\right]=0,\label{008}
\end{eqnarray}
where the last equality follows by noticing $X^{{\bar{\pi}}}_{\tau^{\bar{\pi}}}=0$ and $V_{b^*}(0)=0$.
By using the monotone convergence twice we have
\begin{align}
&\lim_{t\rightarrow\infty}\lim_{n\rightarrow\infty}\E_x\left[\int^{\tau^{\bar{\pi}} \wedge
\tau_n\wedge t}_0e^{-(\lambda+\delta) s}(\Lambda+(\gamma-\beta) \bar{\pi}_s)\dif s\right]\nonumber\\
=&\;\E_x\left[\int^{\tau^{\bar{\pi}}}_0e^{-(\lambda+\delta) s}(\Lambda+(\gamma-\beta)\bar{\pi}_s)\dif s\right]=\E_x\left[\int^{\tau^{\bar{\pi}}\wedge \eta_0}_0e^{-\delta s}(\Lambda+(\gamma-\beta)\bar{\pi}_s)\dif s\right],\label{00811}
\end{align}
where the last equality follows from using \cite[Eq.(A.1)]{ZhuSiuYang2020}. We then have
\begin{align}
\E_x\left[\int^{\tau^{\bar{\pi}}}_0\lambda\alpha e^{-(\lambda+\delta)s} V_{b^*}^E(X^{\bar{\pi}}_{s})\dif s\right]=\E_x\left[\alpha e^{-\delta \eta_0} V_{b^*}^E(X^{\bar{\pi}}_{\eta_0})I\{\eta_0\le \tau^{X^{\bar{\pi}}}\}\right]. \label{0019819-1}
\end{align}
By letting $t\rightarrow\infty$ and $n\rightarrow\infty$ on both sides of \eqref{22819-1}, and then using \eqref{008} - \eqref{0019819-1}, we conclude
that for $x\ge 0$,
\begin{align*}
V_{b^*}(x)&\ge \E_x\left[\int^{\tau^{\bar{\pi}}\wedge \eta_0}_0
e^{-\delta s}(\Lambda+(\gamma-\beta)\bar{\pi}_s)\dif s+\alpha e^{-\delta \eta_0} V_{b^*}^E(X^{\bar{\pi}}_{\eta_0})I\{\eta_0\le \tau^{\bar{\pi}}\}\right]\nonumber\\
&=\E_x\left[\int^{\tau^{\bar{\pi}}\wedge \eta_0}_0
e^{-\delta s}(\Lambda+(\gamma-\beta){l}_s)\dif s+\alpha e^{-\delta \eta_0} V_{b^*}^E(X^{\bar{\pi}}_{\eta_0})I\{\eta_0\le \tau^{\bar{\pi}}\}\right]\nonumber\\
&=\E_x\left[\int^{\tau^{\bar{\pi}}\wedge \eta_0}_0
e^{-\delta s}(\Lambda+(\gamma-\beta){l}_s)\dif s+\alpha e^{-\delta \eta_0} V_{b^*}^E(X^{\bar{\pi}}_{\eta_0})I\{\eta_0\le \tau^{\bar{\pi}}\}\right]\nonumber\\
&=\E_x\left[\int^{\tau^{L\wedge \eta_0}}_0
e^{-\delta s}(\Lambda+(\gamma-\beta){l}_s)\dif s+\alpha e^{-\delta \eta_0} V_{b^*}^E(X^{L}_{\eta_0})I\{\eta_0\le \tau^{L}\}\right]=\mathcal{P}(x;L,L^{b^*}),  
\end{align*}
where the second-to-last equality follows by noticing $\bar{\pi}_s=l_s$ for $s<\eta_0$ and the last equality by \eqref{opt-problem}. By the arbitrariness of $L$, we conclude $V_{b^*}(x)=\mathcal{P}(x;L,L^{b^*})$ by virtue of $V_{b^*}(x)=\mathcal{P} (x;L^{b^*},L^{b^*})(x)$ (see  \eqref{Vb-25}). Hence, the threshold strategy $L^{b^*}$ is a Markov perfect equilibrium strategy, and the associated value function is ${V_{b^*}}(x)$ for $x \ge 0$. The explicit expression for ${V_{b^*}}(x)$ is readily available from Remark~\ref{11216-1}.
\hfill $\square$

\section{Derivations for Subsection \ref{subsec:BM-nonexp}}
\label{appendix:BM-nonexp}

Recall that $V_{b}(\cdot )$ is the bounded and continuously differentiable
solution to the following equations:
$\frac{\sigma ^{2}}{2}g^{\prime \prime }(x)+\mu g^{\prime }(x)-(\lambda+\delta)
g(x)+\lambda \alpha V_{b}^{E}(x)+\Lambda =0$ for $0< x< b,$ and $\frac{\sigma ^{2}}{2}g^{\prime \prime }(x)+(\mu -\bar{l})g^{\prime
}(x)-(\lambda+\delta) g(x)+\lambda \alpha V_{b}^{E}(x)+(\gamma-\beta) \bar{l}+\Lambda =0$ for 
$x> b$ with $g(0)=0$.
Note that $e^{\tilde{\theta} _{1} x}$ and $e^{-\tilde{\theta} _{2}x}$ form a set of linearly independent solutions to $\frac{\sigma ^{2}}{2}g^{\prime \prime }(x)+\mu g^{\prime }(x)-(\lambda+\delta) g(x)=0$.
Using the method of variation of parameters, we obtain $P_3(x; b)$ (see \eqref{P325}) as a particular solution to  $\frac{\sigma ^{2}}{2}g^{\prime \prime }(x)+\mu g^{\prime }(x)-(\lambda+\delta)g(x)+\lambda\alpha V_{b}^{E}(x)+\Lambda =0$: 
\begin{equation*}
P_{3}(x;b)=-e^{\tilde{\theta} _{1}x}\frac{2}{\sigma ^{2}}\int_{0}^{x}\frac{\left(\lambda \alpha V_{b}^{E}(s)+\Lambda \right) e^{-\tilde{\theta} _{2}s}}{(\tilde{\theta}_{1}+\tilde{\theta} _{2})e^{(\tilde{\theta} _{1}-\tilde{\theta} _{2})s}}\mathrm{d}s+e^{-\tilde{\theta} _{2}x}\frac{2}{\sigma ^{2}}\int_{0}^{x}\frac{\left( \lambda \alpha V_{b}^{E}(s)+\Lambda \right) e^{\tilde{\theta} _{1}s}}{(\tilde{\theta} _{1}+\tilde{\theta}_{2})e^{(\tilde{\theta} _{1}-\tilde{\theta} _{2})s}}\mathrm{d}s.\end{equation*}
Similarly, 
a particular solution to $\frac{\sigma ^{2}}{2}g^{\prime \prime }(x)+(\mu-\bar{l})g^{\prime }(x)-(\lambda+\delta) g(x)+\lambda \alpha V_{b}^{E}(x)+(\gamma-\beta) \bar{l}+\Lambda =0$ is given by $P_{4}(x;b)$: 
\begin{align*}
P_{4}(x;b)&=-e^{\tilde{\theta} _{3}x}\frac{2}{\sigma ^{2}}\int_{0}^{x}\frac{\left(\lambda \alpha V_{b}^{E}(s)+\Lambda +(\gamma-\beta) \bar{l}\right) e^{-\tilde{\theta} _{4}s}}{(\tilde{\theta} _{3}+\tilde{\theta} _{4})e^{(\tilde{\theta} _{3}-\tilde{\theta}_{4})s}}\mathrm{d}s\nonumber\\&+e^{-\tilde{\theta} _{4}x}\frac{2}{\sigma ^{2}}\int_{0}^{x}\frac{\left( \lambda \alpha V_{b}^{E}(s)+\Lambda +(\gamma-\beta) \bar{l}\right) e^{\tilde{\theta}_{3}s}}{(\tilde{\theta}_{3}+\tilde{\theta} _{4})e^{(\tilde{\theta}_{3}-\tilde{\theta} _{4})s}}\mathrm{d}s.\end{align*}
Therefore, we have
\begin{equation}
V_{b}(x)=
\begin{cases}
N_{1}(b)e^{\tilde{\theta} _{1}x}+N_{2}(b)e^{-\tilde{\theta} _{2}x}+P_{3}(x;b), & 0\leq x<b, \\
N_{3}(b)e^{\tilde{\theta} _{3}x}+N_{4}(b)e^{-\tilde{\theta}_{4}x}+P_{4}(x;b), & x\geq b,%
\end{cases}\label{311025-1}
\end{equation}
where $N_1(b)- N_4(b)$ are those fulfilling
$V_{b}(0)=N_{1}(b)+N_{2}(b)=0$ ,$V_{b}(b-)=V_{b}(b+)$, $V^{\prime}_{b}(b-)=V^{\prime}_{b}(b+)$, and $V_{b}(x)\le \frac{(\gamma-\beta)\bar{l}+\Lambda}{\delta} \frac{\lambda \alpha+\delta}{\lambda+\delta}$.

From \eqref{sol1-000}, we can rewrite $V_{b}^{E}(x)$ as
\begin{equation*}
V_{b}^{E}(x) =
\begin{cases}
M_{1}(b)e^{\theta _{1}x}+M_{2}(b)e^{-\theta _{2}x}+M_{3}(b), & 0\leq x<b ,\\
M_{4}(b)e^{-\theta _{4}x}+M_{5}(b), & x\geq b,%
\end{cases}%
\end{equation*}%
where $M_1(b)-M_5(b)$ are given in \eqref{M125}-\eqref{M525}. %
 Note for $0\le x<b$,
\begin{align*}
\int_{0}^{x}V_{b}^{E}(s)e^{-\tilde{\theta} _{1}s}\mathrm{d}s
=&\int_{0}^{x}M_{1}(b)e^{-(\tilde{\theta}_1-\theta_1)s}+M_{2}(b)e^{-(\tilde{\theta} _{1}+\theta_{2})s}+M_{3}(b)e^{-\tilde{\theta} _{1}s}\mathrm{d}s \\
=&\frac{M_{1}(b)}{\tilde{\theta}_1-\theta_1}(1-e^{-(\tilde{\theta}_1-\theta_1)x})+\frac{M_{2}(b)}{\tilde{\theta} _{1}+\theta _{2}}(1-e^{-(\tilde{\theta}_{1}+\theta _{2})x})+\frac{M_{3}(b)}{\tilde{\theta}_{1}}(1-e^{-\tilde{\theta} _{1}x}),\\
\int_{0}^{x}V_{b}^{E}(s)e^{\tilde{\theta} _{2}s}\mathrm{d}s
=&\int_{0}^{x}M_{1}(b)e^{(\theta _{1}+\tilde{\theta}_{2})s}+M_{2}(b)e^{(\tilde{\theta}_2-\theta_2)s}+M_{3}(b)e^{\tilde{\theta} _{2}s}\mathrm{d}s \\
=&\frac{M_{1}(b)}{\theta _{1}+\tilde{\theta} _{2}}(e^{(\theta _{1}+\tilde{\theta}_{2})x}-1)+\frac{M_{2}(b)}{\tilde{\theta}_2-\theta_2}(e^{(\tilde{\theta}_2-\theta_2)x}-1)+\frac{M_{3}(b)}{\tilde{\theta}_{2}}(e^{\tilde{\theta}_{2}x}-1).
\end{align*}

\noindent Thus, we can obtain the expression for $P_3(x; b)$ as in \eqref{P325}, along with its derivative:
\begin{align*}
P_{3}^{\prime }(x;b)=&\frac{2\Lambda }{\sigma ^{2}(\tilde{\theta} _{1}+\tilde{\theta} _{2})}(-e^{\tilde{\theta} _{1}x}+e^{-\tilde{\theta} _{2}x})
-\frac{2\lambda \alpha }{\sigma ^{2}(\tilde{\theta} _{1}+\tilde{\theta} _{2})}\left( \frac{\tilde{\theta} _{1}M_{1}(b)}{\tilde{\theta} _{1}-\theta _{1}}+\frac{\tilde{\theta} _{1}M_{2}(b)}{\tilde{\theta} _{1}+\theta _{2}}+M_{3}(b)\right) e^{\tilde{\theta}_{1}x} \\
&+\frac{2\lambda \alpha }{\sigma ^{2}(\tilde{\theta} _{1}+\tilde{\theta} _{2})}\left( \frac{\tilde{\theta} _{2}M_{1}(b)}{\theta _{1}+\tilde{\theta} _{2}}+\frac{\tilde{\theta} _{2}M_{2}(b)}{\tilde{\theta} _{2}-\theta _{2}}+M_{3}(b)\right) e^{-\tilde{\theta} _{2}x}\\
&+\frac{2\lambda \alpha }{\sigma ^{2}(\tilde{\theta} _{1}+\tilde{\theta} _{2})}\left[\left( \frac{M_{1}(b)}{\tilde{\theta} _{1}-\theta _{1}}+\frac{M_{1}(b)}{\theta _{1}+\tilde{\theta} _{2}}\right) \theta_1 e^{\theta_{1}x}
-\left( \frac{M_{2}(b)}{\tilde{\theta} _{1}+\theta _{2}}+\frac{M_{2}(b)}{\tilde{\theta} _{2}-\theta_{2}}\right) \theta_2 e^{-\theta_{2}x}\right].
\end{align*}

\noindent In the same way, we have
\begin{align*}
P_{4}(x;b)=&-\frac{2((\gamma-\beta) \bar{l}+\Lambda)}{\sigma ^{2}}\frac{e^{\tilde{\theta} _{3}x}-1}{\tilde{\theta}_{3}(\tilde{\theta} _{3}+\tilde{\theta} _{4})}
+\frac{2((\gamma-\beta) \bar{l}+\Lambda)}{\sigma ^{2}}\frac{1-e^{-\tilde{\theta} _{4}x}}{\tilde{\theta} _{4}(\tilde{\theta} _{3}+\tilde{\theta} _{4})}
+\frac{2\lambda \alpha }{\sigma ^{2}\tilde{\theta}  _{3}\tilde{\theta}  _{4}}M_{5}(b) \\
&-\frac{2\lambda \alpha }{\sigma ^{2}(\tilde{\theta} _{3}+\tilde{\theta} _{4})}\left[\left( \frac{M_{4}(b)}{\tilde{\theta} _{3}+\theta _{4}}+\frac{M_{5}(b)}{\tilde{\theta} _{3}}\right) e^{\tilde{\theta}_{3}x}
+\left( \frac{M_{4}(b)}{\tilde{\theta} _{4}-\theta_{4}}+\frac{M_{5}(b)}{\tilde{\theta}_{4}}\right) e^{-\tilde{\theta} _{4}x}\right] \\
&+\frac{2\lambda \alpha }{\sigma ^{2}(\tilde{\theta} _{3}+\tilde{\theta} _{4})}\left( \frac{M_{4}(b)}{\tilde{\theta} _{3}+\theta _{4}}+\frac{M_{4}(b)}{\tilde{\theta} _{4}-\theta _{4}}\right) e^{-\theta_{4}x}.
\end{align*}

\noindent To ensure the boundedness of $V_{b}$, we need  the coefficient of $e^{\tilde\theta_3 x}$ in \eqref{311025-1} to be $0$, which yields 
\[N_3(b)=\frac{2((\gamma-\beta) \bar{l}+\Lambda)}{\sigma ^{2}\tilde{\theta}_{3}(\tilde{\theta} _{3}+\tilde{\theta} _{4})}+\frac{2\lambda \alpha }{\sigma ^{2}(\tilde{\theta} _{3}+\tilde{\theta} _{4})}\left( \frac{M_{4}(b)}{\tilde{\theta} _{3}+\theta _{4}}+\frac{M_{5}(b)}{\tilde{\theta} _{3}}\right).\]
As a result, all the $e^{\tilde\theta_3 x}$ terms in $V_{b}$ disappear and the condition $V_{b}(0)=0$ implies $N_{2}(b)=-N_{1}(b)$. Hence,
\begin{equation*}
V_{b}(x)=%
\begin{cases}
N_{1}(b)e^{\tilde{\theta} _{1}x}-N_{1}(b)e^{-\tilde{\theta} _{2}x}+P_{3}(x;b), & 0\leq x<b, \\
N_{4}(b)e^{-\tilde{\theta} _{4}x}+P_{5}(x;b), & x\geq b,
\end{cases}
\end{equation*}
where $P_{5}(x;b)$ is given in \eqref{P525}.

  With $V_{b}(b-)=V_{b}(b+)$ and $V_{b}^{\prime }(b-)=V_{b}^{\prime}(b+)$, we can solve $N_1(b)$ and $N_4(b)$ (see \eqref{N125} and \eqref{N425}). 
Note that{
\begin{align*}
&P_{5}^{\prime}(x;b)=\frac{2((\gamma-\beta) \bar{l}+\Lambda)}{\sigma ^{2}(\tilde{\theta} _{3}+\tilde{\theta} _{4})}e^{-\tilde{\theta} _{4}x}
+\frac{2\lambda \alpha }{\sigma ^{2}(\tilde{\theta} _{3}+\tilde{\theta} _{4})}\left( \frac{\tilde{\theta} _{4}M_{4}(b)}{\tilde{\theta} _{4}-\theta_{4}}+M_{5}(b)\right) e^{-\tilde{\theta} _{4}x} \\
& \quad \quad \quad \quad \quad -\frac{2\lambda \alpha }{\sigma ^{2}(\tilde{\theta} _{3}+\tilde{\theta} _{4})}\left( \frac{M_{4}(b)}{\tilde{\theta} _{3}+\theta _{4}}+\frac{M_{4}(b)}{\tilde{\theta} _{4}-\theta _{4}}\right) \theta_{4} e^{-\theta_{4}x},
\end{align*}
and 
$ V_{b}^{\prime }(b)=-\tilde{\theta}_4 N_{4}(b)e^{-\tilde{\theta} _{4}b}+P_{5}^{\prime }(b;b)$. 
The threshold $b^*$ can be obtained by solving $-\tilde{\theta}_4 N_{4}(b)e^{-\tilde{\theta} _{4}b}+P_{5}^{\prime }(b;b)={\gamma-\beta}$. 

\section{Proofs of Section \ref{expectation}}\label{aD}

\begin{Lemma}
 For any fixed $s$, $\widetilde{L^b}(x;s)$ is the unique continuously differentiable bounded solution to the following  problem:
\begin{align}
&\frac{\sigma^2(x)}{2}g^{\prime\prime}(x) +
\mu(x)g^\prime(x)-s g(x)=0 \mbox{ for $0< x< b^*$},\quad g(0)=1, \label{1819-100}\\
&\frac{\sigma^2(x)}{2}g^{\prime\prime}(x) +
(\mu(x)-\bar{l})g^\prime(x)-s g(x)+\bar{l}(\gamma-\beta)=0 \mbox{ for $ x> b^*$}.\label{1819-200}
 \end{align}
\end{Lemma}
 \proof  Following the same lines as in Lemma \ref{28525-1}, 
 we can prove the existence and uniqueness of the stated solution  by constructing a solution and then verifying its uniqueness.

Let $g$ represent the solution. Note that $g$ is continuously differentiable on $(0,\infty)$ and it is not hard to see that $g$ is twice continuously differentiable on $(0,b)\cup(b,\infty)$ by expressing the second derivative in terms of the first derivative and the function itself by using \eqref{1819-100} and \eqref{1819-200}. Recall that $L^b$ represent the strategy with the production rate at time $t$ be $l_t^b=\bar{l}I\{X_{t}^b\ge b\}$.  Note the following dynamics for the optimally controlled process:
$  \dif X_t^b=(\mu(X_{t}^b)-\bar{l}I\{X_{t}^b\ge b\})\dif t+
  \sigma(X_{t}^b)\dif W_t$ for $t\ge 0.
$
By applying \cite[Lemma A.1]{ZhuSiuYang2020} we can obtain that for $x\ge 0$ and any $t>0$, and for some sequence of stopping times  $\{\tau_n\}$ with $\lim_{n\rightarrow\infty}\tau_n=\infty$,
\begin{align}    
  & \mathbb{E}_{x}\bigg[e^{-s(\tau^b\wedge t\wedge {\tau_n})}g(X^b_{\tau^b\wedge t\wedge {\tau_n}})\bigg] \nonumber\\
   =&\;g(x)
    +\mathbb{E}_{x}\bigg[\int_0^{t\wedge \tau^b\wedge {t_n}}e^{-st}\left(\frac12\sigma^2(X_u^b)g^{\prime\prime}(X_u^b)+(\mu(X_u^b)-\bar{l}I\{X_{u}^b\ge b^b\})g^{\prime}(X_u^b)-sg(X_u^b)\right)du\bigg]\nonumber\\
=&\;g(x),\label{7525-2}
     \end{align}
 where the last equality follows since $g$ satisfies  \eqref{1819-100} and \eqref{1819-200}.
Equation \eqref{7525-2} can be rewritten as
 \begin{align}    
 g(x)= & \mathbb{E}_{x}\bigg[e^{-s\tau^b}g(X^b_{\tau^b})I\{ t\wedge \tau_n>\tau^b\}\bigg]+\mathbb{E}_{x}\bigg[e^{-s(t\wedge \tau_n)}g(X^b_{ t\wedge \tau_n})
  I\{ t\wedge t_n\le \tau^b\}\bigg]\nonumber\\
  =& \mathbb{E}_{x}\bigg[e^{-s\tau^b}I\{ t\wedge \tau_n>\tau^b\}\bigg]+\mathbb{E}_{x}\bigg[e^{-s(t\wedge \tau_n)}g(X^b_{ t\wedge \tau_n})
  I\{ t\wedge \tau_n\le \tau^b\}\bigg],\label{241123-1}
 \end{align}
 where the last equality is obtained by using $g(0)=1$.  Furthermore, $g$ is bounded and thus $\lim_{n\rightarrow \infty}\lim_{t\rightarrow \infty}\mathbb{E}_{x}\bigg[e^{-s(t\wedge \tau_n)}g(X^b_{ t\wedge \tau_n})
  I\{ t\wedge \tau_n\le \tau^b\}\bigg]=0$. Therefore,  by letting $t\rightarrow \infty$ and $n\rightarrow \infty$ (which implies $\tau_n\rightarrow \infty$) on both sides of \eqref{241123-1} we arrive at $    
 g(x)= \mathbb{E}_{x}\bigg[e^{-s\tau^b}\bigg]=L_b(x;s)$ for $x\ge 0.$
\hfill $\square$\\

\noindent \textbf{Proof of Theorem \ref{tm:laplace}.}  
For any $s$, let $v_4(\cdot; s)$ and $v_5(\cdot;s)$ represent the solutions  to 
$
      \frac{\sigma^2(x)}{2}g^{\prime\prime}(x)
+\mu(x)g^\prime(x)-s g(x)=0$, 
under the following two sets of initial values respectively, 
$v_4(0;s)=0$ and $ v_4^\prime(0;s)=1$, and $v_5(0;s)=1$, and $v_5^\prime(0;s)=1.$ 
    The existence and uniqueness of $v_4$ and $v_5$ are guaranteed by using Theorem 5.4.2. of \cite{Krylov1996}. 
Let $v_6(x;s)$ and $u(x;s)$ be  bounded solutions to $\frac{\sigma^2(x)}{2}g^{\prime\prime}(x)+
(\mu(x)-\bar{l})g^\prime(x)-s g(x) =0$ on $[0,\infty)$ with initial value $g(0)=0$ and $g(0)=1$, respectively.
The existence of $v_6(\cdot;s)$ and $u_2(\cdot;s)$ can be  proven  by extending  the differential equation to $(-\infty,-1)\cup (0,\infty)$ and adding the boundary condition $g(-1)=1$, and then using Corollary 8.1 of  \cite{Pao1992}.
 Therefore, $\widetilde{L_b}(x;s)$ can be determined by solving the above initial value problem, which has the following representation:
 \begin{eqnarray*}
\widetilde{L_b}(x;s)=\left\{\begin{array}{ll}
C_4(b;s)v_4(x;s)+C_5(b;s)v_5(x;s)&0\le x< b,\\
C_6(b;s) v_6(x;s)+u(x;s)&x\ge b,
\end{array}\right.
\end{eqnarray*}
where $C_4(b; s)$, $C_5(b; s)$, and $C_6(b; s)$ 
are determined by solving the system
$C_4(b;s)v_4(0;s)+C_5(b;s)v_5(0;s)=1$, $C_4(b;s)v_4(b;s)+C_5(b;s)v_5(b;s)=C_6(b;s)v_6(b;s)+u(b;s)$, and $C_4(b;s)v_4^\prime(b;s)+C_5(b;s)v_5^\prime(b;s)
=C_6(b;s)v_6^\prime(b;s)+u^\prime(b;s)$,  and we have the representations $C_5(b;s) =1$, while the representations for $C_4(b; s)$ and $C_6(b; s)$ result from solving the system and are given in \eqref{C425} and \eqref{C625}, respectively.\hfill $\square$

\end{document}